\documentclass[
prb,
twocolumn,
superscriptaddress,
 amsmath,amssymb,
 aps,
longbibliography,
]{revtex4-2}

\usepackage[utf8]{inputenc}
\usepackage[T1]{fontenc}
\usepackage{graphicx}
\usepackage{dcolumn}
\usepackage{bm}

\usepackage{hyperref}

\usepackage{nameref,zref-xr}
\zxrsetup{toltxlabel}
\zexternaldocument*{supplementary_material} 

\usepackage{comment}
\usepackage{xcolor}
\usepackage{bm}

\newcommand*{\state}{\ensuremath{\nu}}
\newcommand*{\statep}{\ensuremath{ {\nu'} }}
\begin{document}

\title{Viscous heat backflow and temperature resonances in extreme thermal conductors}

\author{Jan Dragašević}
\altaffiliation{Current address: Niels Bohr Institute, University of Copenhagen (DK)}

\author{Bogdan Rajkov}
\affiliation{Theory of Condensed Matter Group, Cavendish Laboratory, University of Cambridge (UK)}

\author{Michele Simoncelli}
\email{michele.simoncelli@columbia.edu}
\affiliation{Theory of Condensed Matter Group, Cavendish Laboratory, University of Cambridge (UK)}
\affiliation{Department of Applied Physics and Applied Mathematics, Columbia University, New York (USA)}

\begin{abstract}
We demonstrate that non-diffusive, fluid-like heat transport, such as heat backflowing from cooler to warmer regions, can be induced, controlled, and amplified in extreme thermal conductors such as graphite and hexagonal boron nitride. We employ the viscous heat equations, i.e., the thermal counterpart of the Navier-Stokes equations in the laminar regime, to show with first-principles quantitative accuracy that a finite thermal viscosity yields steady-state heat vortices, and governs the magnitude of transient temperature waves. Finally, we devise strategies that exploit devices' boundaries and resonance to amplify and control heat hydrodynamics, paving the way for novel experiments and applications in next-generation electronic and phononic technologies.
\end{abstract}

\maketitle

\textit{Introduction.---}Crystals with ultrahigh thermal conductivity, such as graphite \cite{schmidt_pulse_2008,balandin_thermal_2011,fugallo_thermal_2014,machida_phonon_2020} and monoisotopic layered hexagonal boron nitride (h$^{11}$BN) \cite{jiang_anisotropic_2018,yuan_modulating_2019}, are critical for copious thermal-management applications in, e.g., electronics and phononics \cite{qian_phonon-engineered_2021,moon_hexagonal_2023}. These materials are also of fundamental scientific interest, since they can host heat-transport phenomena that violate Fourier’s diffusive law \cite{chen_non-fourier_2021,huang_observation_2022,Huberman2019,Jeong2021}.
For example, striking hydrodynamic-like phenomena such as temperature waves---where heat transiently backflows from cooler to warmer regions---have recently been observed in graphite up to $\sim$200~K \cite{Ding2022}\footnote{Specifically, Ref.~\cite{Ding2022} observed temperature oscillations at temperatures as high as 200 K in isotopically purified graphite, while the pioneering works \cite{Huberman2019,Jeong2021} observed temperature oscillations around 80-100~K in graphite at natural isotopic abundance}. While these phenomena hold great potential for heat-management technologies \cite{qian_phonon-engineered_2021,chen_non-fourier_2021}, they are weak and challenging to observe.
Thus, technologically exploiting heat hydrodynamics requires 
unraveling the fundamental physics determining its emergence and how to control it.

\begin{figure*}
\vspace*{-2mm}
	\includegraphics[width=\textwidth]{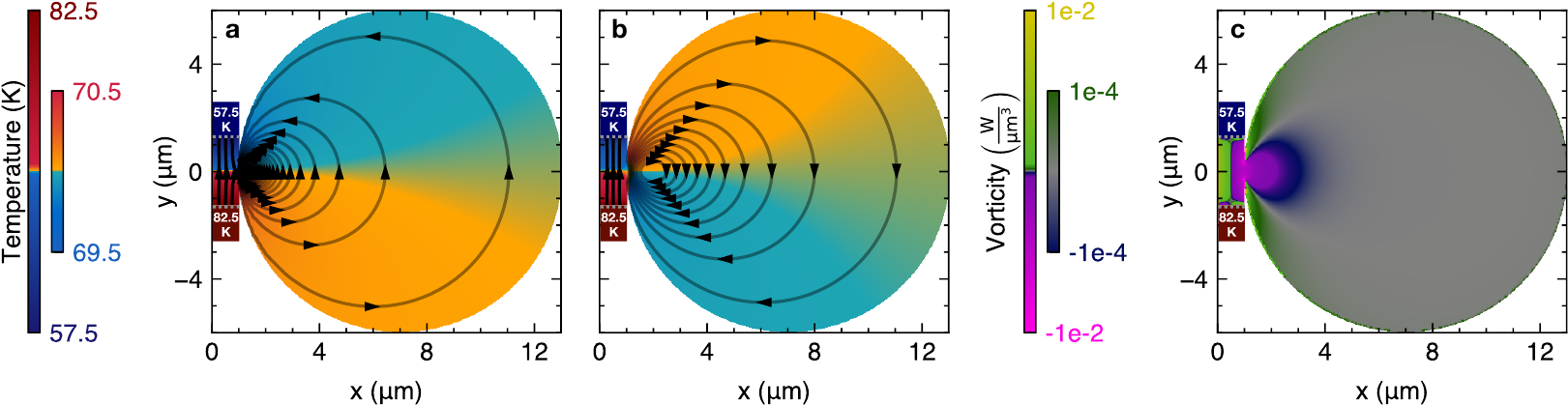}\\[-4mm]
\caption{\label{fig:1_vortex}
\textbf{Viscous heat backflow and temperature inversion.} 
In-plane ($x{-}y$) heat flow (streamlines) and temperature (colormap) in a tunnel-chamber device made of graphite. Panel \textbf{a} (\textbf{b}) shows the solution of Fourier's equation (VHE) in the presence of a temperature gradient applied to the tunnel ($T=70{\pm}12.5$ K at $y{=}{\mp} 1.25\mu m$); the other boundaries are adiabatic, i.e. $\nabla T{\cdot} \bm{\hat n}{=}0$ (and in VHE also $\bm{u}{\cdot} \bm{\hat n}{=}0$), where $\bm{\hat n}$ is the unit vector normal to the boundary. In the VHE case we also impose finite-slip boundary conditions with slip length 0.4 $\mu m$\cite{bocquetFlowBoundaryConditions2007}, corresponding to diffusive phonon-boundary scattering (see end notes for details). In Fourier's case (\textbf{a}), the direction of the temperature gradient in the chamber mirrors that in the tunnel. In contrast, the VHE (\textbf{b}) allows for the emergence of viscous backflow, whose hallmark is a temperature gradient in the chamber reversed compared to the tunnel. Panel~\textbf{c}, vorticity of the VHE heat flux, $\nabla {\times}\bm{Q}^{\rm TOT}$; the vorticity for Fourier's flux (not reported) is trivially zero. 
The density of streamlines is proportional to the magnitude of the heat flux: in Fourier (VHE) case the grey streamlines in the chamber are 100 (10) times closer to each other compared to black streamlines in the tunnel.
}
\end{figure*}

Hitherto, the theoretical analysis of heat hydrodynamics has been done relying on the linearized Peierls-Boltzmann transport equation (LBTE) \cite{peierls1955quantum} and on first-principles simulations \cite{lindsay_perspective_2019,qian_phonon-engineered_2021,chen_non-fourier_2021,chaput_direct_2013,fugallo_ab_2013,app_Lee2015,cepellotti_thermal_2016}. 
These works have provided microscopic insights on hydrodynamic heat transport in three-dimensional \cite{fugallo_thermal_2014,app_Ding2018,guo_size_2021,Ding2022,li_reexamination_2022,huang_mapping_2022,jiang_anisotropic_2018} and two-dimensional \cite{app_Cepellotti2015,cepellotti_thermal_2016,fugallo_thermal_2014,app_Lee2015,majee_dynamical_2018,raya-moreno_hydrodynamic_2022,guo_phonon_2021,zhang_emergence_2022,han_is_2023} materials, quantitatively discussing how this emerges from the predominance of momentum-conserving (normal) phonons' collisions over momentum-relaxing (Umklapp) collisions.
Under these conditions, the phonon's local equilibrium is described by the drifting distribution, $N^D_{\bm{q},s}[T(\bm{r},t),\bm{u}(\bm{r},t)]{=}\big[\exp\big(\tfrac{\hbar}{k_B T(\bm{r},t)}[\omega_{\bm{q}s}{-}\bm{q}{\cdot}\bm{u}(\bm{r},t)]\big){-}1\big]^{-1}$ ($\hbar\omega_{\bm{q}s}$ is the energy of the phonon with crystal momentum $\hbar\bm{q}$ and mode $s$), which is parametrized not only by a local temperature $T(\bm{r},t)$ (emerging from the conservation of energy in all phonon collisions), but also by a drift velocity $\bm{u}(\bm{r},t)$ that emerges from the conservation of crystal momentum in the dominant normal collisions.
While the LBTE has provided insights on heat hydrodynamic phenomena such as second sound (temperature oscillations) \cite{app_Cepellotti2015,Huberman2019,Ding2022} and Poiseuille-like heat flow \cite{app_Lee2015,cepellotti_boltzmann_2017,machida_observation_2018,sendra_hydrodynamic_2022,li_reexamination_2022,huang_observation_2022}, 
its complexity and high computational cost (stemming from microscopically resolving how every phonon mode $\bm{q}s$ interacts and contributes to transport) hinder its use in exploring new hallmarks of heat hydrodynamics.
Recent research has focused on developing mesoscopic models that capture the physics governing heat hydrodynamics at reduced complexity and cost; in practice, these are obtained by marginalizing the microscopic (phonon) degrees of freedom in the LBTE, thus describing only the evolution of local-equilibrium fields ($T$ and $\bm{u}$) \cite{li_role_2018,guo_nonequilibrium_2018,shang_heat_2020,Simoncelli2020,sendra_derivation_2021}.

\textit{Viscous heat equations.---}Ref.~\cite{Simoncelli2020} has shown how the symmetries of the microscopic LBTE imply that two transport coefficients
are needed to describe thermal transport from the diffusive to the laminar hydrodynamic regime. The first one is the well known thermal conductivity tensor $\kappa^{ij}$ ($i,j$ are Cartesian indexes), which originates from the odd-parity part of the LBTE's solution, and quantifies the response of the crystal to a temperature perturbation. 
The second one, the thermal viscosity tensor $\eta^{ijkl}$, emerges from the even-parity part; this is usually neglected but becomes relevant in the hydrodynamic regime, quantifying the response to a drift-velocity perturbation.
Conductivity and viscosity parametrize two coupled viscous heat equations (VHE) describing the evolution of $T({\bm r}, t)$ and $\bm{u}({\bm r}, t)$ in the linear regime close to equilibrium (repeated indexes are summed):
\begin{align}
&C\frac{\partial T({\bm r}, t)}{\partial t} +\alpha^{ij}\frac{\partial u^j({\bm r}, t)}{\partial r^i} - \kappa^{ij}_D \frac{\partial^2 T ({\bm r}, t)}{\partial r^i \partial r^j} = \dot{q}({\bm r}, t),
 \label{viscous_heat_T}\\
&{A}^{ij}\frac{\partial u^{j}({\bm r}, t)}{\partial t} {+} 
\beta^{ij}
\frac{\partial T({\bm r}, t)}{\partial r^j} {-} \eta^{ijkl} \frac{\partial^2 u^k({\bm r}, t)}{\partial r^j \partial r^l} {=} {-} \gamma^{ij} u^j({\bm r}, t),\label{viscous_heat_U}
\raisetag{1mm}
\end{align}\\[-5mm]
where $C$ is the specific heat; $A^{ij}{=}A^{i}\delta^{ij}$ is a matrix having zero off-diagonal elements and the specific momentum in direction $i$, $A^{i}$, on the diagonal;
$\alpha^{ij}$ and $\beta^{ij}$ are coupling tensors that originate from the relation between crystal momentum and energy for phonons; 
$\gamma^{ij}$ describes the dissipation of crystal momentum due to Umklapp processes (this term is weak in the hydrodynamic regime and becomes strong in the diffusive regime, implying that in the latter the VHE reduce to Fourier's law);
$\kappa^{ij}_D=\kappa^{ij}{-}\alpha^{ik}[\gamma^{-1}]^{kl}\beta^{lj}$ is the conductivity contribution from diffusion-damped relaxons, see Supplementary Material (SM) for details;
$\dot{q}({\bm r}, t)$ describes the energy exchange with a localized external heater.
All these parameters can be determined from the LBTE, accounting for the actual material’s phonon band structure and full collision matrix with first-principles accuracy; in the regime where phonon mean free paths are comparable to the device size, they are subject to a Bosanquet-type rescaling, analogously to rarefied fluids \cite{michalis_rarefaction_2010}\footnote{We note that the VHE have a mathematical form analogous to the damped and diffusion-extended Navier-Stokes equations, used to describe rarefied fluids flowing in porous media in isothermal and laminar conditions, with VHE's temperature mapping to fluid's pressure and VHE's drift velocity mapping to fluid velocity. 
To see this, we highlight how the linear damping term $-\gamma^{ij}u^j(\bm{r},t)$ appearing in the second VHE~(\ref{viscous_heat_U}) is analogous to the dissipative term appearing in the linearized ``damped'' Navier-Stokes equations \cite{balasubramanian_darcys_1987,dardis_lattice_1998,bresch_existence_2003,cai_weak_2008,zhang_uniqueness_2011} used, e.g., to describe a fluid flowing through a porous medium.
Then, the diffusive term appearing in the first VHE~(\ref{viscous_heat_T}) $-\kappa_D^{ij}\tfrac{\partial^2 T(\bm{r},t)}{\partial r^i\partial r^j}$ is analogous to the self-diffusion term appearing for pressure in the extended Navier-Stokes equations \cite{brenner_navierstokes_2005,sambasivam_numerical_2014,schwarz_openfoam_2023} used to describe the flow of an isothermal, compressible, and rarefied gas \cite{maurer_second-order_2003,dongari_pressure-driven_2009}.} (see SM).
Finally, we recall that the VHE predict the total heat flux to originate from both
 temperature gradient and drift velocity, i.e., $\bm{Q}^{TOT}{=}\bm{Q}^{\delta}{+}\bm{Q}^{D}$ where ${Q}^{\delta,i}{=}{-} \kappa^{ij}_D\nabla^j T$ and ${Q}^{D,i}{=}\alpha^{ij}u^j$.

The VHE encompass, as special limiting cases, Fou\-rier's equation for heat diffusion, the dual-phase-lag equation (DPLE) \cite{joseph_heat_1989,tzou_unified_1995} for second sound \footnote{The DPLE encompasses Cattaneo's second-sound equation \cite{cattaneo1958form} as a special case, as discussed in SM~\ref{sec:DPLE_derivation} and \cite{tzou_unified_1995}}, and the Guyer-Krumhansl Equation (GKE) \cite{guyer_solution_1966} for phonon hydrodynamics in idealized materials with linear-isotropic phonon dispersion.
Fourier's limit emerges when Umklapp dissipation ($\gamma^{ij}u^j(\bm{r}, t)$) dominates over viscous effects ($\eta^{ijkl}\tfrac{\partial^2 u^k(\bm{r}, t)}{\partial r^j \partial r^l}$) \cite{Simoncelli2020}, while the DPLE describes temperature waves in the transient inviscid limit ($\eta^{ijkl}{=}0$), see SM~\ref{sec:DPLE_derivation} for a proof. Both limits are obtained when viscous effects are negligible, the former in the steady-state and the latter in the transient domain. 
On the other hand, the GKE include a viscous stress term, but they cannot be applied to realistic materials with non-linear phonon dispersions such as those studied here (see SM~\ref{sec:GKE_limit}).
Thus, understanding how thermal viscosity influences hydrodynamic transport signatures in steady-state and transient regimes is an open fundamental question.
Here we address these questions, investigating from first principles how to induce, amplify, and control viscous heat hydrodynamics in graphite at various isotopic purities, as well as in h$^{11}$BN \footnote{We investigate monoisotopic h$^{11}$BN because Refs.~\cite{yuan_modulating_2019,SimoncelliPhD} suggest that heat hydrodynamics in h$^{11}$BN is stronger than in hexagonal boron nitride with natural isotopic-mass disorder (19.9\% $^{10}$B and 80.1\% $^{11}$B)}.

\textit{Steady-state viscous heat backflow.---}In Fig.~\ref{fig:1_vortex} we investigate how viscosity affects steady-state thermal transport by comparing the numerical solution of Fourier's (inviscid) equation (panel \textbf{a}) with that of the viscous VHE (panel \textbf{b}). We consider a graphitic device with a tunnel-chamber geometry \cite{e-vortexes}, which promotes vortical hydrodynamic behavior (more on this later). 
We highlight that the VHE temperature profile in the chamber is reversed compared to the profile in the tunnel, a behavior completely opposite to that predicted by Fourier's law. Panel \textbf{c} shows that this temperature inversion---which in principle can be detected in thermal-imaging experiments \cite{menges_nanoscale_2016,cheng_battery_2022,cahill_nanoscale_2014,braun_spatially_2022,ziabari_full-field_2018,reihaniQuantitativeMappingUnmodulated2022,goblotImagingHeatTransport2024}---occurs in the presence of viscous vortical flow, when heat backflows against the temperature gradient. In SM~\ref{sec:steady_state_viscous_heat_backflow_in_h} we show that heat vortices are not limited to graphite, predicting them also in h$^{11}$BN around 60 K.

To see how the emergence the heat backflow vortex in Fig.~\ref{fig:1_vortex} requires the presence of a finite thermal viscosity, we start from Eqs.~(\ref{viscous_heat_T}, \ref{viscous_heat_U}) in the steady state and focus on geometries for which transport is isotropic (e.g., graphite and h$^{11}$BN in the in-plane directions). Hereafter, Cartesian indexes will be omitted for tensors that are proportional to the identity in the in-plane directions, see SM~\ref{sec:parameters_entering_in_the_viscous_heat_equations}.
Then, considering the inviscid limit, Eq.~(\ref{viscous_heat_U}) reduces to 
$\beta\nabla T({\bm r}, t){=} {-} \gamma \bm{u}({\bm r}, t)$, and using this in Eq.~(\ref{viscous_heat_T}) readily shows that this inviscid limit is governed by a Fourier-like irrotational equation, where the total heat flux is exclusively determined by the temperature gradient and thus, vortices and backflow cannot emerge \footnote{We recall that the curl of a gradient is zero, so Fourier's heat flux $\bm{Q}^{\delta}={-} \kappa_D\nabla T$ is irrotational.}.
In contrast, when a nonzero thermal viscosity tensor is considered in Eq.~(\ref{viscous_heat_U}), the drift velocity is no longer proportional to the temperature gradient, 
and thus the total heat flux $\bm{Q}^{TOT}{=}\bm{Q}^{\delta}{+}\bm{Q}^{D}{=}{-} \kappa_D\nabla T{+}\alpha \bm{u}$ cannot be simplified to an irrotational expression. 
This demonstrates that having a nonzero viscosity is necessary to have nonzero vorticity and observe steady-state viscous heat backflow.
However, having nonzero viscosity is not sufficient to observe heat backflow; in fact, one also needs a device's geometry and boundary conditions that ensure the presence of nonzero second derivative of the drift velocity, i.e., of a total heat flux with nonzero vorticity. 
In this regard, we discuss in SM~\ref{sec:BC_steady_state} how the tunnel-chamber geometry can drive non-homogeneities in the 
drift velocity that are sufficient to produce viscous heat backflow and temperature inversion, and show in the end note how the magnitude of the temperature inversion depends on the boundary conditions for the mesoscopic fluid velocity. 
In SM~\ref{sec:effects_of_temperature_and_isotopic_disorder_on_viscous_heat_backflow} we discuss how such backflow depends on device's size, average temperature, and isotopic disorder. 

\begin{figure}[t!]
	\centering
	\includegraphics[width=\columnwidth]{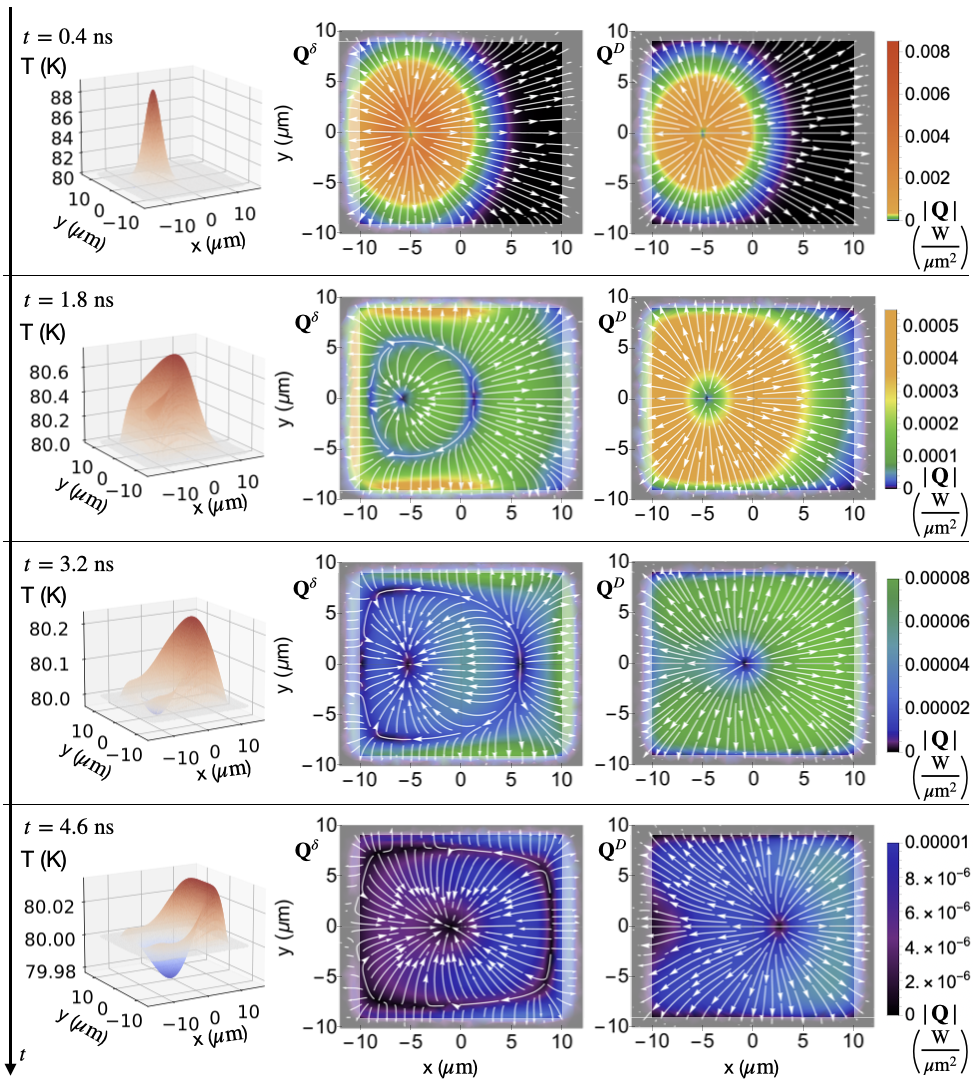}\\[-3mm]
	\caption{\textbf{Transient viscous heat backflow.}
	Columns show temperature (left), temperature-gradient heat flux ($\bm{Q}^{\delta}$, center), and drifting heat flux ($\bm{Q}^{D}$, right) at different times (relaxation starts at $t{=}0.4$ ns, see text), 
	and are obtained solving the VHE in a graphitic device thermalised at 80 K in the shaded boundary regions. 
	In the 2D plots, color is the heat-flux modulus, and white streamlines show its direction.\vspace*{-3mm}
	}
	\label{fig:fig2}
\end{figure}

\textit{Transient viscous heat backflow.---}Recent experiments in graphite have observed heat backflowing against the temperature gradient only in the time-dependent domain, in the form of second sound \cite{Huberman2019,Ding2022} or lattice cooling \cite{Jeong2021}. The pioneering theoretical analyses have been performed relying on the microscopic LBTE \cite{cepellotti_transport_2017,Huberman2019,Ding2022,zhang_transient_2021,Jeong2021}
without resolving effects induced by the thermal viscosity, or relying on the inviscid mesoscopic DPLE \cite{joseph_heat_1989,tzou_unified_1995,xu_thermal_2002,ordonez-miranda_exact_2010,kang_method_2017,gandolfi_accessing_2019,xu_thermal_2021,mazza_thermal_2021}. 
It is therefore natural to wonder how the viscous heat backflow emerging from the VHE behaves in the time domain, and, more precisely, if there is a relationship between temperature waves and transient viscous heat backflow.
Therefore, we perform the time-dependent simulation shown in Fig.~\ref{fig:fig2}. We consider a rectangular device that is initially at equilibrium ($T{=}80$ K \footnote{In this time-dependent simulation we choose an equilibrium temperature of 80 K to match the temperature at which transient hydrodynamic heat propagation has been observed in recent experiments \cite{Huberman2019,Jeong2021}.} and $\bm{u}{=}0$ everywhere); we perturb it with a heater localized at $(x_c=5\mu m,0)$ for $0{<}t{<}t_{\rm heat}{=}0.4\ \mathrm{ns}$ ($\dot{q}(\bm{r},t){=}\mathcal{H}{\cdot}\theta(t_{\rm heat}{-}t) \exp\left[{-}\tfrac{(x+x_c)^2}{2\sigma_x^2}{-}\tfrac{y^2}{2\sigma_y^2}\right]$, see Eq.~(\ref{viscous_heat_T}) and note \cite{footnote_param}); at $t{=}t_{\rm heat}$ we switch off the heater and monitor the relaxation to equilibrium. The device is always thermalised at the boundaries ($T{=}80K$ and $\bm{u}{=}\bm{0}$), see Ref.~\cite{braun_spatially_2022} for an experimental example of this boundary condition, and SM~\ref{sec:boundary_conditions_for_time_dependent_simulations} for details on how boundary conditions (average temperature and thermalisation lengthscale), size, and isotopic disorder affect the relaxation. 
The VHE evolution of the temperature field (first column in Fig.~\ref{fig:fig2}) shows an oscillatory behavior, with the local appearance of temperature values lower than the initial temperature. In contrast, we show in SM~\ref{sec:effects_of_viscosity_on_lattice_cooling} that, in Fourier's law, a positive temperature perturbation relaxes remaining non-negative with respect to the initial equilibrium value (a consequence of the smoothing
property of the diffusion equation \cite{skinner_university_nodate}).

\begin{figure}[b!]
\includegraphics[width=0.9\columnwidth]{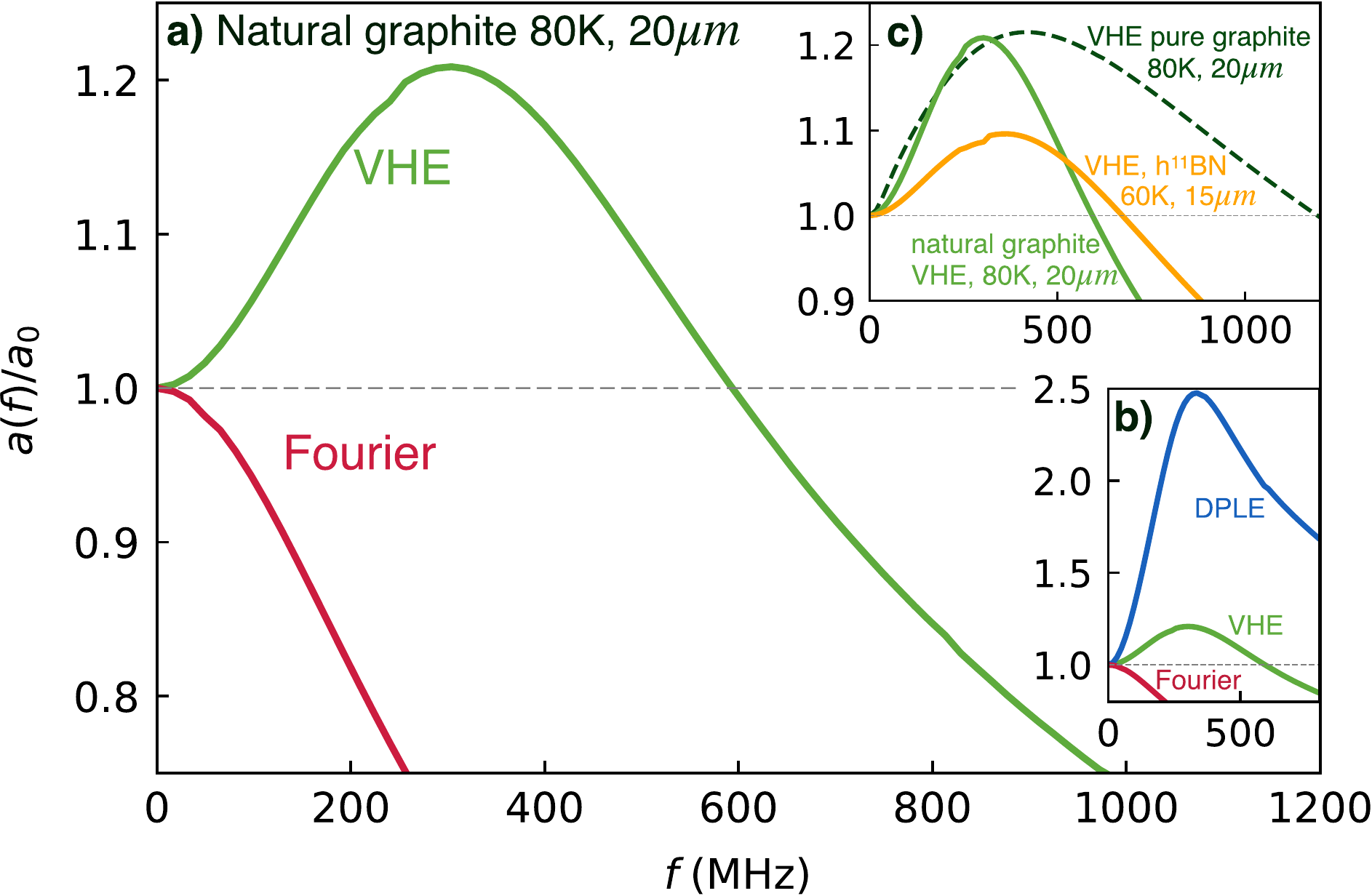}\\[-3mm]
\caption{\textbf{Resonant amplification of temperature waves} in graphite at natural isotopic disorder (98.9\% $^{12}C$, 1.1\% $^{13}C$) and around $T{=}80 K$ (\textbf{a}); the prediction from the viscous VHE is green, red is Fourier's law, and the DPLE is reported in inset (\textbf{b}). Inset (\textbf{c}), the VHE predict that isotopically purified graphite (99.9\% $^{12}C$, 0.1\% $^{13}C$, dashed dark green) features a stronger resonant amplification compared to natural graphite (solid green); in addition, analogous signatures are predicted to appear in h$^{11}$BN, around $T{=}60 K$ in 15$\mu m$-long devices.
\label{fig:rect_freq}}
\end{figure}

\begin{figure*}
\vspace*{-3mm}
\includegraphics[width=\textwidth]{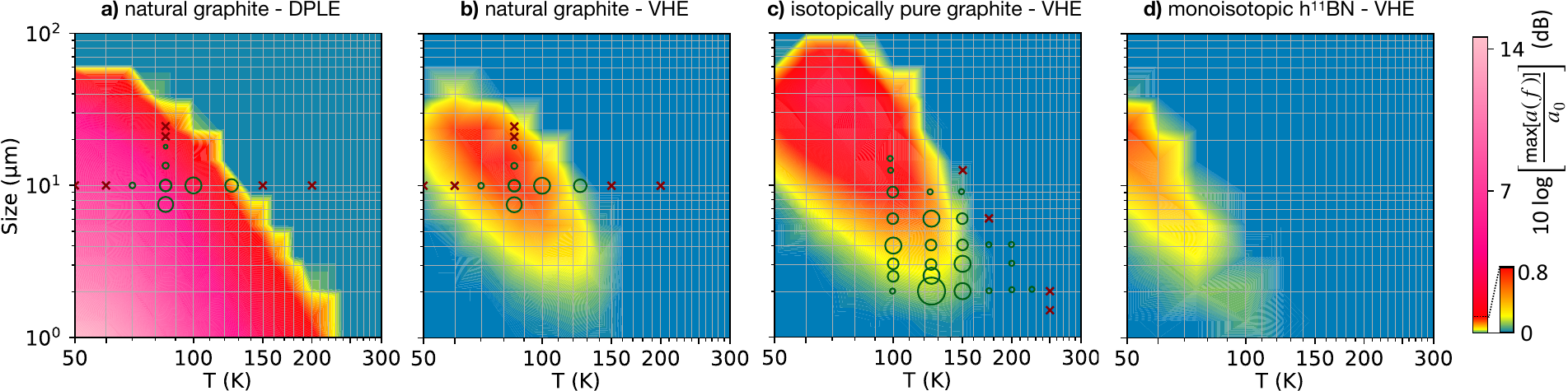}\\[-3mm]
\caption{\label{fig:freq_VHE_vs_DPLE}\textbf{Maximum resonant amplification as a function of size, temperature, and isotopic disorder.} 
The colormaps show the maximum resonant amplification (see text) as a function of the characteristic size of the rectangular device \cite{footnote_legend} and of the temperature around which the perturbation is applied. Panels \textbf{a,b)} refer to natural graphite, \textbf{a)} is the DPLE and \textbf{b)} the VHE.
Empty green circles have an area proportional to the hydrodynamic strength \cite{footnote_fig4} measured in experiments by \citet{Huberman2019} for natural samples, and by \citet{Ding2022} for isotopically purified samples; the red crosses correspond to diffusive behavior (no hydrodynamics). The VHE (\textbf{b}) capture the temperatures and lengthscales at which hydrodynamic behavior appears; in contrast, the inviscid DPLE departs from experiments at low temperature.
Panel \textbf{c)}, VHE hydrodynamic behavior in isotopically purified graphite, in broad agreement with experiments \cite{Ding2022}. Panel \textbf{d)}, VHE resonant behavior in h$^{11}$BN.
}
\end{figure*}

To understand what drives the emergence of heat backflow, we analyze the evolution of the heat-flux components $\bm{Q}^{\delta}$ and $\bm{Q}^{D}$.
Their streamlines in the second and third column of Fig.~\ref{fig:fig2} show that $\bm{Q}^{\delta}$ and $\bm{Q}^{D}$ can assume opposite directions during the relaxation, with the drifting flux $\bm{Q}^{D}{\propto}\bm{u}$ backflowing against the temperature-gradient flux $\bm{Q}^{\delta}{\propto}{-}\nabla T$. This is a consequence of the delayed coupling between $\bm{Q}^{\delta}$ and $\bm{Q}^{D}$ in the VHE~(\ref{viscous_heat_T},\ref{viscous_heat_U}), which, in the inviscid limit ($\eta{=}0$),
reduces exactly to the lagged relationship between heat flux and temperature gradient described by the dual-phase-lag model \cite{tzou_unified_1995}. Specifically, we show in SM~\ref{sec:DPLE_derivation} that in the inviscid limit $\bm{Q}^{TOT}(\bm{r},t{+}\tau_Q){=}{-}\kappa\nabla T(\bm{r},t{+}\tau_T)$, where $\tau_Q{=}A/\gamma$ is the delay between the application of a temperature gradient and the appearance of a heat flux,
and $\tau_T{=}\alpha\beta A/(\kappa\gamma^2)$ is the time needed to generate the temperature gradient from an established heat flux. 
This also shows that while steady-state heat backflow can emerge exclusively as a consequence of a finite viscosity, time-dependent heat backflow does not necessarily require viscosity to appear (SM~\ref{sec:effects_of_viscosity_on_lattice_cooling}). Nevertheless, we will show that it is necessary to consider viscous effects to rationalize experiments.

\textit{Resonant amplification of temperature waves.---}The temperature waves observed in Fig.~\ref{fig:fig2} have a small amplitude and consequently are difficult to detect. Nevertheless, they are expected to exhibit resonant amplification when driven by a perturbation periodic in time, and this could be exploited to facilitate their observation and control. 
Therefore, we investigate the behavior of the device in Fig.~\ref{fig:fig2} when driven by a periodic perturbation. 
Considering the analogies between temperature and mechanical waves, we applied to the device in Fig.~\ref{fig:fig2} a perturbation mathematically similar to the $(1,2)$ mode of a loaded rectangular membrane, i.e. $\dot{q}(\bm{r},t){=}\mathcal{H}{\cdot} [\sin(2\pi f t){+}1] \exp\big[{-}\tfrac{(x{+}x_c)^2}{2\sigma_x^2}{-}\tfrac{y^2}{2\sigma_y^2}\big]{+}$ $ {+} \mathcal{H}{\cdot} [\sin(2\pi f t {+}\pi){+}1] \exp\big[{-}\tfrac{(x-x_c)^2}{2\sigma_x^2}{-}\tfrac{y^2}{2\sigma_y^2}\big]$.
This perturbation is always non-negative, representing the laser heating employed in experiments \cite{Huberman2019,Jeong2021,Ding2022}.
Then, we monitored how the amplitude of the temperature oscillation, $a$, varies as a function of frequency, $f$. 
Fig.~\ref{fig:rect_freq} shows that the solution of the periodically driven VHE in natural graphite displays resonant behavior, i.e., plotting the oscillation amplitude as a function of frequency, $a(f)$, we see a peak reminiscent of that observed in the frequency response of a driven underdamped oscillator. We highlight how the resonant behavior obtained from the viscous VHE is weaker than that obtained from the inviscid DPLE \cite{barletta_hyperbolic_1996,xu_thermal_2002,xu_thermal_2021}, while Fourier's law does not show a resonant response (analogously to an overdamped oscillator). 
Finally, the insets show that reducing isotope disorder in graphite yields a stronger VHE resonant response, and that 
analogous signatures are predicted in h$^{11}$BN around $60 K$ in 20$\mu m$-long devices.

Next, we systematically investigate how the maximum resonant amplification vary as a function of device size, average temperature, and type of material. 
We performed simulations analogous to Fig.~\ref{fig:rect_freq} varying device's size \footnote{All the simulations shown in Fig.~\ref{fig:rect_freq} were performed accounting for grain-boundary scattering as in Ref. \cite{Simoncelli2020}, and considering a grain size of 20 $\mu$m.
This value was estimated considering the largest grains in Fig.~S3 of Ref.~\cite{Huberman2019}, Fig.~S1 of Ref. \cite{Jeong2021}, Fig.~S11 of Ref.~\cite{Ding2022}, and Fig. 1e of Ref.~\cite{yuan_modulating_2019}, which are all in broad agreement with 20 $\mu$m.} and equilibrium temperature, computing for every simulation the maximum resonant amplification as ${\rm max}_f[a(f)]/a_0$ (where $a_0{=}\lim_{f{\to} 0}a(f)$, see Fig.~\ref{fig:rect_freq}).
In Fig~\ref{fig:freq_VHE_vs_DPLE}\textbf{a,b} we compare the inviscid DPLE (\textbf{a}) with the VHE (\textbf{b}) in natural graphite. We see that the temperatures and lengthscales at which the VHE predict the emergence of hydrodynamic resonant behavior in natural samples are in broad agreement with the temperatures and lengthscales for the emergence of heat hydrodynamics discussed by \citet{Huberman2019}; in contrast, the DPLE fails to capture the reduction of hydrodynamic behavior as temperature is decreased below 100 K.
Turning our attention to isotopically purified graphite (Fig.~\ref{fig:freq_VHE_vs_DPLE}\textbf{c}), we see that here the VHE resonant response is stronger than in natural graphite, and it also persists up to larger lengthscales and higher temperatures, in broad agreement with the experiments by \citet{Ding2022}\footnote{Experiments \cite{Ding2022} for isotopically purified graphite are available only at temperatures higher than 100 K, preventing us from comparing VHE and DPLE in the low-temperature limit, where Fig.~\ref{fig:freq_VHE_vs_DPLE}\textbf{a,b} show that viscous effects are largest.}.
Finally, Fig.~\ref{fig:freq_VHE_vs_DPLE}\textbf{d} predicts that resonant behavior for viscous temperature waves also occurs in h$^{11}$BN, with a magnitude slightly weaker than in natural graphite.

\begin{figure*}
	\centering
	\includegraphics[width=\textwidth]{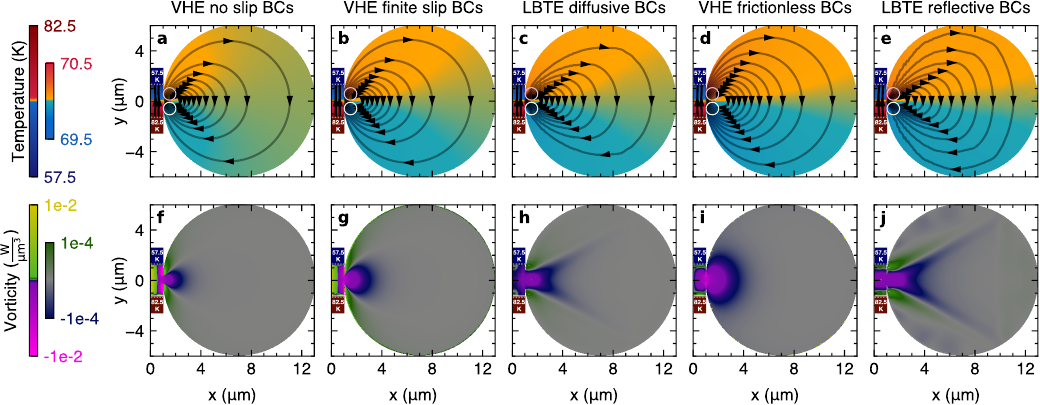}
	\caption{\textbf{Viscous heat backflow and temperature inversion from VHE \& LBTE, and dependence on BCs.}
	Top row: in-plane ($x{-}y$) heat flow (streamlines) and temperature (colormap); bottom row: vorticity. All panels show a tunnel-chamber device made of natural graphite having a temperature difference applied to the tunnel (i.e., tunnel's boundaries at $y{=}{\mp} 1.25\mu m$ are thermalized at $70\pm 12.5$ K). 
	The non-thermalized boundaries are always adiabatic---i.e., $\bm{Q}^{TOT}\cdot \bm{\hat n}=0$, where $\bm{\hat n}$ is the unit vector orthogonal to the boundary---and each column shows the solution of VHE or LBTE for BCs exerting different friction on the total heat-flux component tangential to the boundary:
	\textbf{a}, idealized no-slip BCs $\bm{u}=0$ within VHE; 
	\textbf{b}, more realistic, 0.4 $\mu m$ finite-slip-length BCs within mesoscopic VHE, corresponding to microscopic diffusive phonon-boundary scattering \cite{raya-moreno_bte-barna_2022} within LBTE (\textbf{c}, see text);
	\textbf{d} and \textbf{e} are idealized, frictionless boundaries, corresponding to zero-shear-stress BCs within VHE (\textbf{d}) and reflective phonon-boundary scattering \cite{raya-moreno_heat_nodate} within LBTE (\textbf{e}).
	The thistle and beige circle of diameter 1 $\mu m$ in the top and bottom parts of the chamber (centered at $x=1.02\mu m, \;y=\pm 0.58\mu m$), respectively, are temperature probes that contain the local maxima (thistle) and minima (beige) of the temperature profile, highlighting that the emergence of viscous backflow implies a temperature gradient in the chamber inverted compared to the tunnel, with inversion being stronger as the boundaries' friction is reduced from left to right. We highlight how VHE and LBTE yield a heat flux with closed streamlines, a sufficient condition to have nonzero vorticity, as confirmed by the bottom row.
	}
	\label{fig:BTE_revision}
\end{figure*}

\textit{Comparison between VHE and microscopic LBTE.---}We recall that the mesoscopic VHE are obtained 
coarse-graining the LBTE for the microscopic phonon distribution function and relying on homogeneous linear-response to determine the transport coefficients (i.e., neglecting gradients in space and time of the out-of-equilibrium phonon distribution) \cite{Simoncelli2020}. Therefore, the VHE are increasingly more accurate as these gradients become weaker. 
To quantitatively evaluate the accuracy of the VHE at the length- and time-scales in focus here, we compare the solution of the VHE with that of the LBTE with the full collision operator \cite{raya-moreno_bte-barna_2022}, both in the steady-state and time-dependent domains. Starting from the steady state, we show in end note Fig.~\ref{fig:BTE_revision} that the solution of the space-dependent LBTE in the geometry of Fig.~\ref{fig:1_vortex} yields a temperature inversion compatible with that predicted by the VHE.
Moreover, in SM~\ref{sec:comparison_between_vhe_and_bte} we show that the VHE and LBTE yield compatible predictions for the time-dependent relaxation of a micrometer-sized temperature perturbation occurring over the $\sim 10$ ns timescale. These analyses confirm that the length- and time-scales in focus here are sufficiently large to allow us to employ the VHE to describe heat hydrodynamics.

\textit{Conclusion.---}We have shed light on the fundamental physics determining the emergence of viscous heat hydrodynamics, discussing
with quantitative first-principles accuracy how in extreme thermal conductors such as graphite and layered h$^{11}$BN it is possible to induce temperature inversion from steady-state heat vortices, and viscous temperature waves. 
We have demonstrated that these phenomena can be amplified by engineering the device’s geometry, boundary conditions, 
or exploiting resonance, paving the way for applications in next-generation technologies.
We have provided novel, fundamental insights on temperature waves, showing that the viscous temperature waves emerging from the VHE differ fundamentally from the inviscid heat waves discussed in past works \cite{joseph_heat_1989,tzou_unified_1995,barletta_hyperbolic_1996,xu_thermal_2002,xu_thermal_2021}. We have quantitatively shown that viscous effects determine the hydrodynamic relaxation timescales \cite{Jeong2021} and lengthscales \cite{Huberman2019,Ding2022} measured in pioneering experiments. 
From a technological perspective, this work identifies the conditions under which heat conduction violates the smoothing property of Fourier's equation, opening new avenues to engineer thermal signals that mimic non-diffusive neuron transfer function for heat-based neuromorphic computing \cite{nataf_using_2024,torres_thermal_2023}.
Our results share fundamental common underpinnings with other quasiparticle’s fluid-like transport phenomena in solids---involving, e.g., electrons \cite{e-vortexes,palm_observation_2024} coupled with phonons \cite{coulter_microscopic_2018,vool_imaging_2021,yang_evidence_2021,jaoui_formation_2022,huang_electron-phonon_2021,protik_elphbolt_2022,levchenko_transport_2020,coulterCoupledElectronphononHydrodynamics2025a}, magnons \cite{wei_giant_2022}, skyrmions \cite{PhysRevLett.130.106703}---thus may inspire analogous developments and applications.

\begin{acknowledgements}
We thank Dr Miguel Beneitez, Dr Gareth Conduit, and Dr Orazio Scarlatella for commenting the manuscript.
We gratefully acknowledge Prof Samuel Huberman, Dr Aleksei Sokolov, and Enrico Di Lucente for the useful discussions.
M. S. acknowledges support from: Gonville and Caius College; the SNSF project P500PT\_203178; the Sulis Tier 2 HPC platform (funded by EPSRC Grant EP/T022108/1 and the HPC Midlands+consortium); the Kelvin2 HPC platform at the NI-HPC Centre (funded by EPSRC and jointly managed by Queen's University Belfast and Ulster University).
B. R. acknowledges support from Trinity College, Cambridge.
J. D. thanks Prof Hrvoje Jasak for the hospitality in Cambridge.
\end{acknowledgements}

\newpage

\begin{center}
  \textbf{Appendix}
\end{center}
\textbf{Temperature inversion \& backflow from LBTE}\\

In this section we solve the space-dependent Linearized phonon Boltzmann Transport Equation (LBTE) with full scattering operator in the tunnel-chamber geometry of Fig.~\ref{fig:1_vortex}. We rely on the deviational Monte Carlo solver provided in the \texttt{BTE-Barna} software package \cite{raya-moreno_bte-barna_2022}, computing the scattering operator in the three-dimensional Brillouin zone of graphite, and simulating a device effectively infinite in the off-plane direction (i.e., with transport taking place in-plane and with temperature and heat flux depending only on the in-plane coordinates $x,y$), as in Fig~\ref{fig:1_vortex}. 
Computational details are reported in SM.

Fig.~\ref{fig:BTE_revision} shows that the viscous temperature inversion and vorticity predicted by the VHE around $\bar{T}=70 $ K (Fig.~\ref{fig:1_vortex}\textbf{b}) is predicted also by the space-dependent solution of the LBTE with full collision matrix. 
The LBTE is solved starting from a uniform temperature $T(x,y)=\bar{T}$ in the device interior, applying a temperature difference of 25 K at the boundaries of the tunnel (i.e., 82.5 K at the tunnel’s boundary
at $y = {-}1.25 \mu m$, and 57.5 K at $y = 1.25 \mu m$), while all other boundaries are treated as adiabatic (i.e., heat cannot cross the boundary) and with a specified phonon-momentum reflectivity (more on this later). As time evolves, heat flows into the tunnel and enters into the chamber, and the temperature profile reaches the steady state. 
As shown in Fig.~\ref{fig:slip_cont} in SM, the system is effectively in steady state for $t\gtrsim 1.5$ ns, and Fig.~\ref{fig:BTE_revision} reports the temperature and heat flux averaged over the steady-state window, i.e., $t\in[2,5.5]$ ns.

The aforementioned adiabatic condition imposed on boundaries not in contact with thermal reservoirs is not sufficient to fully determine the problem; an additional assumption about phonon-boundary scattering is required. 
We therefore simulate the two limiting cases: diffusive and reflective boundaries \cite{raya-moreno_heat_nodate,raya-moreno_hydrodynamic_2022}. 
The diffusive case (Fig.~\ref{fig:BTE_revision}\textbf{c,h}) does not conserve phonon momentum, as the final phonon state is randomly selected from a Lambert cosine-law distribution \cite{sabattiSimulationPhononTransport2016a,raya-moreno_hydrodynamic_2022}. 
The reflective case (Fig.~\ref{fig:BTE_revision}\textbf{e,j}), in contrast, conserves the tangential component of momentum while reflecting the normal component \footnote{Implemented in \texttt{BTE-Barna} following Sec.~4.2.2.1 of Ref.~\cite{raya-moreno_heat_nodate}}. 
Comparing panels \textbf{c} and \textbf{e}, it is evident that reflective BCs yield a stronger temperature inversion than diffusive BCs (see Fig.~\ref{fig:slip_cont} in the SM), which is also accompanied by the reflective case (\textbf{j}) having vorticity larger than the diffusive case (\textbf{h}).

At the mesoscopic VHE level, the adiabatic boundary condition can be imposed by requiring that the temperature-gradient and drift-velocity heat-flux components orthogonal to the boundary vanish. The first condition on the temperature gradient ($\nabla T\cdot \bm{\hat n}=0$, where $\bm{\hat n}$ is the unit vector orthogonal to the boundary) is ubiquitously used when solving Fourier’s equation, while the second condition on the drift velocity requires additional discussion. 
In our previous work \cite{Simoncelli2020} we considered the ``no-slip'' condition for the drift velocity, i.e., $\bm{u}=\bm{0}$ at the boundary, corresponding to phonon-boundary scattering completely dissipating the crystal momentum. 
More generally, one can satisfy the adiabatic BCs by imposing $\bm{u}\cdot \bm{\hat n}=0$ (no hydrodynamic heat flux across the boundary, corresponding to the impermeability condition in fluid dynamics) while allowing the drift-velocity component tangential to the boundary to be nonzero, which microscopically corresponds to phonon-boundary scattering that does not fully dissipate crystal momentum. 

A convenient approach to simulate partially slipping boundaries with a finite slip length $b$ is to introduce a lubrication layer \cite{bocquetFlowBoundaryConditions2007} between the outer edge of the device and its interior, as in Fig. \ref{fig:device_ll}. The width of the lubrication layer $w_{\rm ll}$ is much smaller than the shortest lengthscale of the device. Partial slipping is then simulated by: (i) reducing shear and rotational components of the viscosity ($\eta_{\rm shr} = (\eta^{ijij}+\eta^{ijji})/2$, $\eta_{\rm rot} = (\eta^{ijij}-\eta^{ijji})/2$; see SM \ref{sub:finite_size} for details) within the lubrication layer by the following slip length-dependent factor \cite{bocquetFlowBoundaryConditions2007} 
\begin{equation}
	\frac{\eta_{\rm shr}}{\eta_{\rm shr, ll}} = \frac{\eta_{\rm rot}}{\eta_{\rm rot, ll}} = \frac{b}{w_{\rm ll}} + 1
	\label{eq:slip_length}
\end{equation}
and (ii) applying no-slip BCs on the side of the layer facing the device exterior. 
In practice, in this work for the tunnel-chamber geometry we choose a lubrication layer width of $w_{\rm ll}{=}0.02\,\mu\mathrm{m}$, and by changing $\eta_{\rm shr, ll}$ ($\eta_{\rm rot, ll}$) from $\eta_{\rm shr}$ ($\eta_{\rm rot}$) to 0 we progressively transition from no-slip ($b{=}0$, \textbf{a} and \textbf{f}) to finite-slip ($b{=}0.4\,\mu\mathrm{m}$, \textbf{b} and \textbf{g}) to frictionless ($b=\infty$, \textbf{e} and \textbf{j}) BCs. 
No-slip BCs are realistic when the mean free path (MFP) of the microscopic excitations is orders of magnitude smaller than the device size; finite-slip boundary conditions instead must be considered in rarefied fluids that do not satisfy this criterion (e.g., when the MFP is comparable to the device size) \cite{laurentLargeVariationBoundaryCondition2011}. 

Since the mesoscopic drift velocity originates from the average phonon momentum, the frictionless BCs can be regarded as the mesoscopic counterpart of the microscopic reflective condition. 
Comparing panels \textbf{d} and \textbf{e}, and Fig.~\ref{fig:slip_cont}\textbf{a} in the SM, we see that the temperature profiles emerging from the frictionless VHE and reflective LBTE are in quantitative agreement, confirming this. 

The slip length $0.4\,\mu\mathrm{m}$ used in panel \textbf{b} was obtained by comparing the difference in temperature inversion between the reflective (\textbf{e}) and diffusive LBTE (\textbf{c}) cases, and progressively increasing $\eta_{\rm ll}$ in Eq.~(\ref{eq:slip_length}) from zero to a value yielding the same difference between the frictionless VHE (\textbf{d}) and finite-slip VHE (\textbf{b}). 

We conclude by noting that the vorticity predicted by the finite-slip VHE (\textbf{g}) and the diffusive LBTE (\textbf{h}) is in closer agreement than that between the frictionless VHE and reflective LBTE (\textbf{h},\textbf{j}). 
In particular, the differences between frictionless VHE and reflective LBTE vorticities are more pronounced near the chamber opening, where the geometry and magnitude of the heat flux are most non-uniform. 
These vorticity discrepancies might originate from discarding spatial gradients of the out-of-equilibrium phonon distribution when coarse-graining the LBTE into the VHE \cite{Simoncelli2020}; they have a weak ($\lesssim 10\%$) effect on the agreement between the VHE and LBTE predictions for the experimentally measurable temperature inversion (Fig.~\ref{fig:slip_cont} in the SM).

Even though spatial gradients of the out-of-equilibrium phonon distribution are discarded in the VHE derivation, the VHE still encode a weakly nonlocal constitutive relation between heat flux and temperature gradient.
Specifically, we show in SM~\ref{sec:GKE_limit} that the VHE encompass the GKE as a special limit, and Ref.~\cite{gurcan_transport_2013} shows that the GKE yield a weakly nonlocal constitutive relation in the presence of a nonzero non-local length $l_\eta=\sqrt{\eta/\gamma}$.  Ref.~\cite{sendra_hydrodynamic_2022} discussed how the weakly nonlocal GKE can capture nonlocal effects emerging in nanoscale devices. 
It is also worth mentioning that nonlocal effects are not necessarily related to a nonzero viscosity; they can also emerge in the ballistic regime from the term $\nabla n^\delta_\nu(\bm{r},t)$ that is not retained in the VHE derivation \cite{allen_analysis_2018}. For the materials, geometries, and temperatures considered here, we show that accounting for the weakly nonlocal effects emerging from viscosity is sufficient to obtain good agreement with the full LBTE solution.

Finally, the derivation of the VHE focuses on the close-to-equilibrium linear regime and therefore neglects nonlinear advection terms. These have so far been discussed only (i) within the Callaway-approximated LBTE for homogeneous temperature waves in two-dimensional materials \cite{shang_unified_2022}, and (ii) in a recent study which benchmarked analytical predictions against a numerical Monte Carlo solution of the LBTE under the relaxation-time approximation \cite{yadavDerivationGeneralizedHeat2025a}. To the best of our knowledge, accounting for these nonlinear effects analytically and numerically while using the full LBTE collision matrix remains an open research problem, which is beyond the scope of this work.

\clearpage

\setcounter{figure}{0}
\renewcommand{\thefigure}{SF \arabic{figure}}
\renewcommand{\theequation}{SE \arabic{equation}}

\begin{center}
  {\Large\textbf{Supplementary Material}}
\end{center}

\section{Momentum and diffusion-damped conductivity contributions}
\label{sec:mom_contribution_k}

The starting point is the expression for thermal conductivity in terms of relaxons $|\theta^{\xi}_{\nu}\big>$ (Eq.~(13) of Ref.~\cite{cepellotti_thermal_2016}), which are eigenvectors of the full LBTE scattering matrix $\widetilde{\Omega}_{\nu,\nu'}= \widetilde{\Omega}^U_{\nu,\nu'}+\widetilde{\Omega}^N_{\nu,\nu'}$ (we highlight that $\widetilde{\Omega}_{\nu,\nu'}$ accounts for both Umklapp and normal scattering through $\widetilde{\Omega}^U_{\nu,\nu'}$ and $\widetilde{\Omega}^N_{\nu,\nu'}$, respectively):
\begin{equation}
\begin{split}
  \kappa^{ij}
&= C  \sum_{\xi=1}^{\mathcal{N}}
\big<\theta^{0}_{\nu} | v^i_{\nu} |\theta^{\xi}_{\nu}\big>\tau^\xi
\big<\theta^{\xi}_{\nu'} | {v}^j_{\nu'} | \theta^{0}_{\nu'}\big> \\
&= C 
\big<\theta^{0}_{\nu} | v^i_{\nu} |
{{\breve{\Omega}}^{-1}_{\nu,\nu'}} | {v}^j_{\nu'} | \theta^{0}_{\nu'}\big>, 
\end{split}
\label{eq:k_tot_b}
\end{equation}
where 
${{\breve{\Omega}}^{-1}_{\nu',\nu''}}=\sum_{\xi=1}^{\mathcal{N}}|\theta^{\xi}_{\nu}\big>\tau^\xi
\big<\theta^{\xi}_{\nu'} |$ is the inverse of the scattering matrix restricted to the non-null space, i.e., with the energy eigenvector $\big<\theta^{0}_{\nu}\big|\propto \omega_{\nu}$ projected out, as indicated by 
the sum starting from the relaxon index $\xi=1$.
The fact that the energy eigenvector does not contribute to Eq.~(\ref{eq:k_tot_b}) is discussed in detail in Refs.~\cite{Simoncelli2020,spohn_phonon_2006}\footnote{One can also directly see that $\big<\theta^{0}_{\nu}\big| v^i_\nu \big|\theta^{0}_{\nu}\big>\propto \sum_s\int_{\mathfrak{B}}\omega_{\bm{q}s}^2 v^i_{\bm{q}s} d^3q =0$, because obtained by integrating an odd-parity integrand over the symmetric Brillouin zone $\mathfrak{B}$.}.
The capability of Eq.~(\ref{eq:k_tot_b}) to predict the thermal conductivity in perfect agreement with other exact approaches to solving the LBTE with full collision operator, such as the variational method \cite{fugallo_ab_2013}, has been verified numerically in Ref.~\cite{cepellotti_thermal_2016}, as well as in this work \footnote{The relaxon \cite{cepellotti_thermal_2016} and variational \cite{fugallo_ab_2013} methods always provide results that are numerically compatible within 0.1\%. For example, in natural graphite at 80 K the bulk thermal conductivity obtained from the variational method \cite{fugallo_ab_2013} at 80 K is 5735.08 W/mK while with the relaxon formalism it is 5735.03 W/mK; in isotopically purified graphite (99.9\% $^{12}$C, 0.1 \%$^{13}$C) the bulk thermal conductivity at 80 K obtained from the variational method is 27484.96 W/mK while with the relaxon formalism it is 27485.09 W/mK; for h$^{11}$BN the variational method at 60 K yields 3944.20 W/mK, while the result of the relaxon method is 3944.23 W/mK.}. We also note that, for the sake of clarity, in this section we denote the sums running over relaxons explicitly.
The eigenvectors $\{\big| 
\theta^{\xi}_{\nu} \big>\}$ of the full collision matrix $\widetilde{\Omega}_{\nu,\nu'}$, and the eigenvectors $\{\big|\phi^{\iota}_{\nu} \big>\}$ of the normal scattering matrix $\widetilde{\Omega}^N_{\nu,\nu'}$ both form a complete basis set for the space of phonon distributions, hence we can express the completeness relation in two equivalent ways:
\begin{equation}
\begin{split}
\hat{1}_{\nu,\nu'} &=
\big|\theta^{0}_{\nu} \big> \big<\theta^{0}_{\nu'}  \big|+
\sum_{\xi=1}^{\mathcal{N}}\big|
\theta^{\xi}_{\nu} \big> \big<\theta^{\xi}_{\nu'} \big|\\
&=\big|\phi^{0}_{\nu} \big> \big<\phi^{0}_{\nu'}  \big|+
\sum_{l=1}^3\big|\phi^{l}_{\nu} \big> \big<\phi^{l}_{\nu'}  \big|+
\sum_{\iota=4}^{\mathcal{N}}\big|\phi^{\iota}_{\nu} \big> \big<\phi^{\iota}_{\nu'} \big|,
\end{split}
\label{eq:expansion_identity}
\raisetag{20mm}
\end{equation}
where $\mathcal{N}=N_{\bm{q}}{\cdot} 3 {\cdot} N_{\rm at}$ (with $N_{\bm{q}}$ being the number of wavevectors $\bm{q}$ used to uniformly sample the Brillouin zone, and $3 N_{\rm at}$ the number of phonon bands $s$, we recall that $\nu=(\bm{q},s)$ is a combined phonon index) is the linear size of the scattering matrix $\widetilde{\Omega}^N_{\nu,\nu'}$.
Because both the full and normal collision matrices conserve energy, the energy eigenvector is common to the two matrices, i.e., $\big|\theta^{0}_{\nu} \big>=\big|\phi^{0}_{\nu} \big>$.

This energy eigenvector describes local equilibrium \cite{allen_temperature_2018,Simoncelli2020,spohn_phonon_2006} and, therefore, does not contribute to thermal conductivity \cite{spohn_phonon_2006} or viscosity \cite{Simoncelli2020}. To determine the linear-response solution of the LBTE using the relaxon formalism, the full scattering matrix is restricted to the orthogonal complement of the null space (spanned by the energy eigenvector), and then inverted \cite{cepellotti_thermal_2016,Simoncelli2020}.
To see this, it is useful to introduce the projector $\hat P^{\perp \rm 0}_{\nu,\nu'}$ in the orthogonal complement of the energy eigenvector:
\begin{equation}
\begin{split}
\hat P^{\perp \rm 0}_{\nu,\nu'} = \sum_{\xi=1}^{\mathcal{N}}\big|
\theta^{\xi}_{\nu} \big> \big<\theta^{\xi}_{\nu'} \big|&=
\sum_{l=1}^3\big|\phi^{l}_{\nu} \big> \big<\phi^{l}_{\nu'}  \big|+
\sum_{\iota=4}^{\mathcal{N}}\big|\phi^{\iota}_{\nu} \big> \big<\phi^{\iota}_{\nu'} \big|\\
&=\hat P^{M}_{\nu,\nu'}+\hat P^{D}_{\nu,\nu'}, 
\label{eq:relation_projection}
\end{split}
\raisetag{5mm}
\end{equation}
where in the last equality we have further decomposed $\hat P^{\perp \rm 0}_{\nu,\nu'}$ into a sum of projectors onto the three-dimensional ``momentum'' subspace $\hat P^{M}_{\nu,\nu'}$ and onto the $(\mathcal{N}{-}4)$-dimensional ``diffusion-damped'' subspace $\hat P^{D}_{\nu,\nu'}$ (the origin of this name will be clear later, as we will see that the eigenvectors in this space always determine transport coefficients for the Laplacian term in the VHE, describing diffusive damping). 
To understand how the momentum eigenvectors contribute to the conductivity~(\ref{eq:k_tot_b}), we insert the projectors $\hat P^{M}$ and $\hat P^{D}$ into Eq.~(\ref{eq:k_tot_b}), obtaining:
\begin{equation}
\begin{split}
  \kappa^{ij}
&= C \big<\theta^{0}_{\nu} | v^i_{\nu} |\hat P^{\perp \rm 0}_{\nu,\nu'}| {{\breve{\Omega}}^{-1}_{\nu',\nu''}} |\hat P^{\perp \rm 0}_{\nu'',\nu'''} | {v}^j_{\nu'''} | \theta^{0}_{\nu'''}\big>,\\
&\approx C \big<\theta^{0}_{\nu} | v^i_{\nu} |\hat P^{M}_{\nu,\nu'}| {{\breve{\Omega}}^{-1}_{\nu',\nu''}} |\hat P^{M}_{\nu'',\nu'''} | {v}^j_{\nu'''} | \theta^{0}_{\nu'''}\big>\\
&+ C \big<\theta^{0}_{\nu} | v^i_{\nu} |\hat P^{D}_{\nu,\nu'}| {{\breve{\Omega}}^{-1}_{\nu',\nu''}} |\hat P^{D}_{\nu'',\nu'''} | {v}^j_{\nu'''} | \theta^{0}_{\nu'''}\big>,
\end{split}
\label{eq:k_tot_c}
\end{equation}
where the contributions from the off-diagonal quadrants $MD$ and $DM$ can be neglected because of the following physical considerations: (i) within the LBTE, zero eigenvalues are associated with conserved quantities \cite{Simoncelli2020,ziman1960electrons}; (ii) because the normal scattering matrix conserves exclusively energy and momentum, the matrix $\breve{\Omega}^N=\hat P^{\perp \rm 0} \widetilde{\Omega}^N\hat P^{\perp \rm 0}$ must have a finite, nonzero spectral gap $g$ between the zero eigenvalues associated to conservation of crystal momentum (subspace $M$) and the rest of the spectrum (subspace $D$); otherwise, other quantities would be conserved; (iii) due to the spectral gap between the momentum eigenvectors in $M$ and the other eigenvectors in $D$, and the perturbative character of $\breve{\Omega}^{U}$  compared to the gap $g$ in the spectrum of $\breve{\Omega}^{N}$ (denoting $\|\breve{\Omega}^{U}\|=\lambda$, we are in the regime  $\frac{\lambda}{g}\ll 1$)
it is possible to perform a unitary Schrieffer-Wolff (SW) transformation \cite{winklerSpinOrbitCouplingEffects2003,bravyiSchriefferWolffTransformation2011} to separate the fast normal-process timescale from the slow Umklapp scattering timescale and
greatly simplify the problem.
In particular, to apply the SW transformation, we determine the generator $S$ by solving $[\breve{\Omega}^N,S]=-\breve{\Omega}^U_{od}$ (here $\breve{\Omega}^U_{od}{=}\begin{bmatrix}
0 &  \breve{\Omega}^{U}_{MD}\\[2pt]
\breve{\Omega}^{U}_{DM} & 0
\end{bmatrix}$ is the off-diagonal part of $\breve{\Omega}^U$), obtaining
$S{=}
\begin{bmatrix}
0 &  \breve{\Omega}^{U}_{MD}[\breve{\Omega}^{N}_{DD}]^{-1} \\[2pt]
-[\breve{\Omega}^{N}_{DD}]^{-1}\breve{\Omega}^{U}_{DM} & 0
\end{bmatrix}
$, which satisfies $S^\dagger=-S$ and therefore $U=e^{-S}$ is unitary. 
Then, applying the SW transformation to  ${\breve{\Omega}}_{\nu',\nu''}{=}{\breve{\Omega}}^{N}_{\nu',\nu''}{+}{\breve{\Omega}}^{U}_{\nu',\nu''}$ yields
$\breve{\Omega}'=e^{-S} [\breve{\Omega}^N+\breve{\Omega}^U]e^{S}$ 
which is approximately block-diagonal in $MM$ and $DD$ (it can be verified that off-diagonal blocks $MD$ and $DM$ are $\mathcal{O}(\frac{\lambda^2}{g})$ and therefore can be neglected when $\lambda/g \ll 1$):\\[-5mm]
\begin{equation}
	\breve{\Omega}'=
\begin{bmatrix}
\breve{\Omega}^{U}_{MM}+\mathcal{O}(\frac{\lambda^2}{g}) & \mathcal{O}(\frac{\lambda^2}{g})\\[2pt]
\mathcal{O}(\frac{\lambda^2}{g}) & \breve{\Omega}^{N}_{DD}+\breve{\Omega}^{U}_{DD}+\mathcal{O}(\frac{\lambda^2}{g})
\end{bmatrix}.
\label{eq:full_matrix}
\end{equation}
We highlight how 
Umklapp processes lift the zero eigenvalues in 
the $M$ block, which becomes $[\breve{\Omega}^{U}_{MM}]^{ij}=D_U^{ij}\sim\mathcal{O(\lambda)}$ (here,
$D^{ij}_U=\big<\phi^i_\nu|\breve{\Omega}^{U}_{\nu,\nu'}|\phi^j_{\nu'}\big>=\frac{1}{V^2}\sum_{\nu,\nu'} \phi^i_\nu \breve{\Omega}^{U}_{\nu,\nu'} \phi^j_{\nu'}$
is the Umklapp dissipation rate introduced in Ref.~\cite{Simoncelli2020}). In the special case of isotropic material properties, degeneracies are preserved with $D_U^{ij}$ proportional to the identity, so that the momentum eigenvectors of $\breve{\Omega}^{N}$ are still eigenvectors of $\breve{\Omega}'$ up to corrections $\mathcal{O}(\lambda^2/g)$.
More generally, for anisotropic materials such as graphite or hBN, we find that $\breve{\Omega}^{U}_{MM}$ is practically diagonal (the magnitude of diagonal elements is $\gtrsim 10^2$ times that of the off-diagonal elements) in the basis of the momentum eigenvectors, with in-plane eigenvalues remaining degenerate and clearly distinct from the out-of-plane eigenvalue (see, e.g., Tab. I and Fig. 9 in Ref.~\cite{Simoncelli2020}).
Focusing on the $D$ block, we see that: it contains a $\mathcal{O}(1)$ term that remains nonzero also for $\lambda\to 0$; in general, it is not diagonalized by the eigenstates of $\breve{\Omega}^{N}$, but as we will see later, this is not a problem for our purposes. 

The SW-transformed collision matrix~(\ref{eq:full_matrix}) is symmetric, positive definite, and defined in partitioned form; its inverse can be determined using the expression for the inverse of a partitioned matrix \cite{hornTopicsMatrixAnalysis1991}
\footnote{
Specifically, Eq.~(0.7.3.1) of Ref.~\cite{hornTopicsMatrixAnalysis1991} shows that for a partitioned nonsingular matrix A, 
\begin{equation}A=\left[\begin{array}{ll}
A_{11} & A_{12} \\
A_{21} & A_{22}
\end{array}\right]
\end{equation}
its inverse can be written in partitioned form as:
  {\tiny
\begin{equation}
A^{-1}{=}\left[\begin{array}{lc}
\left(A_{11}-A_{12} A_{22}^{-1} A_{21}\right)^{-1} & A_{11}^{-1} A_{12}\left(A_{21} A_{11}^{-1} A_{12}-A_{22}\right)^{-1} \\
A_{22}^{-1} A_{21}\left(A_{12} A_{22}^{-1} A_{21}-A_{11}\right)^{-1} & \left(A_{22}-A_{21} A_{11}^{-1} A_{12}\right)^{-1}
\end{array}\right]
\end{equation}}}, 
obtaining,
\begin{widetext}
\begin{equation}
	\begin{split}
\breve{\Omega'}^{-1}&=\!\!\left[\begin{array}{lc}
\left(\breve{\Omega'}_{MM}{-}\breve{\Omega'}_{MD} [\breve{\Omega'}_{DD}]^{{-}1} \breve{\Omega'}_{DM}\right)^{{-}1} & [\breve{\Omega'}_{MM}]^{{-}1} \breve{\Omega'}_{MD}\left(\breve{\Omega'}_{DM} [\breve{\Omega'}_{MM}]^{{-}1} \breve{\Omega'}_{MD}{-}[\breve{\Omega'}_{DD}]\right)^{{-}1} \\
{[\breve{\Omega'}_{DD}]^{{-}1} }\breve{\Omega'}_{DM}\left(\breve{\Omega'}_{MD}[\breve{\Omega'}_{DD}]^{{-}1} \breve{\Omega'}_{DM}{-}\breve{\Omega'}_{MM}\right)^{{-}1} & \left(\breve{\Omega'}_{DD}{-}\breve{\Omega'}_{DM} [\breve{\Omega'}_{MM}]^{{-}1} \breve{\Omega'}_{MD}\right)^{{-}1}
\end{array}\right]\\
&=\left[\begin{array}{lc}
  [\breve{\Omega'}_{MM}]^{-1}
+\mathcal{O}(\frac{\lambda^2}{g^3}) & \mathcal{O}(\frac{\lambda}{g^2})\\
\mathcal{O}(\frac{\lambda}{g^2}) & {[\breve{\Omega'}_{DD}]^{{-}1}}+\mathcal{O}(\frac{\lambda^3}{g^4})
\end{array}\right],
\end{split}
\raisetag{5mm}
\end{equation}
\end{widetext}
where $[\breve{\Omega'}_{MM}]^{{-}1} {=} D_U^{-1}{+}\mathcal{O}(g^{-1})$ and 
$[\breve{\Omega'}_{DD}]^{{-}1} {=} \mathcal{O}(g^{-1})$.
In summary, relying on the SW transformation, we have rigorously justified Eq.~(\ref{eq:k_tot_c}), showing that when the timescale of Umklapp processes is long compared to the spectral-gap timescale  $\lambda/g \ll 1$, in the conductivity decomposition into momentum and diffusion-damped contributions the off-diagonal blocks $MD$ and $DM$ are negligible compared to the diagonal ones. From a mathematical viewpoint, the SW transformation allows us to replace the full matrix $\breve{\Omega}^{-1}$ with the matrix containing only its diagonal blocks $MM$ and $DD$, $[\breve{\Omega'}]^{-1}$, neglecting $\mathcal{O}(\lambda/g^2)$ corrections.

We can now estimate the thermal conductivity originating from the momentum subspace appearing in Eq.~(\ref{eq:k_tot_c}), obtaining:
\begin{equation}
\begin{split}
\kappa^{ij}_{MM}&=C \big<\theta^{0}_{\nu} | v^i_{\nu} |\hat P^{M}_{\nu,\nu'}| {{\breve{\Omega}}^{-1}_{\nu',\nu''}} |\hat P^{M}_{\nu'',\nu'''} | {v}^j_{\nu'''} | \theta^{0}_{\nu'''}\big>\\
&= \sum_{k,l=1}^3 C 
W^i_{0k} W^j_{l0} \Big([D^{-1}_U]^{kl}+[\mathcal{O}(g^{-1})]^{kl}\Big)\\
&= \kappa^{ij}_{M} +\Delta \kappa^{ij}_{M},
\label{eq:k_momentum_0}
\end{split}
\end{equation}
where we relied on the momentum-relaxon velocity defined in \cite{Simoncelli2020}, $W^i_{0k}=\big<\theta^0_\nu|v^i_\nu|\phi^k_\nu\big>=\frac{1}{V}\sum_\nu \theta^0_\nu v^i_\nu \phi^k_\nu$ (here $V=\mathcal{V}N_{\bm{q}}$, where $\mathcal{V}$ is the unit cell volume and $N_{\bm{q}}$ the number of $\bm{q}$ points used to sample the Brillouin zone).
Importantly, in Eq.~(\ref{eq:k_momentum_0}) $\kappa^{ij}_{M}{\sim} \mathcal{O}(\lambda^{-1})$ and $\Delta \kappa^{ij}_{M}{\sim} \mathcal{O}(g^{-1})$, implying that the first term dominates over the second, i.e.,
\begin{equation}
  \kappa^{ij}_{MM}=\kappa^{ij}_{M}\left(1+ \frac{\Delta \kappa^{ij}_{M}}{\kappa^{ij}_{M}} \right)= \kappa^{ij}_{M} + \mathcal{O}\left(\frac{\lambda}{g}\right).
  \label{eq:k_momentum_1}
\end{equation}
It follows that when the strength of Umklapp process $\lambda$ is weak compared to the spectral gap $g$, $\frac{\lambda}{g}{\ll} 1$, one can approximate 
\begin{equation}
\kappa^{ij}_{MM}{\approx}\kappa^{ij}_{M}=\sum_{k,l=1}^3\alpha^{ik}[\gamma^{-1}]^{kl}\beta^{lj},
\label{eq:k_momentum}
\end{equation}
where we used the definitions $\alpha^{ij}{=}W_{0j}^i\sqrt{\overline T A^jC}$, $\beta^{ij}{=}\sqrt{\frac{CA^i}{\overline T}} W_{i0}^j$, and $\gamma^{ij}{=}\sqrt{A^i A^j} D_U^{ij}$ introduced to simplify the notation in Eqs.~(\ref{viscous_heat_T}, \ref{viscous_heat_U}) (see Sec.~\ref{sec:parameters_entering_in_the_viscous_heat_equations}).

The remaining conductivity component from the diffusion-damped subspace is determined following Eq.~(\ref{eq:k_tot_c}), 
by taking the difference between the full conductivity $\kappa^{ij}$ (discussed in Refs.~\cite{cepellotti_thermal_2016,Simoncelli2020}) and the momentum contribution~(\ref{eq:k_momentum}), i.e., 
\begin{equation}
	{\kappa}^{ij}_{D, \rm bulk}={\kappa}^{ij}-\kappa^{ij}_{M}.
	\label{eq:kappaD_def}
\end{equation}
In summary, by relying on the SW transformation we have shown that it is possible to rigorously separate the total conductivity into contributions from momentum ($\kappa^{ij}_{M}$) and diffusion-damped ($\kappa^{ij}_{D}$) subspaces.
The numerical evaluation of these two thermal-conductivity contributions requires only the full collision matrix (its eigenvectors and eigenvalues), the three momentum eigenvectors, and the phonon group velocities.

When these considerations are taken into account in the derivation of the VHE, they yield the same mathematical form but a more accurate estimate of the coefficient, solving the mismatch mentioned in Ref.~\cite{Simoncelli2020} between the value of effective conductivity resulting from the VHE, and the one used in Fourier's law. 
It is also worth mentioning that, 
because the separation between momentum and diffusion-damped conductivity (\ref{eq:kappaD_def}) preserves, by construction, the value of the total conductivity, it can also be applied in the regime of strong Umklapp scattering where the diffusion-damped conductivity dominates and the VHE reduce to Fourier's law.

To see this, we repeat the key steps of the derivation of the VHE reported in Appendix D1 of Ref.~\cite{Simoncelli2020}.
After symmetrizing the LBTE (Eq.~(1) in Ref.~\cite{Simoncelli2020}), we write the phonon  distribution function $\tilde{n}_{\state}(\bm{r},t)$ appearing in it in the basis of the eigenvectors of the normal scattering operator $\big|\phi_{\state}^{\beta}\big>$, i.e., $\tilde{n}_{\state}(\bm{r},t) = \sum_\beta z_{\beta}(\bm{r},t) \big|\phi_{\state}^{\beta}\big>$, obtaining:
\begin{align}
     \sum_\beta & \frac{\partial  z_\beta(\bm{r},t) }{\partial t}\big|\phi_{\state}^\beta\big> +\bm{v}_{\state}\cdot \bigg(\sum_\beta \nabla z_\beta(\bm{r},t) \big|\phi_{\state}^\beta\big> \bigg) = \nonumber \\
     =& - \sum_{\beta>3} \frac{z_\beta(\bm{r},t)}{\tau^N_\beta} \big|\phi_{\state}^\beta\big> -\frac{1}{V} \sum_{\statep, \beta>0} z_\beta(\bm{r},t) \tilde{\Omega}^U_{\state \statep}\big|\phi_{ \statep}^\beta\big> \;.
     \label{eq:LBTE_normal_eigenvectors}
     \raisetag{2mm}
\end{align}
Then, we recall from Eq.~(4) in Ref.~\cite{Simoncelli2020} that the phonon population expansion contains special terms related to the local-equilibrium temperature $T(\bm{r})$ and drift velocity $\bm{u}$:
\begin{align}
\label{eq:fundamental_projections_2}
  \tilde{n}_{\state}&(\bm{r},t)
  = \tilde{n}^T_{\state}(\bm{r},t) + \tilde{n}^D_{\state}(\bm{r},t) + \tilde{n}^{\delta}_{\state}(\bm{r},t) \nonumber \\
  =& \sqrt{\frac{C}{k_B \bar{T}^2}} \big|\phi^0_{\state}\big> (T(\bm{r},t)-\bar{T}) \nonumber
  + \sum_{i=1}^3 \sqrt{\frac{A^i}{k_B\bar{T}}} \big|\phi^i_{\state}\big> u^i (\bm{r},t)  \\
  &+ \sum_{\beta>3} \big|\phi^\beta_{\state}\big> z_\beta (\bm{r},t)  \;,
\end{align}
where we see that coefficients $z_\beta$ in front of the energy and momentum eigenvectors are $z_0 (\bm{r},t) = \sqrt{\frac{C}{k_B\bar{T}^2}} (T(\bm{r},t)-\bar{T})$ and 
    $z_i (\bm{r},t) = \sqrt{\frac{A^i}{k_B\bar{T}}} u^i(\bm{r},t), \; i=1,2,3 $ respectively.
Taking the scalar product between the LBTE~(\ref{eq:LBTE_normal_eigenvectors}) and the energy eigenvector $\big<\phi^0_{\state}\big|$, we obtain \cite{Simoncelli2020} 
\begin{align}
  \sqrt{\frac{C}{k_B\bar{T}^2}} \frac{\partial T(\bm{r},t)}{\partial t} 
  &+ 
  \sum_{i,j=1}^3 \sqrt{\frac{A^i}{k_B\bar{T}}} W_{0i}^j \frac{\partial u^i(\bm{r},t)}{\partial r^j} + \nonumber \\
  &+ 
  \sum_{\beta>3} \bm{W}_{0 \beta}\cdot \nabla z_\beta(\bm{r},t) = 0 \;.
\label{eq:zero_eigenvector_proj}
\end{align}
where we used the above expressions for $z_0 (\bm{r},t)$ and $z_i (\bm{r},t), \; i=1,2,3$.

In the linear-response regime that yields Eq.~(\ref{eq:k_tot_b}), one has that the coefficients in front of the eigenvectors in the subspace $D$ are proportional to the temperature gradient \cite{cepellotti_thermal_2016},
$z_\beta(\bm{r},t)= \sum_i \nabla^i T f^i_\beta $. 
It follows that we can rewrite the last term in Eq.~(\ref{eq:zero_eigenvector_proj}) as:
\begin{equation}
  \begin{split}
&\sum_{\beta>3} \bm{W}_{0 \beta}\cdot \nabla z_\beta(\bm{r},t)=
\sum_{\beta>3}\sum_{i,j=1}^3 {W}^i_{0 \beta}\cdot f^j_\beta \nabla^i\nabla^j T\\
&{=}
\sum_{i,j=1}^3\!\!\left[\!\sum_{\alpha>0}C {w}^i_{0 \alpha}{w}^j_{0 \alpha} \tau_\alpha {-}
\sum_{k,l=1}^3 C{W}^i_{0 k}{W}^j_{l0} [D_U^{-1}]_{k,l}\!\right]\!\!\nabla^i\nabla^j T,      
  \end{split}
  \raisetag{22mm}
  \label{eq:compare}
\end{equation}
where $w^i_{0\alpha}= \langle \theta^0_\nu \vert v^i_\nu \vert \theta^\alpha_\nu\rangle$ and $\tau_\alpha$ are the standard relaxon velocities and lifetimes \cite{cepellotti_thermal_2016,Simoncelli2020}.
Comparing Eq.~(\ref{eq:compare}) with Eqs.~(\ref{eq:k_momentum},\ref{eq:kappaD_def}), we see that this is exactly equal to $-\sum_{i,j=1}^3 \kappa^{ij}_D \frac{\partial^2 T(\bm{r},t)}{\partial r^i\partial r^j} $. 
Therefore, we have seen that the mathematical form of the VHE remains unchanged but the coefficient that has to be used in the VHE equation for temperature~(\ref{viscous_heat_T}) is $\kappa^{ij}_D$ defined in Eq.~(\ref{eq:kappaD_def}), i.e.,
\begin{equation}
\begin{aligned}
C \frac{\partial T(\bm{r},t)}{\partial t} 
  &+ \sum_{i,j=1}^3  W_{0j}^i \sqrt{\bar{T}A^j C} \frac{\partial u^j(\bm{r},t)}{\partial r^i} \\
  & - \sum_{i,j=1}^3 \kappa^{ij}_D \frac{\partial^2 T(\bm{r},t)}{\partial r^i\partial r^j} = 0 \;.
\label{eq:mesoscopic_energy}
\end{aligned}
\end{equation}
The derivation of the VHE for the drift velocity~(\ref{viscous_heat_U}), and of the coefficients appearing in it, are both unaffected by this reasoning.
Importantly, we note that the refinement discussed here solves the problem mentioned in Ref.~\cite{Simoncelli2020} of having an effective VHE conductivity numerically different from the one used in Fourier's law, even in the diffusive regime.
To see this, we consider the drift-velocity~(\ref{viscous_heat_U}) equation in the high-temperature limit, where the Umklapp dissipation term $\propto D_U^{ij}$ dominates over viscous effects, obtaining
$\beta^{ij} \frac{\partial T({\bm r}, t)}{\partial r^j} {=} {-} \gamma^{ij} u^j({\bm r}, t)$. Recalling that $\beta^{ij}{=}\sqrt{\frac{CA^i}{\overline T}} W_{i0}^j$ and $\gamma^{ij}{=}\sqrt{A^i A^j} D_U^{ij}$ \cite{Simoncelli2020}, and inserting this relation between $\bm{u}$ and $\nabla T$ in Eq.~(\ref{eq:mesoscopic_energy}), we obtain
\begin{equation}
\begin{split}
  C \frac{\partial T(\bm{r},t)}{\partial t} 
  &- \sum_{i,j=1}^3 \bigg[C\sum_{k,l=1}^3 W_{0k}^i W_{l0}^j [D_U^{-1}]_{k,l}\bigg]\frac{\partial^2 T(\bm{r},t)}{\partial r^i\partial r^j} \\
  & - \sum_{i,j=1}^3 \kappa^{ij}_D \frac{\partial^2 T(\bm{r},t)}{\partial r^i\partial r^j} = 0 \;.
\label{eq:mesoscopic_energy_diffusive}
\end{split}
\raisetag{8mm}
\end{equation}
and by direct comparison, we verify that the term in square brackets is exactly $\kappa^{ij}_M$ defined in Eq.~(\ref{eq:k_momentum}), and its sum with the diffusion-damped conductivity $\kappa^{ij}_D$ yields the total conductivity $\kappa^{ij}$ used in Fourier's law (Eq.~\ref{eq:kappaD_def}). This implies that in the regime of strong Umklapp dissipation, Eq.~(\ref{eq:mesoscopic_energy}) analytically reduces to Fourier's heat equation. In formulas, it is straightforward to verify that Eq.~(\ref{eq:mesoscopic_energy}) is $C \frac{\partial T(\bm{r},t)}{\partial t} -  \sum_{i,j=1}^3 \kappa^{ij} \frac{\partial^2 T(\bm{r},t)}{\partial r^i\partial r^j} = 0$.

We conclude by noting that a more general discussion of this derivation is reported in Ref.~\cite{coulterCoupledElectronphononHydrodynamics2025a}, and shows that this separation between momentum and diffusion-damped eigenspaces is necessary to obtain Gurzhi's equation for electrons' hydrodynamics \cite{gurzhiHYDRODYNAMICEFFECTSSOLIDS1968,bandurin_negative_2016} from the relaxon representation of the electron-phonon Boltzmann transport equation.

\section{DPLE as inviscid limit of the VHE}
\label{sec:DPLE_derivation}
In this section we show analytically that the dual-phase-lag equation (DPLE) \cite{joseph_heat_1989,tzou_unified_1995}, widely used to describe temperature waves \cite{xu_thermal_2002,ordonez-miranda_exact_2010,kang_method_2017,gandolfi_accessing_2019,xu_thermal_2021,mazza_thermal_2021}, emerges from the viscous heat equations (VHE, see Eqs. (\ref{viscous_heat_T},~\ref{viscous_heat_U}) and \cite{Simoncelli2020}) in the limit of vanishing viscosity.
To see this, we start from the VHE without the viscosity and source term,
\begin{eqnarray}
C\frac{\partial T(\bm{r},t)}{\partial t} {+} \alpha^{ij}\frac{\partial u^j(\bm{r},t)}{\partial x^i}{-}  \kappa_D^{ij}\frac{\partial^2}{\partial x^i \partial x^j} T(\bm{r},t) {=} 0,\hspace*{6mm}\label{eq:e1}
\\
A_k\frac{\partial u_k(\bm{r},t)}{\partial t} + \gamma^i_k u^i(\bm{r},t) + \beta^i_k\frac{\partial T(\bm{r},t)}{\partial x^i}= 0,\hspace*{10mm}\label{eq:e2}
\end{eqnarray}
where the indexes $i,k$ denote Cartesian directions.
Then we consider a setup where the tensors appearing in Eqs.~(\ref{eq:e1},\ref{eq:e2}) can be considered diagonal and isotropic (as is the case for a device mode of graphite or h$^{11}$BN with non-homogeneities exclusively in the basal plane, see Sec.~\ref{sec:parameters_entering_in_the_viscous_heat_equations}).
Thus, writing the second equation by components for the $2$-dimensional case, we have:
\begin{eqnarray}
C\frac{\partial T}{\partial t}+ \alpha\left(\frac{\partial u_x}{\partial x} + \frac{\partial u_y}{\partial y}\right) - \kappa_D\left(\frac{\partial^2 T}{\partial x^2} + \frac{\partial^2 T}{\partial y^2}\right) = 0,\label{eq:d1} \hspace*{8mm}\\
A\frac{\partial u_x}{\partial t} + \gamma u_x + \beta\frac{\partial T}{\partial x} = 0,\label{eq:d2}\hspace*{20mm}\\
A\frac{\partial u_y}{\partial t} + \gamma u_y + \beta\frac{\partial T}{\partial y} = 0.\label{eq:d3}\hspace*{20mm}
\end{eqnarray}
Differentiating both sides of Eq.~(\ref{eq:d2}) with respect to $x$, and both sides of Eq.~(\ref{eq:d3}) with respect to $y$, we obtain:
\begin{eqnarray}
\left(A\frac{\partial}{\partial t} + \gamma\right) \frac{\partial u_x}{\partial x}  = - \beta\frac{\partial^2 T}{\partial x^2} \label{eq:bf1}; \\
\left(A\frac{\partial}{\partial t} + \gamma\right) \frac{\partial u_y}{\partial y}  = - \beta\frac{\partial^2 T}{\partial y^2}\label{eq:bf2}.
\end{eqnarray}

Then, we sum Eq.~(\ref{eq:bf1}) and Eq.~(\ref{eq:bf2}), obtaining:
\begin{eqnarray}
\left(A\frac{\partial}{\partial t} {+} \gamma\right)\!\! \left(\frac{\partial u_x}{\partial x}{+}
\frac{\partial u_y}{\partial y}\right)\!
 {=} {-} \beta\bigg(\frac{\partial^2 T}{\partial x^2}{+}\frac{\partial^2 T}{\partial y^2}\bigg).\hspace*{5mm}
 \label{eq:sum_2eqs}
\end{eqnarray}
Now we notice that applying the operator $\hat{\mathcal O}=\left(A\frac{\partial}{\partial t} {+} \gamma\right)$ to both sides of Eq.~(\ref{eq:d1}), recalling that $\kappa=\kappa_D+\alpha\beta/\gamma$, and by using Eq.~(\ref{eq:sum_2eqs}), allows us to write an equation for temperature only:
\begin{equation}
\begin{gathered}
\frac{CA}{\gamma} \frac{\partial^2 T}{\partial t^2} {+} C\frac{\partial T}{\partial t} {-} \kappa\!\left[\!\frac{\partial^2 T}{\partial x^2} {+}\frac{\partial^2 T}{\partial y^2}\!\right]\!  {-}\frac{\alpha\beta A}{\gamma^2}\frac{\partial}{\partial t} \!\left[\frac{\partial^2 T}{\partial x^2} {+} \frac{\partial^2 T}{\partial y^2}\!\right]\! {=} 0.\label{eq:DPLE_final}
\end{gathered}
\end{equation}
Eq.~(\ref{eq:DPLE_final}) is exactly the DPLE equation discussed by \citet{joseph_heat_1989} and \citet{tzou_unified_1995}.
We can now see the relationship between the constants appearing in the VHE, and the time lags appearing in the DPLE \cite{tzou_unified_1995} and discussed in the main text: $\tau_T = \frac{\alpha\beta A}{ \kappa\gamma^2}$, $\tau_Q = \frac{A}{\gamma}$. We conclude by noting that the thermal diffusivity appearing in the DPLE is $\alpha_E = \frac{\kappa}{C}$.
It can be readily seen that in the limit $\tau_T=\tau_Q$, as well as in the steady-state, the DPLE becomes equivalent to Fourier's law, while in the limit $\tau_T=0$ it becomes equivalent to Cattaneo's second-sound equation \cite{tzou_unified_1995}.

In summary, the above derivation demonstrates that the viscous temperature waves emerging from the VHE are fundamentally different from the inviscid DPLE heat waves previously discussed in Refs. \cite{joseph_heat_1989,tzou_unified_1995,barletta_hyperbolic_1996,xu_thermal_2002,xu_thermal_2021}.

\section{Guyer-Krumhansl equations as limiting case of VHE}
\label{sec:GKE_limit}
In this section we discuss the Guyer-Krumhansl equations (GKE) \cite{guyer_solution_1966,sendra_derivation_2021} as a limit of the VHE. The GKE assume that the phonon bandstructure is linear and isotropic, in which case the drift velocity component of the heat flux dominates over the temperature gradient component, $\bm{Q}=\kappa_D\nabla T +\alpha \bm{u}\to \alpha \bm{u}$.
The VHE (Eqs. \ref{viscous_heat_T} and \ref{viscous_heat_U}) in the steady state reduce to:
\begin{align}
\nabla\cdot \bm{u}&=0,
 \label{viscous_heat_T_1}\\
 \alpha\bm{u}({\bm r}, t)&=-\frac{\beta\alpha}{\gamma} \nabla  T({\bm r}, t) {+} \frac{\eta}{\gamma} \nabla^2 \alpha \bm{u}({\bm r}, t),\label{viscous_heat_U_1}
\end{align}
which is analogous to the form of the GKE given in \cite{sendra_derivation_2021}. $\frac{\eta}{\gamma}$ is a squared length that describes the strength of the non-local relation between the temperature gradient and the heat flux and $\frac{\beta\alpha}{\gamma}=\kappa_M$ is the momentum contribution to the thermal conductivity discussed in SM Sec.~\ref{sec:mom_contribution_k}.
By applying the divergence to Eq.~\ref{viscous_heat_U_1} and recalling Eq.~\ref{viscous_heat_T_1}, one obtains that the isotropic steady-state GKE always yields a harmonic temperature field \cite{tzou_unified_1995,kovacs_two-temperature_2022}, i.e., satisfying  $\nabla^2 T({\bm r}, t)=0$.
Harmonic functions have the property that they cannot assume local maxima or minima in the domain interior.

We note that this is an important difference compared to the VHE, which can describe materials with a complex, non-linear dispersion relation where the heat flux has both a diffusive and a drifting component (analogous to that of rarefied fluids \cite{sambasivam_numerical_2014}), which can result in a temperature profile that is not harmonic ($\nabla^2 T\neq 0$). The possibility for the temperature field to violate the harmonic property is further supported by the analytical arguments discussed in Ref.~\cite{coulterCoupledElectronphononHydrodynamics2025a}.

In the tunnel-chamber device made of graphite discussed in Figs.~\ref{fig:1_vortex},\ref{fig:BTE_revision}, both our VHE and LBTE simulations predict the emergence of local minima and maxima in the domains' interior. These are contained in the thistle and beige regions in Fig.~\ref{fig:BTE_revision}, and their exact coordinates are reported in Tab.~\ref{tab:loc_maxmin}. Because the temperature profile predicted by the GKE must satisfy the harmonic property; GKE cannot reproduce the local maxima and minima emerging from the VHE and LBTE.

\begin{table}[t]
\label{tab:loc_maxmin}
\caption{Comparison between positions of the local extrema obtained from different levels of transport theory, and their dependence on boundary conditions. }
\begin{tabular}{l|c|c|c}
  \hline
  \hline
Theory & $x$ ($\mu$m) & $y$ ($\mu$m) & boundary distance ($\mu$m) \\
\hline
VHE no slip            & 0.75 & $\pm$ 0.40  & 0.24 \\
\hline
VHE finite slip        & 0.60 & $\pm$ 0.55  & 0.08 \\
LBTE diffuse    & 0.64 & $\pm$ 0.26  & 0.16 \\
\hline
VHE frictionless               & 0.55 & $\pm$ 0.60  & 0.03 \\
LBTE reflective  & 0.64 & $\pm$ 0.27  & 0.16 \\
\hline
Fourier            & \multicolumn{2}{c|}{no local extrema}    & 0.00 \\
GKE                & \multicolumn{2}{c|}{no local extrema}      & 0.00 \\
  \hline
  \hline
\end{tabular}

\end{table}

\section{Decomposition of viscosity into principal components}
\label{sec:viscosity_decomposition}

Here, we briefly discuss the form of the viscosity tensor relevant for the geometry and materials studied. Firstly, all simulations performed in this work are restricted to the two-dimensional basal planes of graphite and h$^{11}$BN, in which the viscosity tensor is isotropic. This immediately reduces the viscosity tensor to the following form:
\begin{equation}
  \begin{split}
        \eta^{ijkl} {=}& (\eta_{\rm vol} {-} \eta_{\rm shr})\delta^{ij}\delta^{kl} {+} (\eta_{\rm shr} {+} \eta_{\rm rot})\delta^{ik}\delta^{jl} \\&{+} (\eta_{\rm shr} {-} \eta_{\rm rot})\delta^{il}\delta^{jk},
  \end{split}
\label{eq:visc_decomposition}
\end{equation}
where $\eta_{\rm vol}$, $\eta_{\rm shr}$, $\eta_{\rm rot}$ are the volume, shear, and rotational viscosity \footnote{The volume viscosity is also called the bulk viscosity, here we use the term 'bulk viscosity' for the value of the viscosity tensor in the bulk of the crystal, without accounting for finite size effects.}. We collectively refer to these three components as the principal components of the viscosity tensor.
Defining the principal components in this way leads to a convenient and physically intuitive form of the viscous stress tensor $\sigma^{ij} = \eta^{ijkl} \tfrac{\partial u^k}{\partial r^l}$. The viscous stress tensor enters the VHE through its divergence, which can also be written in terms of principal components:
\begin{equation}
\begin{split}
\frac{\partial \sigma^{ij}}{\partial r^j} &= 
\eta^{ijkl} \frac{\partial^2 u^k}{\partial r^j \partial r^l} \\ 
&= \left[\eta_{\rm vol} \nabla (\nabla\cdot\bm{u}) + \eta_{\rm shr} \nabla^2 \bm{u} + \eta_{\rm rot} \nabla \times (\nabla \times \bm{u})\right]^i.
\end{split}
\raisetag{13mm}
\end{equation}
Volume and shear viscosity appear in conventional fluids and describe the stress arising from compressing and shearing the fluid, respectively \cite{landau_fluid}. Meanwhile, rotational viscosity arises only in the special case when the angular momentum related to the fluid velocity is not conserved. Rotational viscosity has previously been discussed in nematic liquid crystals, where angular momentum can flow between the translational and rotational motion of the molecules \cite{groot_thermodynamics}, and in electron fluids with crystal-momentum dissipation \cite{el_rot_viscosity}. The temperature dependence of these three viscosity components for materials investigated in this work is discussed in SM~\ref{sec:parameters_entering_in_the_viscous_heat_equations}.

\section{Finite-size effects}
\label{sub:finite_size}

Finite-size effects can be included through Matthiessen-type corrections to the bulk thermal conductivity $\kappa_{D, \rm bulk}^{ij}$ \cite{ziman1960electrons,Simoncelli2020}, analogous to Bosanquet formulas in rarefied fluids \cite{michalis_rarefaction_2010}:
\begin{equation}
\frac{1}{\kappa^{ij}_D}=\frac{1}{\kappa^{ij}_{D, \rm bulk}}+\frac{1}{\kappa^{ij}_{D,\mathrm{ballistic}}}.
  \label{eq:conductivity_comb}
\end{equation}
where the intrinsic diffusion-damped conductivity $\kappa^{ij}_{D, \rm bulk}$ is given by Eq.~(\ref{eq:kappaD_def}), and its ballistic counterpart is:
\begin{equation}
	\kappa^{{ij}}_{\rm D,ballistic}=
	\left[\frac{1}{V}\sum_\state (\hbar \omega_\state)^2 \frac{\bar{N}_\state(\bar{N}_\state+1)}{k_B \bar{T}^2}v^{{i}}_\state v^j_\state \frac{1}{|\bm{v}_{\state}|}
	\right]
	\cdot L_S\frac{\kappa^{ij}_{D, \rm bulk}}{\kappa^{ij}},
   \label{eq:conductivity_fs}
\end{equation}
where the term in square brackets is the $K^{ij}_S$ prefactor defined in Eq.~(C3) of Ref.~\cite{Simoncelli2020}, $L_S$ is the characteristic lengthscale of the device, and the final rescaling factor $\frac{\kappa^{ij}_{D, \rm bulk}}{\kappa^{ij}}$ takes into account 
that the ballistic estimate of the diffusion-damped conductivity has to be determined only by vibrational excitations in the non-hydrodynamic diffusion-damped subspace, i.e., $\frac{\kappa^{ij}_{D, \rm bulk}}{\kappa^{ij}}=\frac{\kappa^{{ij}}_{\rm D,ballistic}}{K^{ij}_S L_S}$, implying that the ratio between momentum conductivity and total conductivity is the same for both the ballistic and intrinsic conductivities.

The ballistic correction to the viscosity tensor is performed analogously, by applying Matthiessen-type corrections to the principal components of the bulk and the ballistic viscosity tensors (see SM~\ref{sec:viscosity_decomposition} for definitions of the principal components). The ballistic viscosity tensor is defined in Eq.~(C4) of Ref.~\cite{Simoncelli2020}:
\begin{equation}
  M^{ijkl} = \frac{1}{V} \sum_\state 
    \hbar^2 q^i v^j_\state q^k v^l_\state
    \frac{\bar{N}_\state(\bar{N}_\state+1)}{k_B \bar{T}}
    \frac{1}{|\bm{v}_{\state}|}.
      \label{eq:viscosity_fs}
\end{equation}
In the basal plane of graphite, where $M$ is isotropic, we can perform the same decomposition as for the bulk viscosity, where $M_{\rm shr} = \tfrac{M^{ijij}+M^{ijji}}{2}$, $M_{\rm rot} = \tfrac{M^{ijij}-M^{ijji}}{2}$, $M_{\rm vol} = M^{iiii} - M_{\rm shr}$. We then correct the principal components of the viscosity tensor (Eq. \ref{eq:visc_decomposition}) accounting for ballistic effects:
\begin{equation}
  \begin{split}
      \frac{1}{\eta_{\rm cmp}} = \frac{1}{\eta_{\rm cmp, bulk}} + \frac{1}{L_S M_{\rm cmp}};\ \rm cmp = vol, shr, rot.
  \end{split}
  \label{eq:viscosity_comb}
  \raisetag{12mm}
\end{equation}

The value of $\gamma^{ij}$ is not affected by finite-size effects related to phonon-boundary scattering, since this effect is already accounted for by the boundary conditions \cite{ziman1960electrons}.

We accounted for finite-size effects considering a characteristic size equal to: (i) the chamber's opening $L=0.52\mu m$ for the simulations in Fig.~\ref{fig:1_vortex}, \ref{fig:hBN_vortex}; (ii) the horizontal side of the domain for the time-dependent simulations in Figs.~\ref{fig:fig2},\ref{fig:rect_freq},\ref{fig:freq_VHE_vs_DPLE}.

\section{Steady-state viscous heat backflow in \texorpdfstring{${\rm h}^{11}$BN}{hBN}}
\label{sec:steady_state_viscous_heat_backflow_in_h}
\begin{figure}[htbp]
\includegraphics[width=\columnwidth]{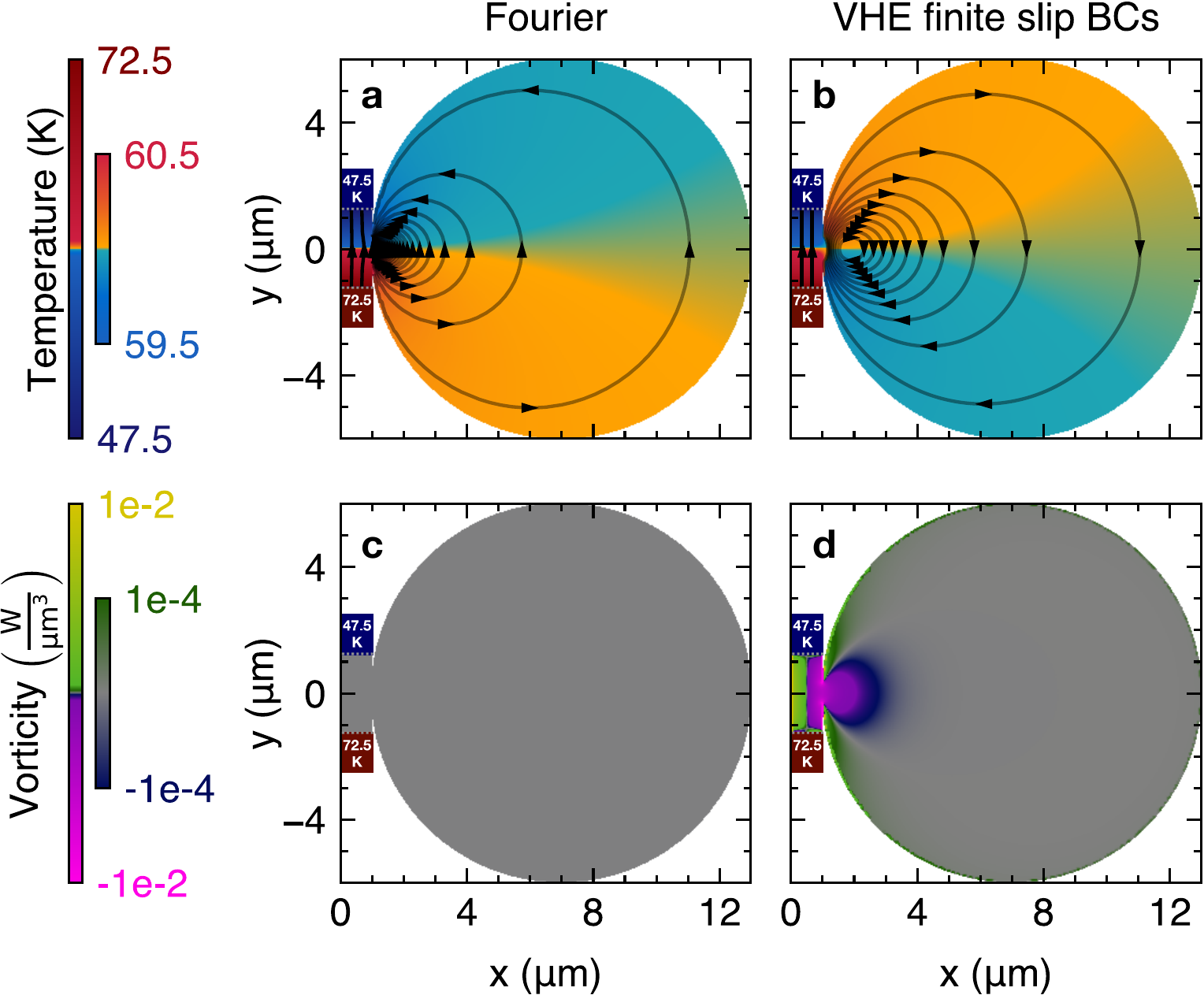}
\caption{\label{fig:hBN_vortex}
\textbf{Signature of viscous heat backflow in h$^{11}$BN.} 
In-plane ($x{-}y$) heat flux (streamlines) and temperature (colormap) in a tunnel-chamber device made of h$^{11}$BN. Panel \textbf{a} (\textbf{b}) shows the solution of Fourier's equation (VHE) in the presence of a temperature gradient applied at the tunnel's boundaries ($60{\pm}12.5$ K at $y{=}{\mp} 1.25\mu m$), and 
considering the other boundaries as adiabatic (i.e. $\nabla T{\cdot} \bm{\hat n}{=}0$, where $\bm{\hat n}$ is the versor orthogonal to the boundary) and, in the VHE, with finite slip length 0.4 $\mu m$ (see end note for details).
In Fourier's case (\textbf{a}), the direction of the temperature gradient in the chamber mirrors that in the tunnel.
In contrast, the VHE (\textbf{b}) account for an additional viscous component for the heat flux---not directly related to the temperature gradient---allowing the emergence of viscous backflow and temperature in the chamber reversed compared to the tunnel.
Panel~\textbf{c)}, the vorticity of Fourier's heat flux is trivially zero; panel \textbf{d)} shows the nonzero vorticity of the total VHE heat flux, $\nabla {\times}\bm{Q}^{\rm TOT}$.}
\end{figure}

In Fig.~\ref{fig:hBN_vortex} we show that heat vortices are not limited to graphite but also appear in devices made of h$^{11}$BN. 

\begin{figure*}
\vspace{-5mm}
\includegraphics[width=0.8\textwidth]{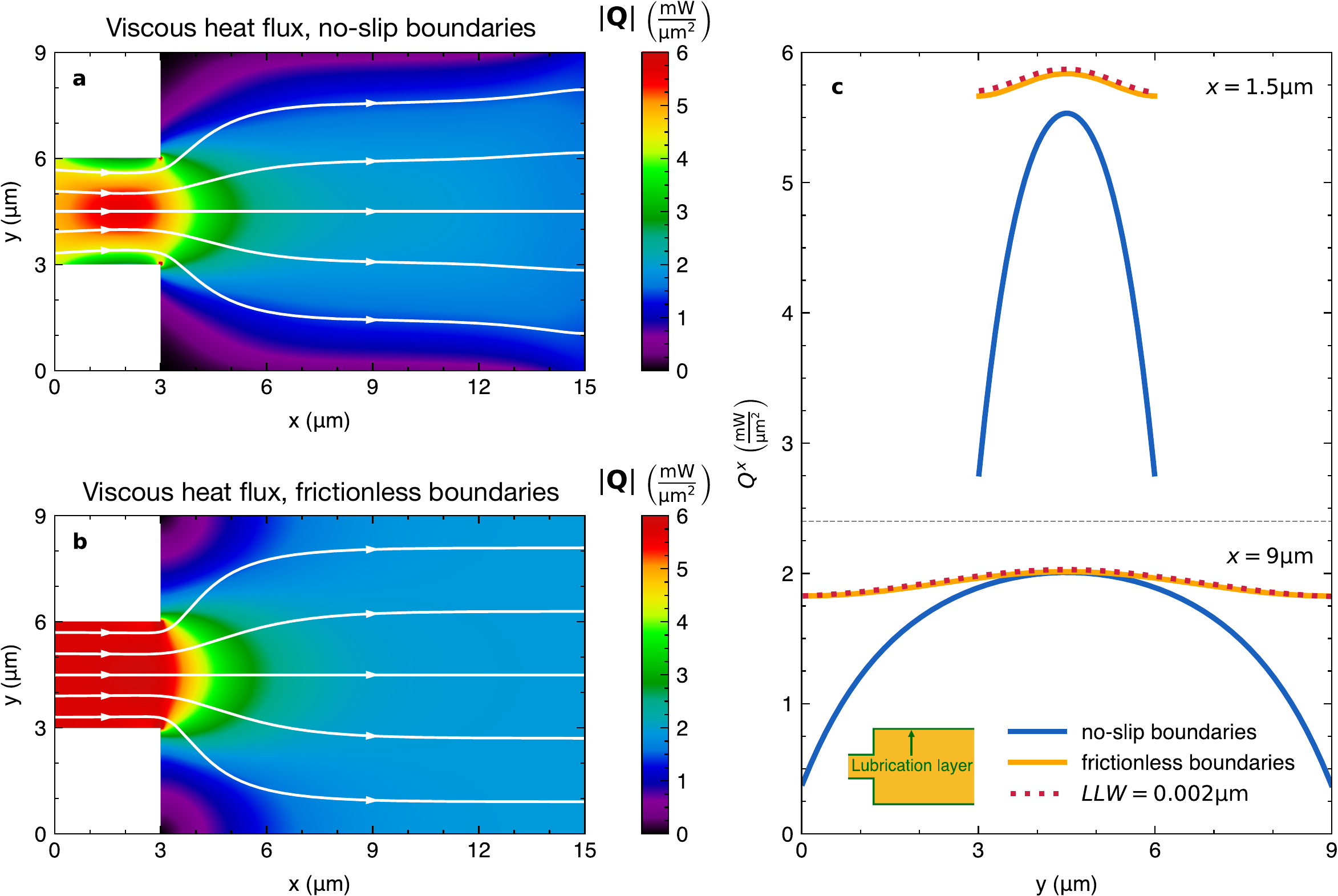}\\[-5mm]
\caption{\label{fig:PRX}
\textbf{Effects of boundary conditions on the heat-flow profile.}
In-plane (x-y) heat flow in graphite, obtained solving the VHE
imposing a temperature of $80\ \mathrm{K}$ on the left side ($x=0\ \mathrm{\mu m}$) and $60\ \mathrm{K}$ on the right side ($x=15\ \mathrm{\mu m}$), assuming all boundaries at $x \neq 0\ \mathrm{\mu m}$ and $x \neq 15\ \mathrm{\mu m}$ to be adiabatic, and using no-slip boundary conditions for the drift velocity ($\bm{u}{=}\bm{0}$ at the boundary) in \textbf{a)} \cite{Simoncelli2020}, and a frictionless boundary condition (see end note for details) in panel \textbf{b)}.
Panel \textbf{c)}, differences between the heat-flow profile along the vertical sections $x=1.5\ \mathrm{\mu m}$ and $x=9\ \mathrm{\mu m}$ for different boundary conditions.
We highlight how the lubrication-layer approach, discussed in the end note and used to model frictionless boundaries in arbitrary geometries, allows to obtain results practically indistinguishable from those obtained using the standard Mathematica ``perfectly slipping'' BCs limited to non-curved geometries. The insets in panel \textbf{c)} show schematically how the lubrication layer is implemented.
}
\end{figure*}

\section{Influence of geometry and boundary conditions on backflow}
\label{sec:BC_steady_state}

As mentioned in the main text, to observe steady-state hydrodynamic deviations from Fourier's law, two conditions are necessary: first, the material must have a non-negligible intrinsic thermal viscosity; second, the device's geometry and boundary conditions (BCs )have to yield a significant (as large as possible) second spatial derivative of the drift velocity.
Here we discuss how the geometry of the sample and the BCs influence hydrodynamic deviations from Fourier's diffusion. In SM \ref{sec:effects_of_temperature_and_isotopic_disorder_on_viscous_heat_backflow}, we show how temperature, size of the sample, and isotopic-impurity concentration affect viscous backflow, while the details on intrinsic material properties are reported in SM~\ref{sec:parameters_entering_in_the_viscous_heat_equations}.

We focus on materials having the same symmetries of those considered in the main text, considering in-plane transport phenomena that allow us to consider several of the tensor coefficients appearing in the VHE~(\ref{viscous_heat_T},\ref{viscous_heat_U}) as proportional to the identity: $\alpha^{ij}=\alpha\delta^{ij}$, $\kappa_D^{ij}=\kappa_D\delta^{ij}$, $\beta^{ij}=\beta\delta^{ij}$, and $\gamma^{ij}=\gamma\delta^{ij}$; see SM~\ref{sec:parameters_entering_in_the_viscous_heat_equations} for details.

\subsection{Lubrication layer and frictionless boundaries}
\label{sub:slipping_boundaries}
First, we discuss the numerical techniques used to simulate different boundary conditions for the drift velocity. 
Specifically, in order to model boundary conditions different from no-slip ($\bm{u}{=}\bm{0}$ at the boundary), which correspond to boundaries that are completely dissipating the crystal momentum, we introduce a "lubrication layer" \cite{bocquetFlowBoundaryConditions2007}, i.e. a region in proximity of the actual boundary in which the shear and rotational viscosity are reduced as discussed by Eq.~(\ref{eq:slip_length}) in the end note.
We show in Fig.~\ref{fig:PRX} that, in analogy to standard fluid-dynamics, no-slip BCs (panel \textbf{a}, also discussed in Ref.~\cite{Simoncelli2020}) yield a Poiseuille-like heat-flow profile. In contrast, using frictionless (infinite slip length) boundaries yield much weaker variations of the heat-flux profile (panel \textbf{b}).
The lubrication-layer approach is particularly advantageous from a numerical viewpoint. In fact, in standard solvers such as \textit{Wolfram Mathematica}, implementing frictionless (i.e., perfectly slipping with infinite slip length) BCs is straightforward in simple geometries such as that in Fig.~\ref{fig:PRX}, where the Cartesian components of the drift velocity are always aligned or orthogonal to the boundaries. However, in complex geometries such as that in Fig.~\ref{fig:1_vortex},
it is not possible to find a coordinate system which is always aligned or orthogonal to the boundaries, rendering the numerical implementation of the frictionless BCs much more complex.
We show in Fig.~\ref{fig:PRX}\textbf{c)} that by choosing a sufficiently small lubrication layer width (LLW=0.002 $\mu m$), and setting the shear and rotational viscosity components to zero in the lubrication layer, one obtains practically indistinguishable results by using the lubrication layer or Mathematica's Cartesian frictionless BCs. 

Thus, the test in Fig.~\ref{fig:PRX} justifies the use of the lubrication layer to simulate frictionless or partially slipping BCs in complex geometries such as the one shown in Fig.~\ref{fig:1_vortex}, which we discuss in detail in the next section. 

\subsection{Tunnel-chamber geometry for viscous backflow}
\label{ssec:geometry_for_viscous_heat_backflow}
Here we discuss how the shape of the tunnel-chamber geometry maximizes the steady-state hydrodynamic signatures of the VHE.
\begin{figure}[b!]
	\centering
	\includegraphics[width=\columnwidth]{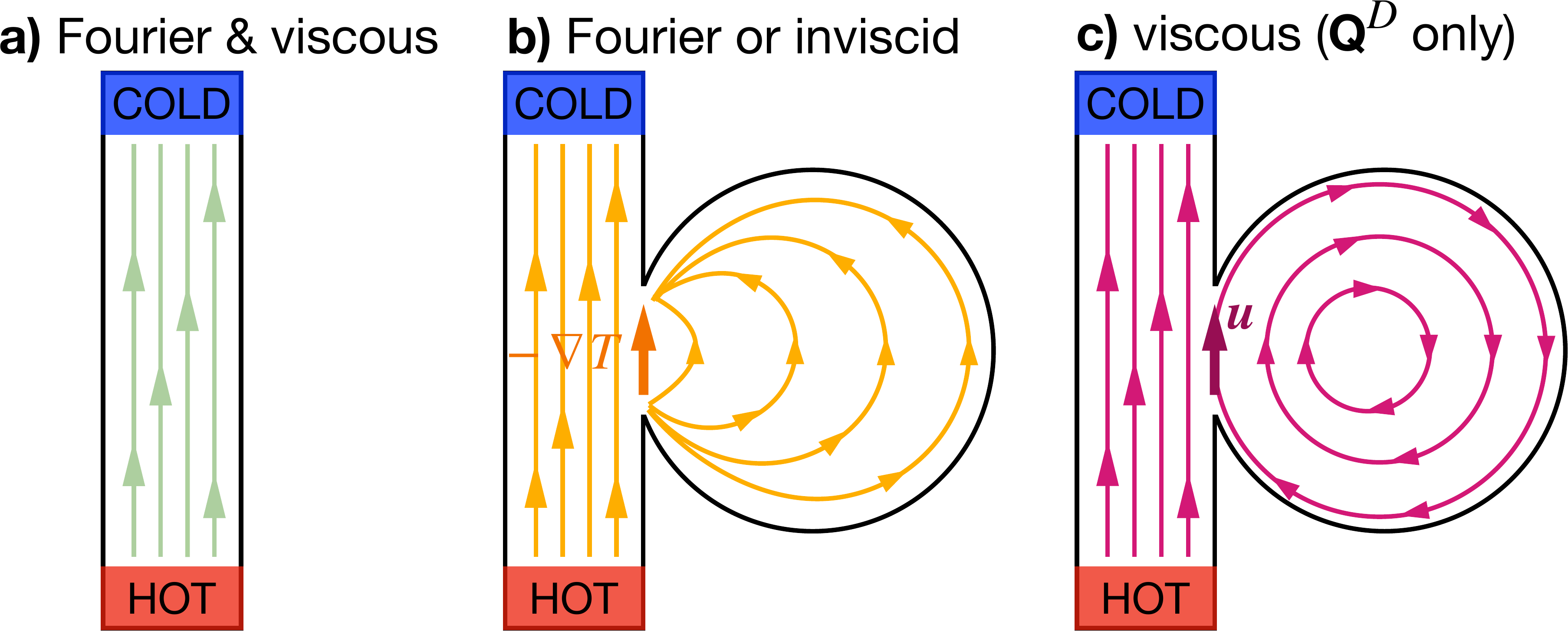}
	\caption{\textbf{Representative streamlines for diffusive and viscous flow.} 
	 \textbf{a)}, streamlines emerging from Fourier or VHE in a rectangular geometry with two thermalized boundaries ($T$ is hot (cold) in the red (blue) box, and $\bm{u}{=}\bm{0}$ in both boxes), and the other boundaries frictionless and adiabatic.
	Drift-velocity and temperature-gradient component of the heat flux are aligned (green streamlines), and no significant differences are observed between VHE and Fourier.
	 \textbf{b)}, Fourier's equation predict that a temperature gradient at the opening of a circular chamber (orange) generates dipole-like, open and counterclockwise heat-flux streamlines.
	 \textbf{c)}, the VHE allow vortical flow and predict that, to minimize viscous stresses, the drifting heat-flux component can form closed streamlines with clockwise direction (opposite to \textbf{b}).}
	\label{fig:intuitive_explanation_tunnel_chamber}
\end{figure}
\begin{figure*}
  \vspace*{-3mm}
	\includegraphics[width=\textwidth]{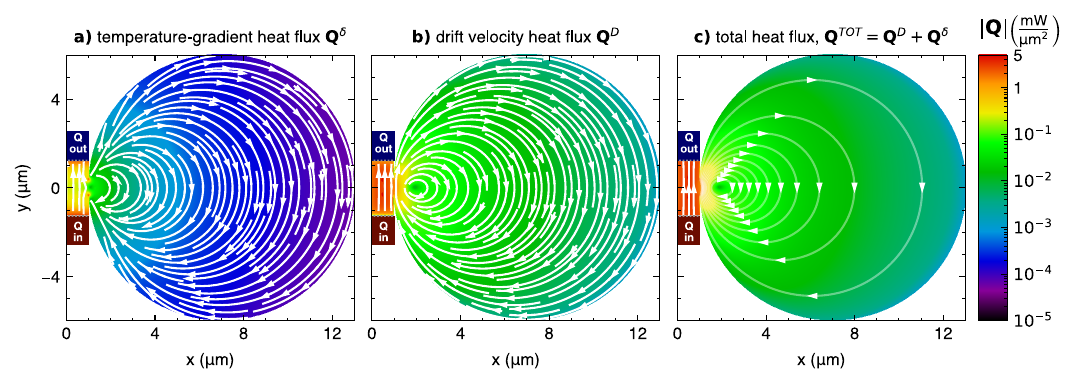}
\caption{\textbf{Heat flux in graphitic tunnel-chamber device,} from the solution of the VHE. Streamlines are white, the colormap shows the magnitude of the heat flux on a log scale. 
\textbf{a}, irrotational temperature-gradient component of the heat flux, $\bm{Q}^{\delta}{=}{-}\kappa_D\nabla T$. \textbf{b)}, drifting (vortical) component of the heat flux, $\bm{Q}^{D}{=}\alpha\bm{u}$. \textbf{c)}, total heat flux $\bm{Q}^{\rm TOT}{=}\bm{Q}^{\delta}{+}\bm{Q}^{D}$, displaying closed streamlines, i.e., vortical behavior. Boundary conditions are as in Fig.~\ref{fig:1_vortex}: a temperature gradient is present in the tunnel ($70{\pm}12.5$ K at $y{=}{\mp} 1.25\mu m$), and all the other boundaries are adiabatic and have a finite-slip BCs with slip length 0.4 $\mu m$.}
	\label{fig:verify}
\end{figure*}
We start by considering a rectangular geometry with two opposite sides thermalized at different temperatures, and all the other sides adiabatic (Fig.~\ref{fig:intuitive_explanation_tunnel_chamber}\textbf{a}). In this case, the temperature gradient will be in the vertical direction and will approximately tend to a constant \footnote{Strictly speaking, in proximity of a perfectly thermalized boundary (temperature fixed at a certain value and $\bm{u}=\bm{0}$), 
the temperature gradient emerging from the VHE is weakly space dependent, due to the coupling between the temperature gradient and the drift velocity in Eq.~(\ref{viscous_heat_U}). In practice, in a rectangular geometry of graphite with gradient applied around 70 K, the temperature gradient and drift velocity reach a practically space-independent value within about 1 $\mu m$ from the boundaries. Analytical details on the mechanisms determining this lengthscale can be found in Appendix G of Ref.~\cite{Simoncelli2020}, and numerical simulations in Fig. 4 of the same reference.}, and from the VHE~(\ref{viscous_heat_T},\ref{viscous_heat_U}) one can see that in the presence of frictionless boundaries (implying no shear stress at the boundaries, a condition imposed using the lubrication layer discussed in the end note and shown in Fig.~\ref{fig:device_ll}) this will create a drift velocity anti-aligned to the temperature gradient, $\bm{u}{=}{-}\tfrac{\beta}{\gamma}\nabla T$, thus aligned diffusive and drifting components of the heat flux ($\bm{Q}^\delta=-\kappa_D\nabla T$ and $\bm{Q}^D=\alpha\bm{u}$, respectively).
If no-slip boundary conditions (corresponding to phonon-boundary scattering dissipating the phonon momentum \cite{ziman1960electrons,Simoncelli2020}) are considered in this geometry, the drift velocity is forced to assume a Poiseuille-like profile along the horizontal direction, a feature that has very small (negligible) influence on the temperature gradient along the vertical direction (see Fig. 4b and Appendix H of Ref.~\cite{Simoncelli2020}).

While the analysis of the perfectly rectangular geometry does not show significant deviations between Fourier and VHE, it shows that it is possible to use temperature gradients to generate nonzero drift velocity in proximity of boundaries. This is of practical interest. In fact, to the best of our knowledge, there are no known experimental techniques that allow us to directly control the phonon drift velocity injected in (or extracted from) a device---for this reason all the analyses done here and in Ref.~\cite{Simoncelli2020} consider always $\bm{u}{=}\bm{0}$ at boundaries in contact with thermal reservoirs.
In the subsection~\ref{subs:BCS_independence} we analytically show that this boundary condition becomes unimportant far from the thermalized boundaries, confirming that it is possible to use a temperature difference at the tunnel's boundaries to induce a nonzero drift velocity in the middle of the tunnel.
This analysis suggests that, by attaching a device at the middle of the rectangular tunnel, one could use the large and homogeneous (robust against perturbations) drift velocity present in the tunnel far from the thermalized boundaries to drive viscous deviations from Fourier's law in the attached circular device. In other words, the drift velocity in the center of the tunnel can be considered as being drift-velocity boundary condition applied at the chamber's opening. These analytical intuitions will be numerically confirmed later.

The most dramatic qualitative steady-state deviation from Fourier's law is a heat flux flowing from a cold to a hot region. This emerges from the VHE when the drifting heat flux $\bm{Q}^D$ has stronger magnitude and opposite direction compared to the temperature-gradient flux $\bm{Q}^\delta$.
We now elucidate how these deviations are promoted in a circular chamber having: a small opening, a small temperature gradient, and constant drift velocity applied to such opening (Figs.~\ref{fig:intuitive_explanation_tunnel_chamber}\textbf{b,c}).
This can be understood by noting that the irrotational form of the temperature-gradient heat flux expression ($\bm{Q}^\delta=-\kappa_D\nabla T$) implies that $\nabla T$ at the opening behaves as the thermal counterpart of an electric dipole, i.e., it generates an irrotational temperature-gradient flux with counterclockwise streamlines that never form closed paths (Fig.~\ref{fig:intuitive_explanation_tunnel_chamber}\textbf{b}). 
An analogous behavior emerges from the inviscid limit of the VHE ($\eta^{ijkl}=0$), since in this case Eq.~(\ref{viscous_heat_U}) implies $\bm{u}{=}{-}\tfrac{\beta}{\gamma}\nabla T$. Intuitively, this corresponds to having zero shear stress in the phonon fluid, allowing the drifting heat-flux component to enter the chamber following the abrupt changes of directions imposed by the irrotational dipole-like behavior of the temperature-gradient field. 
In contrast, for non-negligible shear viscosity, when the drifting heat flux changes its direction to enter the chamber, it is subject to viscous shear; to minimize such shear, it minimizes changes in its direction and thus enters the chamber, forming clockwise streamlines (Fig.~\ref{fig:intuitive_explanation_tunnel_chamber}\textbf{c}). In this regard, the frictionless boundary conditions (no shear stress at the boundary, imposed with the  lubrication-layer approach of Fig.~\ref{fig:device_ll}) ensure that the drifting heat flux does not decrease in modulus as it approaches the chamber's boundary, maximizing the magnitude of the drift velocity at the chamber's opening and, therefore, hydrodynamic effects. 
\begin{figure}[htbp]
\includegraphics[width=\columnwidth]{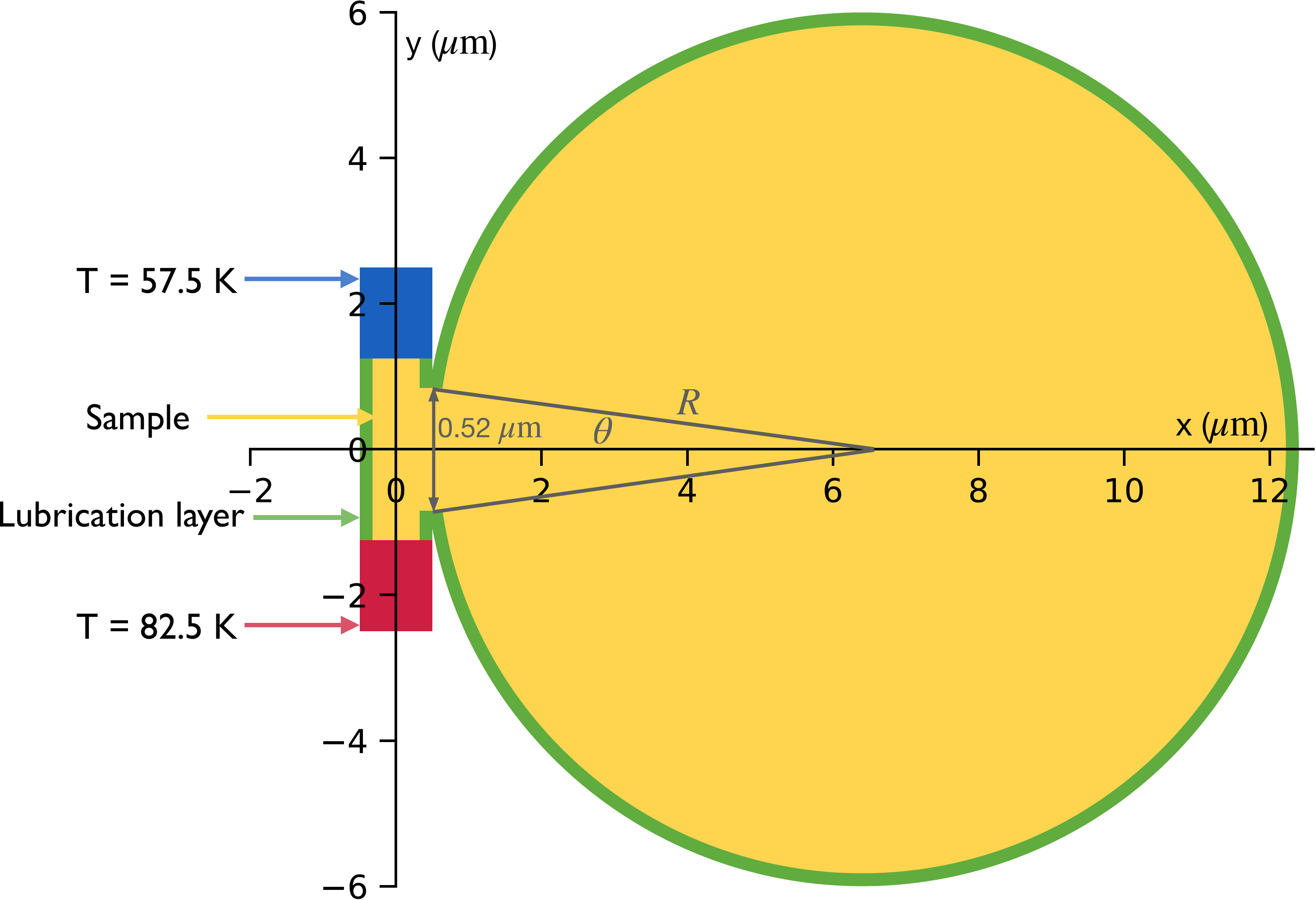}\\[-2mm]
\caption{\label{fig:device_ll} 
\textbf{Tunnel-chamber geometry for viscous heat backflow.}
The simulation domain where all the VHE parameters assume physical values is yellow; the green area is the lubrication-layer region, where the viscosity is reduced to account for partially reflective boundaries (see end note for details).
The width of the lubrication layer (0.02 $\mu m$), and of the opening connecting the tunnel and the chamber (0.52 $\mu m$), are represented in an exaggerated way for graphical clarity. 
The opening angle used here and in Fig.~\ref{fig:1_vortex} is $\theta = 4.9672^\circ$. A temperature of 82.5 K (57.5 K) is applied at the lower (upper) opening of the tunnel, all the other boundaries are adiabatic. }
\end{figure}
We also note that the VHE allow the drifting heat flux $\bm{Q}^{\delta}$ to have nonzero vorticity, which in the circular chamber geometry in focus here is evident from $\bm{Q}^{TOT}$ forming closed streamlines (see ~\ref{fig:verify}c) \footnote{For a well-behaved vector field, the circulation computed on a closed streamline $\vec{\mathcal{S}}$ is nonzero: $\mathcal{C}=\oint_{\mathcal{S}}\bm{Q}^D{\cdot} d\vec{\mathcal{S}}\neq 0$; therefore, one sees from Stokes' theorem that this implies nonzero vorticity: $\mathcal{C}=\int_{\mathcal{A}}(\nabla {\times} \bm{Q}^D){\cdot}\bm{m} d{\mathcal{A}}\neq 0$, where $\bm{m}$ is the unit vector outward-normal to the surface $\mathcal{A}$ enclosed by $\vec{\mathcal{S}}$.}.
When this happens, we have that along the closed streamline, the direction of $\bm{Q}^{TOT}$ is tangential to the line, while the overall change in $T$ is zero. This implies that, in some parts of the streamline, $\bm{Q}^{TOT}$ and $\nabla T$ must be aligned, which is exactly the condition we defined above.
As anticipated, $\bm{Q}^{TOT}$ and $\nabla T$ are anti-aligned in the diffusive or inviscid limit, where $\bm{u}{=}{-}\tfrac{\beta}{\gamma}\nabla T$. Eq.~(\ref{viscous_heat_U}) shows that aligned behavior requires viscous effects to dominate over momentum dissipation ($|\eta^{ijkl} \frac{\partial^2 u^k({\bm r}, t)}{\partial r^j \partial r^l} |\gg |\gamma^{ij} u^j({\bm r}, t)|$); this permits the emergence of the viscous temperature inversion  (Fig.~\ref{fig:1_vortex}\textbf{b}) \footnote{Specifically, in the region where temperature inversion is most pronounced in Fig.~\ref{fig:1_vortex}, one has $\beta\frac{\partial T({\bm r}, t)}{\partial r^y} {\sim} \eta^{yyyy} \frac{\partial^2 u^y({\bm r}, t)}{\partial r_y^2}$, a condition that permits temperature gradient and drift velocity to be aligned, thus temperature-gradient heat flux $\bm{Q}^\delta=-\kappa_D\nabla T$ and drifting heat flux $\bm{Q}^D=\alpha\bm{u}$ to have opposite directions.}.
This illustrative analysis is quantitatively confirmed in Fig.~\ref{fig:verify}, where we show that in the tunnel-chamber geometry of Fig.~\ref{fig:1_vortex} the temperature-gradient flux does not form closed streamlines, while the drifting velocity does, and the total, vortical heat flux shows closed streamlines. 
Since in all the cases considered the total heat flux is divergence free, we ensured a correct representation of closed streamlines by computing them as level curves of the stream function following the matplotlib algorithm\cite{Hunter_2007,streamlines_curve_level_streamfunction}. 

In summary, in the tunnel-chamber geometry features antithetical conditions for the two components of the VHE heat flux at the opening of the chamber: the temperature dipole at the opening drives a heat flux with open streamlines and in counterclockwise direction in the chamber; in contrast, the large nonzero drift velocity at the opening drives drifting heat-flux streamlines that tend to be closed and in clockwise direction. 
The relative strength of these two heat-flux components is affected by material properties, including isotopic content, which determine the coefficients entering the VHE, and by the device's geometrical properties such as diameter of the chamber and the size of the opening.
The effect of the chamber diameter is analyzed in detail in Fig.~\ref{fig:tunnel_cont}. The size of the opening directly influences the strength of the dipole-like temperature gradient field, as well as shear stress applied by the boundary drift velocity to the $\bm{u}$ field inside the chamber. Using larger openings yields stronger dipole-like temperature-gradient fields, which we numerically found to generally decrease temperature inversion in chambers with radius $6~\mu m$ or larger. Here, the opening's size has been fixed to 0.52 $\mu$m (Fig.~\ref{fig:device_ll}), such a value is sufficiently small to yield visible viscous heat backflow, and sufficiently large to be within the regime in which mesoscopic VHE and microscopic LBTE yield compatible predictions (Fig.~\ref{fig:BTE_revision}).

\subsection{Analytical proof that a temperature gradient can drive a drift velocity}
\label{subs:BCS_independence}
Here we generalize the analytical solution reported in Appendix~G of Ref.~\cite{Simoncelli2020}, accounting for momentum dissipation.
Because the experimental control of the drift velocity at the boundaries is an open question, in the main text we solved the VHE under BCs assuming zero phonon drift velocity at the tunnel boundaries in contact with the thermal reservoirs, and exploited the coupling between temperature gradient and drift velocity to drive a nonzero drift velocity at the chamber opening. Here we analytically show that, due to this coupling, a temperature difference applied at the tunnel boundaries induces a nonzero drift velocity in the tunnel interior; most importantly, the drift velocity far from the boundaries does not depend on the drift-velocity BCs.

We start from the VHE~(\ref{viscous_heat_T},\ref{viscous_heat_U}) in the one-dimensional steady-state case (e.g., along the vertical line in the center of the rectangular geometry in Fig.~\ref{fig:intuitive_explanation_tunnel_chamber}\textbf{a}, in the limit of infinite horizontal width and finite vertical length $L$), 
\begin{align}
&\alpha\frac{\partial u(x)}{\partial x} - \kappa_D \frac{\partial^2 T (x)}{\partial x^2} = 0,
 \label{viscous_heat_T_1d}\\
&
\beta
\frac{\partial T(x)}{\partial x} {-} \eta \frac{\partial^2 u(x)}{\partial x^2} {=} {-} \gamma u(x),\label{viscous_heat_U_1D}
\raisetag{1mm}
\end{align}
where all the vectors and tensors in Eqs.~(\ref{viscous_heat_T},\ref{viscous_heat_U}) have been written as scalars determined by the direction considered (e.g., $\alpha=\alpha^{xx}$, $\eta=\eta^{xxxx}$, $u(x)=u^x(x)$, and so on).

In an effectively one-dimensional geometry, the total heat flux $Q^{\rm tot}$ can be determined by integrating Eq.~(\ref{viscous_heat_T_1d}), which implies that $Q^{\rm tot}$ flows along the $x$ axis and is constant in space:
\begin{equation}
  Q^{\rm tot}{=}\alpha {u(x)} {-} \kappa_D \frac{\partial T (x)}{\partial x} {=} {\rm constant}.
  \label{eq:Q_tot_1D}
\end{equation}
By combining Eq.~(\ref{eq:Q_tot_1D}) with Eq.~(\ref{viscous_heat_U_1D}), we obtain:
\begin{equation}
 \eta \frac{\partial^2 u(x)}{\partial x^2} - \Big[\frac{\alpha\beta}{\kappa_D} {+} \gamma\Big]
  {u(x)}
   =-\frac{\beta}{\kappa_D}Q^{\rm tot} ,
   \label{eq:visc_term}
\end{equation}
For the sake of generality, we solve this equation with the following boundary conditions:
\begin{gather}
\begin{aligned}
    T(x=0) &= T + \Delta T, \\
    T(x=L) &= T - \Delta T,
\label{eq:1d_vhe_bc_T}
\end{aligned} \\
    u(x=0) = u(x=L) = u_b.
\label{eq:1d_vhe_bc_u}
\end{gather}
Eq.~(\ref{eq:1d_vhe_bc_T}) corresponds to considering two thermal reservoirs that fix temperature at the boundaries. Eq.~(\ref{eq:1d_vhe_bc_u}) allows us to investigate how the drift velocity influences the solution, and encompasses two special cases: (i) $u_b=0$, which implies that at the boundaries with the reservoirs all the heat flux is determined by a temperature gradient. (ii) $u_b>0$, whose investigation is  motivated by the analytical argument that in the GKE limit, all the heat flux is determined by the drift velocity, and therefore a nonzero $u_b$ is necessary to have non-adiabatic boundaries that exchange heat with a reservoir.
Solving Eq.~(\ref{eq:visc_term}) with these BCs yields:
\begin{gather}
\begin{aligned}
u(x)&=u_p+\big(u_b-u_p\big)\,
\frac{\cosh\!\big(\frac{x-\tfrac{L}{2}}{l_\eta}\big)}
{\cosh\!\big(\frac{L}{2l_\eta}\big)},\\
u_p&= \dfrac{\kappa_M}{\kappa_M+\kappa_D} \dfrac{Q^{\rm tot}}{\alpha},
\label{eq:u_sol}
\end{aligned}
\end{gather}
where $u_p$ is the particular solution, and we defined the viscous length $l_\eta$ as
\begin{equation}
  l^2_\eta= \frac{\eta}{\gamma + \alpha\;\beta / \kappa_D}
  =\left(\frac{1}{l_M^2} + \frac{1}{l_D^2} \right)^{-1}.
     \label{eq:viscous_lenght}
\end{equation}
Here, $l_M=\sqrt{\eta/\gamma}$ is the length scale within which viscous effects originating from the momentum subspace are visible, and $l_D= \sqrt{\eta \;\kappa_D /(\alpha\;\beta)}$ is the length scale within which viscous effects originating from the diffusion-damped subspace are visible.
Intuitively, the viscous length $l_\eta$ characterizes the viscosity-induced variations of $u(x)$ as $x$ moves away from no-slip boundaries. Meanwhile, $\kappa_M$ is the contribution to the heat flux from the momentum subspace in the bulk limit (defined in Sec.~\ref{sec:mom_contribution_k}), i.e., in the regime where the momentum "drifting" heat flux is proportional to the temperature gradient $Q^D=\alpha \bm{u}=-\kappa_M \frac{\partial T}{\partial x}$. Note that $l_M$ and $l_D$ are related as $\frac{l_M^2}{l_D^2} = \frac{\kappa_M}{\kappa_D}$ (here, $\kappa_D$ is the diffusion-damped contribution to the thermal conductivity, also defined in Sec.~\ref{sec:mom_contribution_k}).

Having found $u(x)$ (Eq.~(\ref{eq:u_sol})), we can insert it into Eq.~(\ref{eq:Q_tot_1D}) and solve for $T(x)$ in terms of $Q^{\rm tot}$:
\begin{equation}
  \begin{split}
-\frac{\partial T (x)}{\partial x} {=}\frac{Q^{\rm tot}}{\kappa_M{+}\kappa_D}
{+}\!\left(\!\frac{\kappa_M}{\kappa_D}\frac{Q^{\rm tot}}{ \kappa_M{+}\kappa_D}{-}\frac{\alpha u_b}{\kappa_D }\!\right)\!\frac{\cosh(\frac{x-L/2}{l_\eta})}{\cosh(\frac{L}{2l_\eta})}  \end{split}
\label{eq:1d_vhe_gradT}
\raisetag{16mm}
\end{equation}
In the bulk limit, corresponding to $x$ far away from the boundaries (e.g., $x=L/2 \gg l_\eta$)
and a domain size much larger than the viscous length, $L\gg l_\eta$,
 Eq.~(\ref{eq:1d_vhe_gradT}) reduces to Fourier's law for any value of $u_b$, 
\begin{equation}
\frac{\partial T (x)}{\partial x} =-\frac{Q^{\rm tot}}{\kappa_M{+}\kappa_D}.
\label{eq:Four_limit_u}
\end{equation}
In this limit, the temperature gradient is constant, thus Eq.~(\ref{viscous_heat_T_1d}) implies that also
$u$ is constant and equal to:
\begin{equation}
u_{\rm bulk}={-\frac{\beta}{\gamma} \frac{\partial T (x)}{\partial x} } =\frac{\kappa_M}{\alpha}\frac{2 \Delta T }{L}.
\label{eq:u_bulk}
\end{equation}
It is worth noting that if the boundary drift velocity $u_b$ equals the bulk value~(\ref{eq:u_bulk}), then it is also equal to the particular solution $u_p$ in Eq.~(\ref{eq:u_sol}), and then Eq.~(\ref{eq:u_sol})  reduces to the Fourier limit~(\ref{eq:Four_limit_u}) for any length.

In summary, we demonstrated that it is possible to start from zero drift-velocity BCs at the tunnel's boundaries, and generate a nonzero drift velocity at the opening of the chamber due to the coupling between temperature gradient and drift velocity. Moreover, the specific value of drift velocity at the boundaries does not influence the value of the drift velocity far from the boundaries. 

\section{Effects of temperature, size, and isotopes on heat backflow}
\label{sec:effects_of_temperature_and_isotopic_disorder_on_viscous_heat_backflow}
\begin{figure}[b!]
\vspace*{-1mm}
\includegraphics[width=0.85\columnwidth]{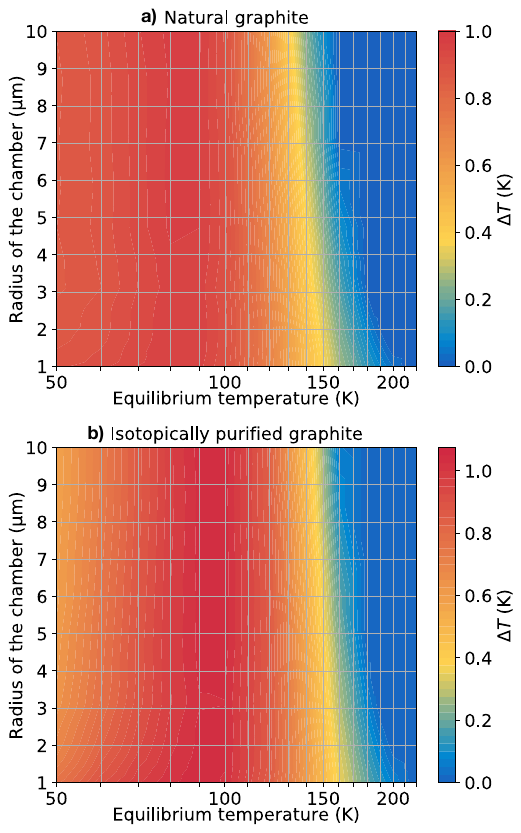}\\[-3mm]
\caption{\label{fig:tunnel_cont}\textbf{Effects of temperature, size, and isotopic disorder on viscous heat backflow.} We show how the temperature inversion, i.e., the difference between the maximum temperature in the upper part of the chamber ($y>0$) and the minimum temperature in the lower part of the chamber ($y<0$), depends on the radius of the chamber and on the average temperature ${T}_{\rm eq}$ of the tunnel (a perturbation of $\delta T={T}_{\rm eq}\pm12.5 K$ is always used at the extremities of the tunnel).
\textbf{a,} refers to samples with natural-abundance isotopic disorder, \textbf{b} to isotopically purified samples.
}
\end{figure}

\begin{figure*}
    \centering
    \includegraphics[width=\textwidth]{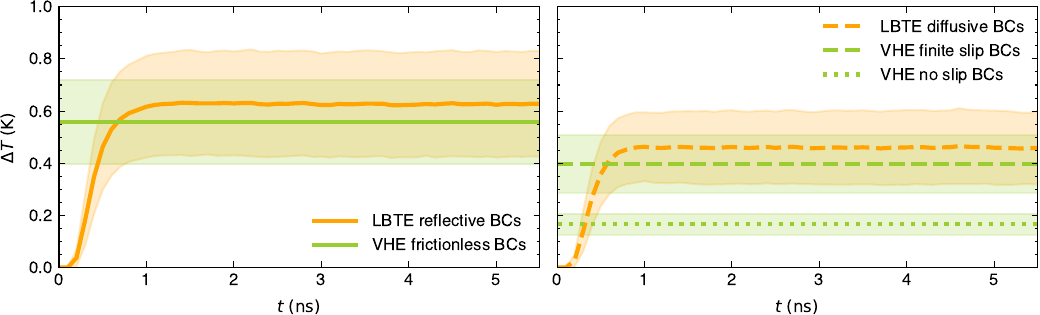}
    \caption{\textbf{Viscous temperature inversion from LBTE and VHE, and its dependence on the boundary conditions.}
	The lines show the 
	difference between the average temperature measured by the probes of diameter 1 $\mu m$ 
	located in the top and bottom part of the chamber ($x=1.02\mu m, \;y=\pm 0.58\mu m$, see the thistle and beige rings in Fig.~\ref{fig:BTE_revision}, respectively). 
	For VHE, solution with frictionless (no shear stress) boundary conditions (BCs) is solid green, with no-slip BCs is dotted green, and with BCs corresponding to finite slip length \cite{bocquetFlowBoundaryConditions2007} $0.4\mu m$ is dashed. 
	For LBTE, phonon-reflective BCs is solid orange (corresponding to frictionless VHE BCs, see text), and phonon-diffusive BCs is dashed orange (corresponding to finite-slip VHE BCs, see text).
	We highlight how in the LBTE the steady state is reached when a time-dependent simulation equilibrates. The shaded areas show $\sigma(\Delta T){=}\sqrt{\rm{var}(T_{up}){+}\rm{var}(T_{down})}$, where $\rm{var}(T_{up})$ ($\rm{var}(T_{down})$) is the variance of the temperature inside the probe region.}
	\label{fig:slip_cont}
\end{figure*}

Here we explore how the temperature inversion due to viscous heat backflow depends on the size of the device, the average temperature around which the temperature gradient is applied, and the presence of isotopic-mass disorder in graphite (this latter affects the parameters such as intrinsic thermal conductivity and intrinsic thermal viscosity).
We show in \ref{fig:tunnel_cont} that the temperature inversion due to heat backflow is maximized at average temperatures below 100 K for natural graphite (98.9\% $^{12}$C, 1.1\% $^{13}$C) and in the range $80\lesssim T\lesssim 110$ K for isotopically purified graphite (99.9\% $^{12}$C, 0.1\% $^{13}$C). Moreover, the temperature inversion is not significantly affected by the chamber's radius, and it is slightly larger in isotopically purified samples compared to natural samples.

\section{Details on steady-state viscous heat backflow from the LBTE} 
\label{sub:steady_state_viscous_heat_backflow_from_the_space_dependent_solution_of_the_full_boltzmann_transport_equation}
In this section we provide additional details on the space-dependent solution of the LBTE with full collision operator discussed in Fig.~\ref{fig:BTE_revision}.
In Fig.~\ref{fig:slip_cont} we show the temperature difference between the thistle and beige probes in Fig.~\ref{fig:BTE_revision} as a function of time, showing that the system effectively reaches the steady state for $t\gtrsim 1.5$ ns.

We also note that the space-dependent heat flux and temperature in Fig.~\ref{fig:BTE_revision} are obtained integrating the microscopic deviational heat flux and energy density, which are both related to the phonon distribution that solves the LBTE, see Ref.~\cite{peraud_efficient_2011} and Eqs. (6) and (7) in Ref.~\cite{raya-moreno_bte-barna_2022}. As discussed in Ref.~\cite{allen_temperature_2018}, temperature and heat flux are well defined within the LBTE with full, energy-conserving collision operator and in the presence of heaters or thermal reservoirs that specify temperature at the boundaries. In particular, the definition of a mesoscopic, space-dependent temperature is related to the constraint of conservation of energy in microscopic phonon collisions---formally this derives from the fact that phonon distributions $n(\bm{q},\bm{r})_s$ (where $\bm{q}$, $s$, and $\bm{r}$ are the wavevector, mode index, and position of the phonon wavepacket, respectively) proportional to the phonon energies $\hbar\omega(\bm{q})_s$ are eigenvectors with zero eigenvalue of the LBTE's full collision matrix \cite{spohn_phonon_2006,allen_temperature_2018}. In other words, the definition of temperature in the VHE follows from the microscopic energy conservation within the LBTE; it is obtained projecting the LBTE into the subspace of the eigenvector with zero eigenvalue proportional to phonon energies \cite{Simoncelli2020}. 

We note that a non-homogeneous, space-dependent temperature is necessary to have heat transport in our setup (Fig.~\ref{fig:device_ll}). Specifically, we recall that we always consider $\bm{u}=\bm{0}$ at non-adiabatic boundaries in contact with thermal reservoirs, since we are not aware about any experimental technique that allows to directly control the drift velocity injected in a device. Under these conditions, non-homogeneities that drive heat transport can be introduced exclusively via reservoirs at different temperature; these imply a nonzero temperature gradient (i.e., $\bm{Q}^\delta$), which then generates drift velocity (i.e., $\bm{Q}^D$) via the coupling between $\nabla T$ and $\bm{u}$ in Eq.~\ref{viscous_heat_U}. 

Overall, the analyses in this section confirm that the size of the device in Fig.~\ref{fig:1_vortex} is sufficiently large for the VHE to be accurate enough to describe the viscous heat backflow and temperature inversion that emerge from the full, space-dependent LBTE solution.
We also note that the mesoscopic description of the hydrodynamic viscous heat backflow provided by the VHE has a computational cost that is more than two orders of magnitude lower than that needed to solve the LBTE \footnote{the VHE solution in Fig.~\ref{fig:1_vortex} was computed in less than two minutes on a laptop, while the corresponding LBTE solution in Fig.~\ref{fig:BTE_revision} took about 72 hours on one node (128 cores) of the Kelvin2 supercomputer from the Northern Ireland High Performance Computing center}.

\subsection{Computational details of LBTE simulation} 
\label{sub:computational_details_for_lbte_simulation}

To convert the harmonic and anharmonic force constants from \texttt{Quantum ESPRESSO}/\texttt{D3Q} format \cite{paulatto2013anharmonic,paulatto2015first} we proceeded as follows. We used the \texttt{Phonopy} code \cite{togo_first-principles_2023,Phonopy_parser} to parse the harmonic force constants from \texttt{Quantum ESPRESSO} to \texttt{FORCE\_CONSTANTS} format, which is readable by the LBTE solver \texttt{BTE-Barna}\cite{raya-moreno_bte-barna_2022}. We have then inverted the code \texttt{import\_shengbte} from the \texttt{D3Q/thermal2} package \cite{thermal2module} to convert the third-order anharmonic force constants from \texttt{Quantum ESPRESSO}-\texttt{D3Q} to \texttt{ShengBTE} \cite{li2014shengbte} format, since the latter is readable by \texttt{BTE-Barna} \cite{raya-moreno_bte-barna_2022}. 
We used the standard diffusive boundary conditions implemented in \texttt{BTE-Barna} to produce Fig.~\ref{fig:BTE_revision}\textbf{c,h}, and to produce Fig.~\ref{fig:BTE_revision}\textbf{e,j} we implemented 
the reflective boundary conditions following Sec. 4.2.2.1. `Specular scattering' of Ref.~\cite{raya-moreno_heat_nodate} (such development will be made publicly available in a next release of BTE-Barna).
The collision matrix was computed using a 17x17x3 mesh, using the adaptive smearing and a \texttt{scalebroad} parameter \cite{li2014shengbte} equal to 0.1, and accounting for isotopic scattering \cite{tamura_isotope_1983} at natural abundance (98.9 \% $^{12}$C and 1.1 \% $^{13}$C). These parameters correspond to the highest accuracy with which the space-dependent solution of the LBTE can be evaluated with the computational resources at our disposal.
Using these parameters in the much less expensive calculation of the homogeneous (bulk) in-plane conductivity with the iterative method in \texttt{BTE-Barna}\cite{raya-moreno_bte-barna_2022} yields $\kappa (17{\times} 17{\times} 3, 70 K){=}$4482.89 W/mK, a value that is compatible within 15 \% with 
the computationally converged bulk conductivity reported in Tab.~I of Ref.~\cite{Simoncelli2020}, $\kappa (49{\times} 49{\times} 3, 70 K){=}$5302.27 W/mK (the latter was computed with the exact diagonalization `relaxon' method). 
The space-dependent solution of the LBTE was determined in a device effectively infinite in the out-of-plane direction and with transport depending only on the in-plane coordinates ($x,y$), i.e., under the same conditions of Fig~\ref{fig:1_vortex} (but still considering the full 3D Brillouin zone of graphite in the calculation of the LBTE scattering operator). In practice, such condition corresponds to considering that the device's aspect ratio is such that the phonon distribution function entering in the LBTE is independent from the out-of-plane ($z$) direction. 
To improve the accuracy of the LBTE solution, we accounted for the mirror symmetry about the x-axis using standard numerical techniques to preserve analytical symmetries \cite{greenshields_openfoam_2024}.

\section{Boundary conditions for time-dependent VHE simulations}
\label{sec:boundary_conditions_for_time_dependent_simulations}

\subsection{Thermalization lengthscale}
\label{sub:modeling_realistic_thermalization}
\begin{figure}[b]
\includegraphics[width=\columnwidth]{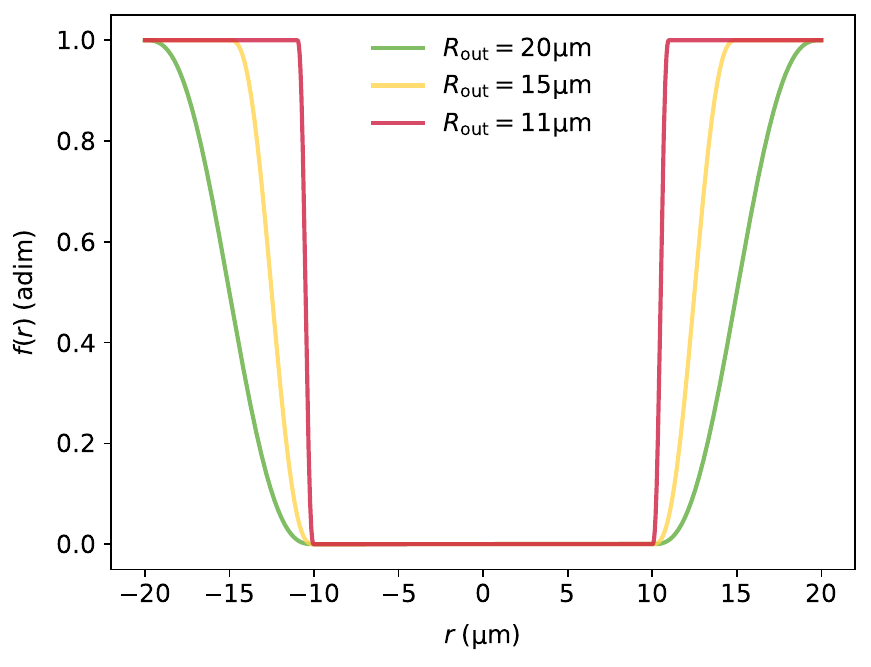}\\[-5mm]
\caption{\label{fig:sig_f}
\textbf{Sigmoid function for realistic thermalization.}
The sigmoid function~(\ref{eq:sigmoid}) is plotted for $R_\text{in}=10\mu m$ and various values of $R_\text{out}$. 
$R_\text{out}=11\mu m$ (red) simulates a nearly ideal thermalization (i.e., occurring over a very short lengthscale), $R_\text{out}=15\mu m$ (orange) is an estimate of a realistic thermalization length \cite{braun_spatially_2022}, and $R_\text{out}=20\mu m$ (green) represents a less efficient thermalization.}
\end{figure}

In actual experimental devices thermalization is not perfectly localized in space, but occurs over a finite lengthscale \cite{braun_spatially_2022}, i.e. there is a smooth transition from the device interior to the thermal bath.
Therefore, we model the boundaries relying on a compact sigmoid function \cite{Bump} to smoothly connect the thermal bath region, where temperature is fixed and drift velocity is zero (since at equilibrium phonons are distributed according to the Bose-Einstein distribution, which has zero drift velocity \cite{Simoncelli2020}), and the device's interior, where the evolution of temperature and drift velocity is governed by the viscous heat equations.
Specifically, we used the smooth-step function \cite{Sigmoid} that is ubiquitously employed in numerical calculations, i.e. a sigmoid Hermite interpolation between $0$ and $1$ of a polynomial of order $5$:
\begin{equation}
  \begin{split}
\hspace{-1mm}f(r){=}
    \begin{cases}
        0, &  \hspace{-8mm}{\rm if\;}\hspace{5mm}d(r) < 0\\
        1 ,& \hspace{-8mm} {\rm if\;}\hspace{5mm}d(r) > 1\\
        {6[d(r)]^5{-}15[d(r)]^4{+}10[d(r)]^3\hspace*{-1mm},} & {\rm otherwise}
    \end{cases}  
    \label{eq:sigmoid}    
      \end{split}
      \raisetag{20mm}
\end{equation}
where $d(r)=\left(\frac{r{-}R_{in}}{R_{out}{-}R_{in}}\right)$ and $r>0$, as well as $R_\text{in}>0$ and $R_\text{out}>0$. The function is second order continuous and saturates exactly to zero and one at the points $R_\text{in}$ and $R_\text{out}$, respectively.
We show in Fig.~\ref{fig:sig_f} the sigmoid function~(\ref{eq:sigmoid}) for $R_\text{in}=10\mu m$ and $R_\text{out}$ ranging from 11 to 20 $\mu m$: $R_\text{out}{\simeq}R_\text{in}$ simulates an ideal thermalization (occurring over a negligible lengthscale), while $R_\text{out}{\gg}R_\text{in}$ simulates a very inefficient (non-ideal) thermalization that occurs over a very large lengthscale.
Then, relying on the sigmoid~(\ref{eq:sigmoid}), we smoothly connected the thermal bath region, where $f(\bm r)=1$, $T({\bm r}, t)={T}_{\rm eq}$, and $\bm{u}({\bm r}, t)=\bm{0}$, with the device's interior region, where $f(\bm r)=0$ and the evolution of $T({\bm r}, t)$ and $\bm{u}({\bm r}, t)$ is governed by the VHE.
In formulas, we have that 
\begin{widetext}
\begin{align}
&f(\bm r)[T({\bm r}, t) - {T}_{\rm eq}]+[1-f(\bm r)]\left(C\frac{\partial T({\bm r}, t)}{\partial t} + \alpha^{ij}\frac{\partial u^j({\bm r}, t)}{\partial r^i} - \kappa_D^{ij} \frac{\partial^2 T ({\bm r}, t)}{\partial r^i \partial r^j} - \dot{q}({\bm r}, t)\right)=0,
 \label{viscous_heat_T_app}\\
&f(\bm r)u^i({\bm r}, t)+[1-f(\bm r)]\left(A^{ij}\frac{\partial u^j({\bm r}, t)}{\partial t} + \beta^{ij}
\frac{\partial T({\bm r}, t)}{\partial r^j} - \eta^{ijkl} \frac{\partial^2 u^k({\bm r}, t)}{\partial r^j \partial r^l} + \gamma^{ij} u^j({\bm r}, t)\right) = 0.\label{viscous_heat_U_app}
\end{align}
\end{widetext}
Details on how the non-ideal thermalization affects the magnitude of signatures of heat hydrodynamics are reported later in this document.
\subsection{Modeling realistic thermalization}
\label{sub:lattice_cooling_boundaries}
\begin{figure}[t]
\includegraphics[width=0.6\columnwidth]{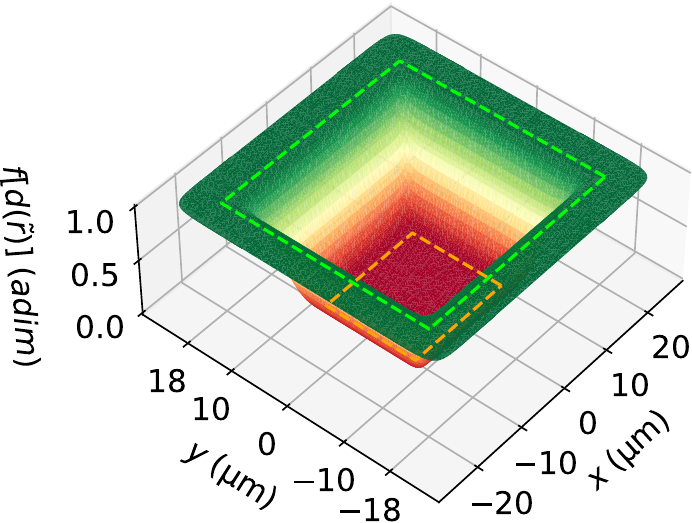}
\caption{\label{fig:sig_3d}
\textbf{Realistic thermalization in rectangular geometry.} Green (outside the dahsed-lime line) is the perfectly thermalized region, where the sigmoid function $f\big(d(\Tilde{r})\big){=}1$, and the thermal reservoir ensures that temperature is perfectly constant and the drift velocity is zero. Red (inside the dashed-orange line) is the inner "free" region where $f\big[d(\Tilde{r})\big]{=}0$ and the evolution of $T$ and $\bm{u}$ is determined by the VHE.}
\end{figure}

In order to apply the realistic-thermalization boundary conditions discussed in Sec.~\ref{sub:modeling_realistic_thermalization} to the rectangular device of Fig.~\ref{fig:fig2}, we employed a smoothed rectangular domain, since we noted that smooth simulation domains yielded better computational performances (faster convergence with respect to the discretization mesh, reduced numerical noise) compared to non-smoothed rectangular domains.
The equation defining the smoothed rectangular domain employed in Figs.~\ref{fig:fig2},\ref{fig:rect_freq},\ref{fig:freq_VHE_vs_DPLE} is:
\begin{equation}
    \left(\frac{|x|}{a_{\rm out}}\right)^{\frac{2a_{\rm out}}{\lambda }}+\left(\frac{|y|}{b_{\rm out}}\right)^{\frac{2b_{\rm out}}{\lambda }}=1,
    \label{eq:recticircle}
\end{equation}
where $a_{\rm out}$ and $b_{\rm out}$ are the sides of the rectangular domain where the thermalization is perfect (outside the dashed-lime region in Fig.~\ref{fig:sig_3d}); $\lambda$ controls the smoothness of the corners, and we used $\lambda=\frac{a_{\rm out}}{3}$ to obtain a domain smooth enough for numerical purposes and sharp enough to practically model a rectangular device.
We note that to employ the sigmoid function~(\ref{eq:sigmoid}) in a rectangular geometry, it has to be adapted. 
Specifically, in Eq.~(\ref{eq:sigmoid}) in place of the distance $r$ we used a distance of the form 
\begin{equation}
	\Tilde{r}(x,y)=\left[\left(\frac{|x|}{a_{\rm out}}\right)^{\frac{2a_{\rm out}}{\lambda }}+\left(\frac{|y|}{b_{\rm out}}\right)^{\frac{2b_{\rm out}}{\lambda }}\right]^{1/p},
	\label{eq:dist}
\end{equation}
where $p$ controls the nonlinearity in which the two terms in Eq.~(\ref{eq:dist}) are combined.
Then, the distance~(\ref{eq:dist}) is nested into the function $d(\Tilde{r})=\left(\frac{\Tilde{r}(x,y){-}R_{in}}{R_{out}{-}R_{in}}\right)$. It is evident that $d(\Tilde{r})< 0$ when $\Tilde{r}(x,y){<}R_{\rm in}$, and $d(\Tilde{r})>1$ when $\Tilde{r}(x,y){>}R_{\rm out}$. Therefore, if one sets $R_\text{in}=\Tilde{r}(a_{\text{in}},b_{\text{in}})$, and $R_\text{out}=\Tilde{r}(a_{\text{out}},b_{\text{out}})$, and inserts $d(\Tilde{r})$ into Eq.~(\ref{eq:sigmoid}), the resulting function $f[d(\Tilde{r})]$ is exactly zero inside the rectangle having sides $a_{\text{in}},b_{\text{in}}$ (red region in Fig.~\ref{fig:sig_3d}, where the evolution of temperature and drift velocity is determined by the VHE) and exactly one outside the rectangle having side $a_{\text{out}},b_{\text{out}}$ (green region in Fig.~\ref{fig:sig_3d}, where the thermal reservoir ensures that temperature is perfectly constant and the drift velocity is zero).
The value of $p$ in Eq.~(\ref{eq:dist}) affects how the sigmoid function goes from zero to one between the dashed-orange and dashed-green rectangles, and the aforementioned value $p=20$ was chosen after checking that the overall transition was sufficiently smooth for numerical purposes (Fig.~\ref{fig:sig_3d}). In the next section we investigate how the thermalization lengthscale, $a_{\rm out}{-}a_{\rm in}$, affects lattice cooling (this investigation will be done varying only $a_{\rm out}$, $b_{\rm out}$ in the above equations and keeping all the other parameters fixed).

\subsection{Dependence of VHE lattice cooling on the boundary conditions }
\label{ssec:dependence_of_lattice_cooling_from_boundary_conditions}
\begin{figure}[b]
\includegraphics[width=\columnwidth]{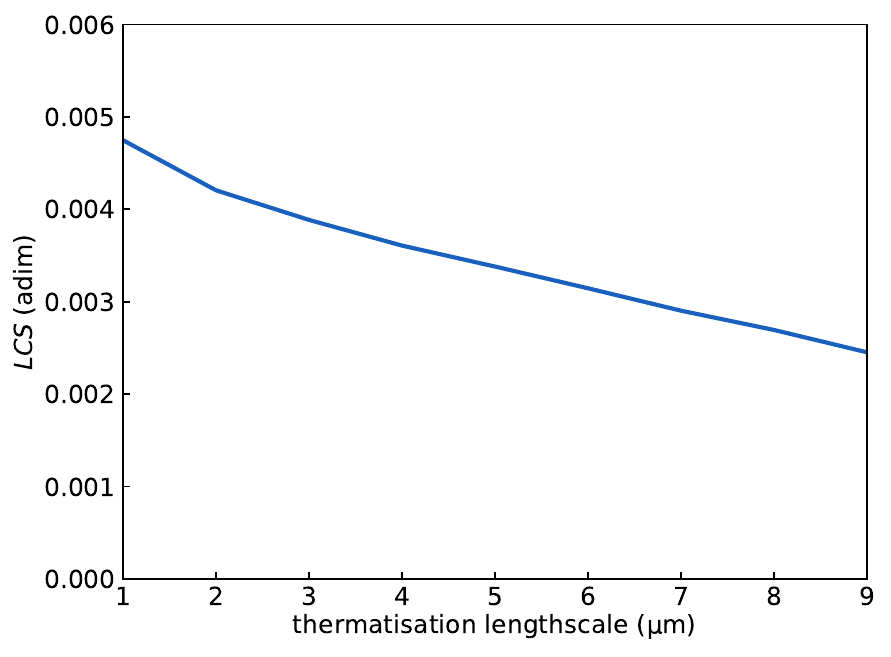}
\caption{\label{fig:sig}\textbf{Dependence of lattice cooling strength on the thermalization lengthscale.} 
We show how LCS in a device having non-thermalized region equal to $20\ \mathrm{\mu m}{\times} 16\ \mathrm{\mu m}$ (as in Fig.~\ref{fig:fig2}) depends on the thermalization lengthscale, i.e., the length over which the sigmoid function shown in Fig.~\ref{fig:sig_f} goes from zero to one.
Even if LCS decreases as the thermalization lengthscale increases, LCS remains appreciable even at thermalization lengthscales as large as 9 $\mu m$.
}
\end{figure}

\setcounter{figure}{10}
\renewcommand{\thefigure}{SF \arabic{figure}}
\begin{figure*}
\centering
\includegraphics[width=\textwidth]{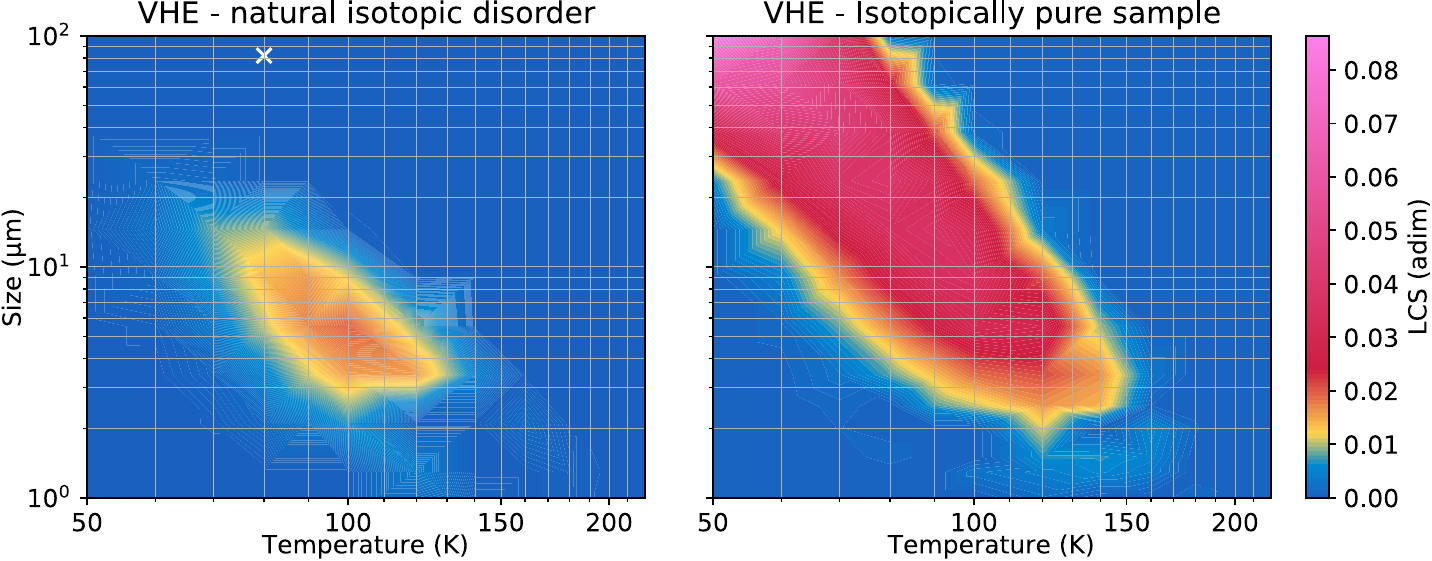}
\caption{\label{fig:iso_pure} 
\textbf{Lattice cooling strength as a function of size, temperature, and isotopic disorder.} 
In natural samples (left) the lattice cooling strength (LCS, Eq.~(\ref{eq:LCS})) is weaker compared to isotopically purified samples (right). The white cross in the left panel represents the simulation which is closest to the single-spot experiment discussed in Sec. 3 of the Supplementary Material of Ref.~\cite{Jeong2021}, i.e., $\sigma_x=6 \mu m$ in Eq.~(\ref{eq:pert_LC}) and simulation boundaries separated by a large distance (80 $\mu m$); the lack of lattice cooling predicted under these conditions agrees with the experiments.}
\end{figure*}
In this section we investigate how the magnitude of the viscous temperature oscillations discussed in Fig.~\ref{fig:fig2} depends on the thermalization lengthscale at the boundaries.
In order to quantify the magnitude of temperature oscillations, we define a descriptor to capture the ``lattice cooling strength'' (LCS) 
\begin{equation}
	LCS= \frac{T_\mathrm{eq}-T_\mathrm{min}}{T_\mathrm{max}-T_\mathrm{min}},
	\label{eq:LCS}
\end{equation}
where $T_\mathrm{max}=\max_t[T(\bm{r}_c,t)]$ and $T_\mathrm{min}=\min_t[T(\bm{r}_c,t)]$ are the maximum and minimum temperatures, respectively, observed during the relaxation in the point $\bm{r}_c{=}(x_c,0)$, where the perturbation
\begin{equation}
	\dot{q}(\bm{r},t){=}\mathcal{H}\;\theta(t_{\rm heat}{-}t) \exp\left[-\tfrac{(x+x_c)^2}{2\sigma_x^2}{-}\tfrac{y^2}{2\sigma_y^2}\right]
	\label{eq:pert_LC}
\end{equation}
is centered. We recall that we used $\mathcal{H}=0.013\tfrac{W}{\mu m^3}$, $t_{\rm heat}=0.4ns$, $x_c=5 \mu m$, $\sigma_x=2\mu m$, $\sigma_y=2.8\mu m$, to ensure that the perturbation created causes variations within 10 \% of the equilibrium temperature ($T_{eq}=80 K$). 
Clearly, LCS is zero if lattice cooling does not take place, and assumes a positive value if lattice cooling occurs.
To gain insights on how lattice cooling is affected by the thermalization lengthscale, we fixed the 
size of the non-thermalized region (where $f(r) = 0$) to $20\ \mathrm{\mu m}\times 16\ \mathrm{\mu m}$ (i.e. exactly as in 
Fig.~\ref{fig:fig2}), and studied the variation of the LCS arising from the solutions of the VHE.
We show in Fig.~\ref{fig:sig} that LCS assumes smaller values as the thermalization lengthscale increases. 
We highlight how lattice cooling persists at thermalization lengthscales as large as $9\ \mu m$ (equal to half the size of the simulated device).
In Figs.~\ref{fig:fig2}, \ref{fig:figDPLE}, \ref{fig:figFourier}, and in the following, we use a thermalization lengthscale of 2 $\mu m$, a value that is expected to be a realistic representation of experimental conditions \cite{braun_spatially_2022}.

\subsection{Effects of isotopic disorder, average temperature, and size on VHE lattice cooling}
\label{sec:LCS_iso_temp_size}
Finally, we discuss how isotopic-mass disorder, sample's size, and temperature affect the LCS emerging from the VHE in graphite. 
The effect of isotopic-mass disorder is taken into account by the parameters appearing in the VHE. These were computed from first principles for graphite with natural concentration of isotopes (98.9 \% $^{12}$C, 1.1 \% $^{13}$C) and for isotopically purified samples (99.9 \% $^{12}$C, 0.1 \% $^{13}$C), see Sec.~\ref{sec:parameters_entering_in_the_viscous_heat_equations} for details. 
The effect of sample size was considered by uniformly rescaling the simulation domain~(\ref{eq:recticircle}) and the perturbation~(\ref{eq:pert_LC}). We accounted for grain-boundary scattering as discussed in Sec.~\ref{sub:finite_size}, considering a grain size of $20 \mu m$ that is realistic for high-quality samples \cite{Jeong2021,Ding2022}. 
The temperature was monitored in the same point where the perturbation was applied.
The effect of equilibrium temperature was taken into account through the temperature dependence of the parameters entering the VHE, as shown by Table I in Ref.~\cite{Simoncelli2020} and Table.~\ref{tab:param_graphite} at the end of this manuscript.

We show in Fig.~\ref{fig:iso_pure} that in natural samples lattice cooling is maximized at temperatures around 70-120 K and is significant in devices having size 5-20 $\mu m$. Importantly, LCS becomes negligible in devices having size larger than 30 $\mu m$.
We note that Ref.~\cite{Jeong2021} performed pump-probe experiments in graphite using heaters with radius equal to 6 $\mu m$ (details are reported in the Supplementary Material of that reference). 
When we used a simulation setup similar to the experiments of Ref.~\cite{Jeong2021}, i.e. $\sigma_x=6 \mu m$ in Eq.~(\ref{eq:pert_LC}) and simulation boundaries separated by a large distance (80 $\mu m$), we did not find temperature oscillations (LCS=0 in the point highlighted with the white cross in the upper panel of Fig.~\ref{fig:iso_pure}), in agreement with the experiments of Ref. \cite{Jeong2021}.
We also simulated the ring-shaped geometry used by \citet{Jeong2021}: in Sec.~\ref{sec:comparison_between_vhe_and_bte} we show that the solution of the VHE in such geometry is in agreement with the corresponding LBTE solution discussed in Ref.~\cite{Jeong2021} (also in agreement with experiments reported in the same reference).

\section{Comparison between viscous, inviscid, and diffusive relaxations}
\label{sec:effects_of_viscosity_on_lattice_cooling}

\subsection{Inviscid (DPLE) relaxation}
\begin{figure}[b]
	\centering
	\includegraphics[width=0.49\textwidth]{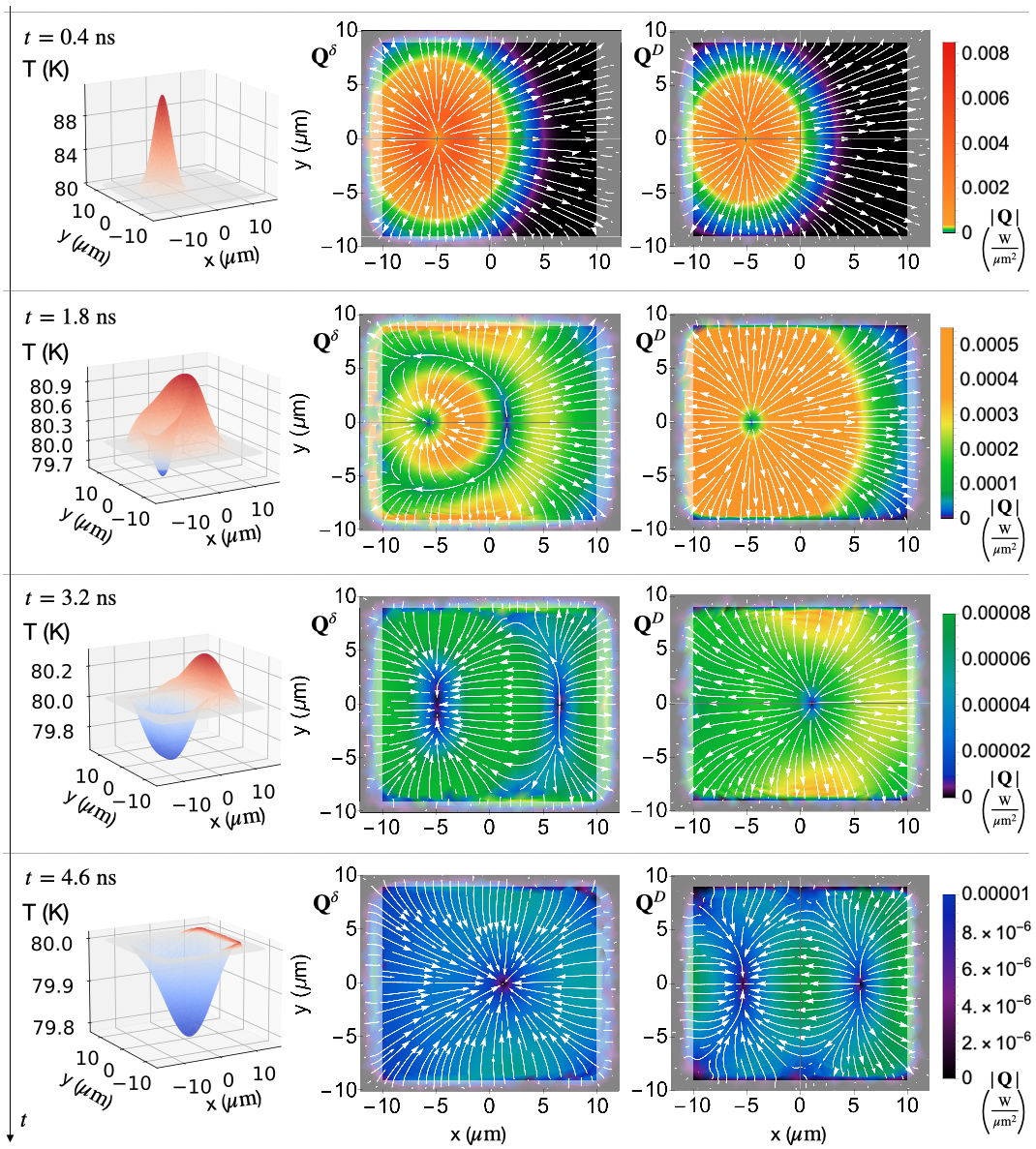}
	\caption{\textbf{Transient hydrodynamic heat backflow and lattice cooling in the DPLE (inviscid) limit.}
	We show the DPLE predictions for the relaxation in time of a temperature perturbation (obtained applying a localized heater to a device in the time interval from 0 to 0.4 ns) in a graphitic device thermalized at 80 K the boundaries (thermalization occurs in shaded regions, see SM~\ref{sub:modeling_realistic_thermalization}).
	Rows show different instants in time for temperature (left column), temperature-gradient heat-flux component ($\bm{Q}^{\delta}$, central column), and drifting heat-flux component ($\bm{Q}^{D}$, right column). 
	The emergence of lattice cooling, i.e. a temperature locally and transiently lower than the initial value $T{=}80$ K, is evident and originates from the lagged coupled evolution of $\bm{Q}^{\delta}$ and $\bm{Q}^{D}$. }
	\label{fig:figDPLE}
\end{figure}
In this section we discuss how the thermal viscosity affects temperature oscillations. 
Fig.~\ref{fig:figDPLE} shows that solving the VHE in the inviscid limit ($\eta=0$)---which is analytically equivalent to solving the DPLE equation, see Sec.~\ref{sec:DPLE_derivation}---yields temperature oscillations with a higher magnitude compared to those obtained in the case of the VHE (Fig.~\ref{fig:fig2}). 
Therefore, Fig.~\ref{fig:figDPLE} shows that a finite viscosity is not necessary to observe lattice cooling, since a lagged response between temperature gradient and heat flux can emerge also in the inviscid limit, as detailed in Sec.~\ref{sec:DPLE_derivation}.
However, we will show later in Sec.~\ref{sec:comparison_between_vhe_and_bte} that accounting for viscosity is necessary to reproduce the relaxation timescales emerging from the time-dependent solution of the microscopic LBTE. Fig.~\ref{fig:freq_VHE_vs_DPLE} in the main text discusses how accounting for thermal viscosity is necessary to capture lengthscales at which hydrodynamic behavior was observed by \citet{Huberman2019} and \citet{Ding2022}.

\subsection{Diffusive (Fourier) relaxation}
\label{sub:diffusive_}
\begin{figure}[htbp]
	\centering
	\includegraphics[width=0.73\columnwidth]{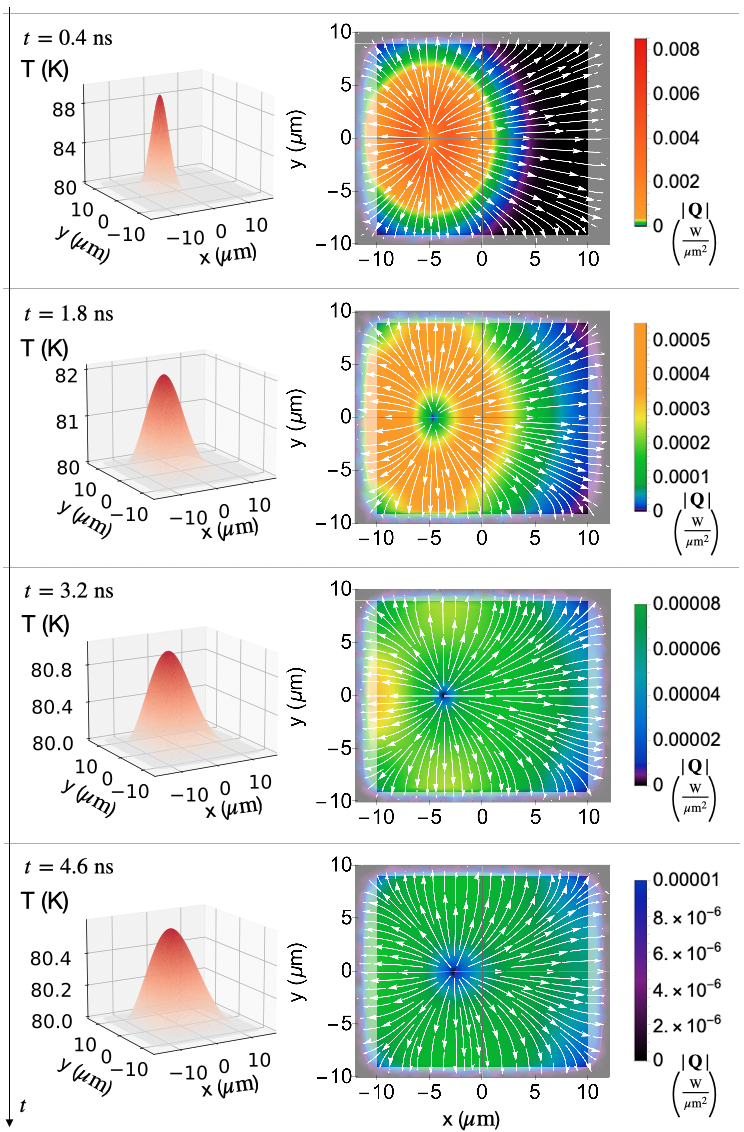}
	\caption{\textbf{Temporal evolution of a localized temperature perturbation according to Fourier's law.} 
Rows show different instants in time for the temperature profile (left) and heat flux (right).}
	\label{fig:figFourier}
\end{figure}

We show in Fig.~\ref{fig:figFourier} that temperature oscillations do not emerge from Fourier's law.
This behavior was trivially expected from an analytical analysis of Fourier's diffusive equation, 
$C\frac{\partial T}{\partial t}{-} \kappa^{ij}\frac{\partial^2 T}{\partial r^ir^j} = 0$, whose smoothing property \cite{skinner_university_nodate} implies that the evolution of a positive temperature perturbation relaxes to equilibrium decaying in a monotonic way.
Consequently, within Fourier's law heat backflow does not emerge.
\subsection{Comparing VHE, DPLE \& Fourier's relaxations}
\label{sub:inviscid}

\begin{figure}[htbp]
	\centering
	\includegraphics[width=0.9\columnwidth]{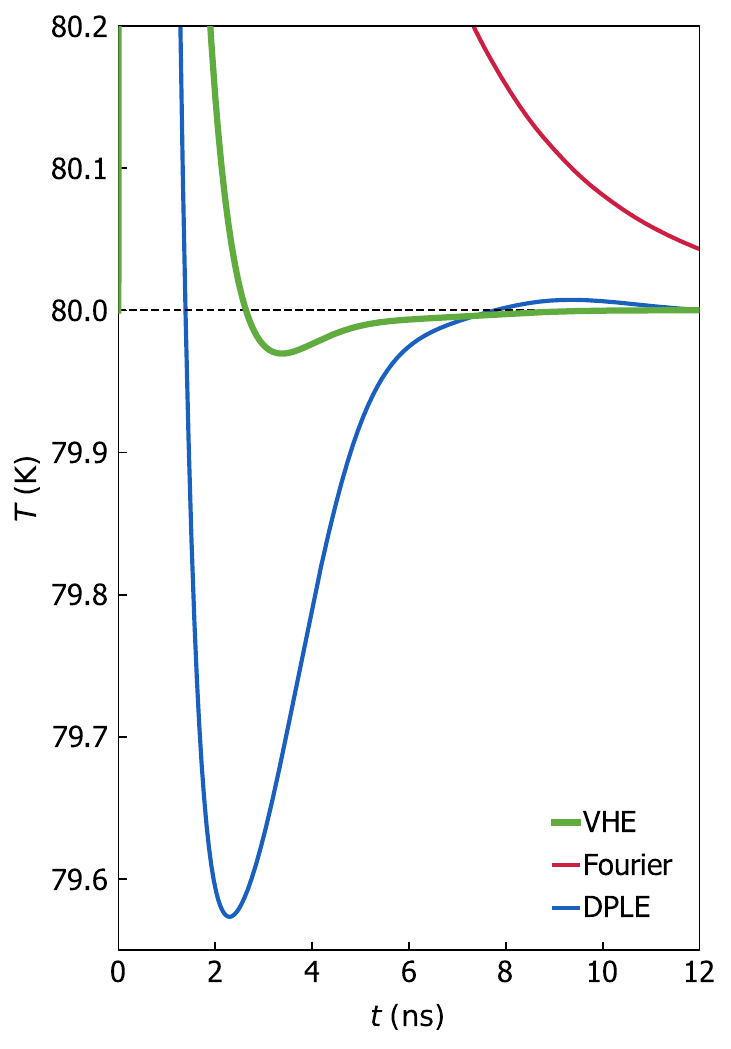}\\[-4mm]
	\caption{\textbf{Temperature relaxation from VHE, DPLE, and Fourier's law.} 
	The temporal evolution of the temperature in the center ($5 \mu m,0\mu m$) of the external heat source (switched off at t=0.4 ns, see text for details) is shown in red for Fourier's law, in blue for the DPLE, and in green for the VHE. Temperature oscillations are visible both in the viscous VHE and inviscid DPLE cases (with a larger amplitude in the inviscid case), and are absent in Fourier's law.}
	\label{fig:compare_all}
\end{figure}

The device geometry, the transient perturbation, and the boundary conditions used in Figs. \ref{fig:fig2}, \ref{fig:figDPLE}, \ref{fig:figFourier} are exactly the same. In Fig.~\ref{fig:compare_all}, we compare the predictions from the viscous VHE, the inviscid DPLE, and the diffusive Fourier's law for the evolution in time of the temperature in the point $\bm{r}_c{=}(5 \mu m,0\mu m$), where the temperature perturbation is centered.
We see that both DPLE and VHE yield a wave-like relaxation for temperature, the inviscid (DPLE) relaxation is faster and yields stronger oscillatory behavior compared to the VHE relaxation. In contrast, an oscillatory relaxation for temperature is absent in Fourier's law.

\section{Comparison between LBTE and VHE in a ring-shaped geometry} 
\label{sec:comparison_between_vhe_and_bte}
We compare the time-dependent solution of the LBTE with full collision matrix discussed in Fig. 4g and Supplementary Fig. S7 of Ref.~\cite{Jeong2021} with the corresponding VHE solution. We consider a device made of pure graphite (100 \% $^{12}$C as in Ref.~\cite{Jeong2021}, parameters in Tab.~\ref{tab:param_graphite_100}). The device is very long (infinite) along the out-of-plane crystalline direction, 
and has a cross-section of $40\mu\mathrm{m}\times40\mu\mathrm{m}$ and is thermalized at $\bar T$ at its boundaries.
A ring-shaped heater of radius 15 $\mu$m, Gaussian profile with FWHM 3 $\mu$m, operates for 0.4 ns at the beginning of the simulation. 
As in Ref.~\cite{Jeong2021}, we monitor the evolution of the temperature in the center of the ring (the temperature is determined by averaging the local temperature at the domain's center with a Gaussian of FWHM $3\mu\mathrm{m}$).

\begin{figure}[b]
  \includegraphics[width=0.95\columnwidth]{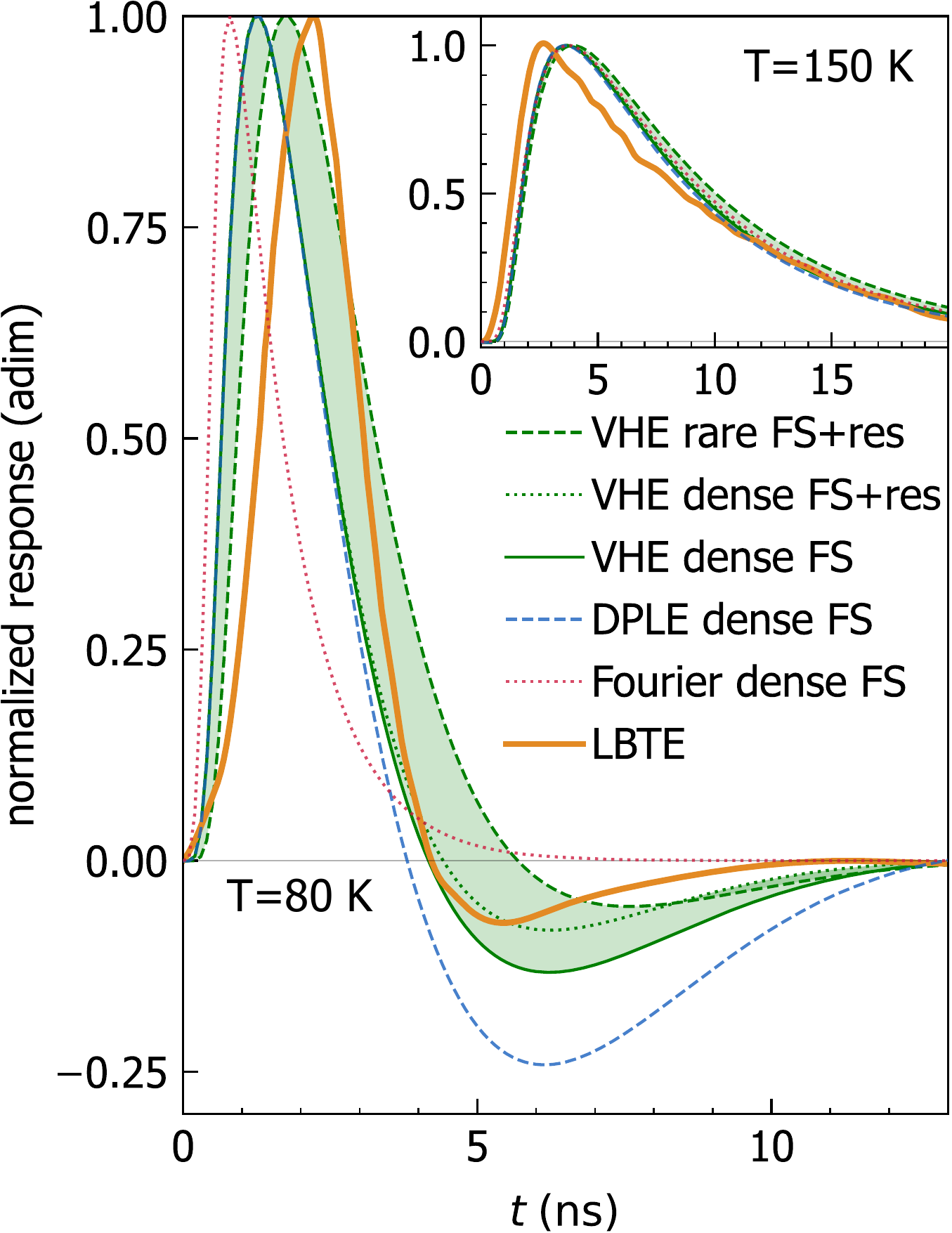}\\[-3mm]
 \caption{\textbf{BTE versus VHE transient heat backflow.} Temporal evolution of temperature at the center of a 15-$\mu$m radius ring-shaped perturbation in a device made of isotopically pure graphite. Around 80 K there is agreement between the lattice cooling predicted by the LBTE (orange, taken from Fig. 4g of Ref.~\cite{Jeong2021}) and by the VHE (green); the inviscid DPLE (dashed blue) overestimates LBTE's lattice cooling, and Fourier's law yields a diffusive (always-positive) relaxation. 
 For the VHE, we show how the solution changes when accounting for finite-size effects in the presence of: (i) dense intrinsic scattering events (solid green); (ii) dense intrinsic scattering and thermalized boundaries that increase the Umklapp dissipation $\gamma$ (dotted green); (iii) rare intrinsic scattering  and thermalized boundaries that increase $\gamma$ (dashed green). Inset: absence of hydrodynamic lattice cooling in the same device around an average temperature of 150 K.}
  \label{fig:relaxation_dynamics_Jeong}
\end{figure}

We show in Fig.~\ref{fig:relaxation_dynamics_Jeong} that around $\bar{T}=80 K$, the full solution of the LBTE from Ref.~\cite{Jeong2021} predicts lattice cooling (normalized response $r(t){=}(T(t){-}\bar{T})/(\max[T(t)]{-}\bar{T})$ assuming negative values) to have a magnitude and evolution timescale that are in agreement with those predicted by the VHE. We discuss different treatments of finite-size (FS) effects. "Dense FS" ("rare FS") refers to using thermal conductivity and viscosity determined by Eqs.~(\ref{eq:conductivity_comb},\ref{eq:viscosity_comb}) with ballistic components computed using Eqs.~(\ref{eq:conductivity_fs},\ref{eq:viscosity_fs}) with the 40 $\mu m$ characteristic lengthscale of the domain (half that length); this corresponds to having intrinsic phonon-phonon scattering events that are dense (rare) compared to phonon-boundary scattering events\cite{carruthers_theory_1961,cepellotti_transport_2017}.
Finally, “res” refers to accounting for an additional damping mechanism also on the momentum contribution to the conductivity. This affects the momentum dissipation \cite{Simoncelli2020} 
 $D^{ij}_{U,{\rm res}}(L_S)=\frac{1}{V}\sum_{\nu} \phi^i_\nu \frac{\bm{v}_\nu}{|L_S|} \phi^j_{\nu}$ and $D^{ij}_{U,{\rm tot}}=D^{ij}_{U}+D^{ij}_{U,{\rm res}}(L_S)$, where $L_S$ is the sample size (divided by two in the limit in which intrinsic phonon–phonon scattering is rare compared to phonon-boundary scattering), and $D^{ij}_U=\frac{1}{V^2}\sum_{\nu,\nu'} \phi^i_\nu \breve{\Omega}^{U}_{\nu,\nu'} \phi^j_{\nu'}$ is the parameter describing intrinsic momentum dissipation defined earlier in this manuscript. We recall that these parameters are related to the parameter $\gamma^{ij}{=}\sqrt{A^i A^j} D_U^{ij}$ appearing in Eq.~(\ref{viscous_heat_U}).
Accounting for this additional damping mechanism might be justified by the fact that the boundaries are thermalized, hence out-of-equilibrium phonon excitation interacting with the boundaries is forced to transition to the Bose-Einstein distribution function, implying that also the phonons that are contributing to the momentum relaxons have an upper-bound on their mean-free path (we note that in the limit of adiabatic boundaries instead, the phonons are scattered by the boundaries but are not constrained to assume the Bose-Einstein distribution corresponding to an externally imposed temperature).
The shaded area highlights how much lattice cooling changes among these different treatments of boundary conditions. Overall, the changes due to different treatment of finite-size effects are smaller than those induced by neglecting the viscosity. In particular, the magnitude of lattice cooling (minimum of the normalized response) is visibly affected by the thermal viscosity, and considering (neglecting) the viscosity using the VHE (DPLE) yields a response in better (worse) agreement with the LBTE one.
In addition, we highlight how Fourier's law predicts a diffusive, always positive response that is in strong disagreement with the oscillatory behavior emerging from the LBTE and confirmed by experiments \cite{Jeong2021}.

While the VHE and LBTE predict a relaxation that is in good agreement at the times at which lattice cooling occurs, there are appreciable differences at very short times. As mentioned in the main text, the VHE have been derived \cite{Simoncelli2020} by coarse-graining the LBTE for the microscopic phonon distribution function into
partial differential equations for local temperature and drift-velocity fields \cite{Simoncelli2020}; such coarse-graining neglected gradients in space and time of the out-of-equilibrium phonon distribution. As such, the VHE are expected to be less accurate at very short timescales (as well as at very short lengthscales, as discussed in Sec.~\ref{sub:steady_state_viscous_heat_backflow_from_the_space_dependent_solution_of_the_full_boltzmann_transport_equation}).
Nevertheless, we highlight how these differences at very short times are practically irrelevant for the phenomenon of lattice cooling occurring at $5\lesssim t \lesssim 10$ ns.

The inset of Fig.~\ref{fig:relaxation_dynamics_Jeong} shows that the response predicted by the time-dependent solution of LBTE is always positive when the perturbation is applied around an average temperature of $\bar{T}=150 K$---i.e., lattice cooling does not occur in this case---and such behavior is in reasonable agreement with the evolution predicted by the VHE.
We also note that in this case the inviscid DPLE and the VHE have practically indistinguishable solutions; this shows that the thermal viscosity has negligible effects on the thermal relaxation around $\bar{T}=150 K$. Finally, at 150 K Fourier's law predicts an always-positive, diffusive relaxation that is qualitatively similar but slightly delayed compared to the relaxations predicted by the LBTE, VHE, and DPLE; this originates from neglecting the delays between temperature gradient and heat flux within Fourier's law.

Overall, these results demonstrate that the VHE capture the salient features of the time-dependent temperature relaxation emerging from the LBTE, showing that the thermal viscosity quantitatively affects transient signatures of heat hydrodynamics around $\bar{T}=80 K$.

\section{Parameters entering the viscous heat equations}
\label{sec:parameters_entering_in_the_viscous_heat_equations}

In order to simplify the notation for the parameters entering the VHE~(\ref{viscous_heat_T},\ref{viscous_heat_U}), we define the following parameters: $\kappa^{ij} = \kappa^{ij}_{P} + \kappa^{ij}_{C}$ (see Refs.~\cite{simoncelli2019unified,simoncelli2021Wigner,caldarelli_many-body_2022,di_lucente_crossover_2023} for details on $\kappa_{\rm C}$), $\alpha^{ij}{=}W_{0j}^i\sqrt{\overline T A^jC}$, $\beta^{ij}{=}\sqrt{\frac{CA^i}{\overline T}} W_{i0}^j$, $\gamma^{ij}{=}\sqrt{A^i A^j} D_U^{ij}$. 
In these expressions, $C$ is the positive-definite specific heat \cite{Simoncelli2020}, 
${W}_{0 j}^i$ is the velocity tensor that arises from the non-diagonal form of the velocity operator in the basis of the eigenvectors of the normal part of the scattering matrix, 
$\bar{T}$ is the reference (equilibrium) temperature on which a perturbation is applied,
$A^i$ is the specific momentum in direction $i$. More details on these parameters can be found in Ref.~\cite{Simoncelli2020}.

We now discuss the symmetries of these tensors:

First, we note that the specific momentum n $A^i=$ is definite poisitive. To show this, we recall its definition, Eq. (A14) of Ref.~\cite{Simoncelli2020}:
\begin{equation}
	A^i=\frac{1}{k_B\bar T V}\sum_\nu \hbar^2 (q^i)^2 N_\nu(N_\nu+1)
	\label{eq:spec_momentum}
\end{equation}
where $k_B$ is Boltzmann's constant, $\bar T$ is the equilibrium temperature, $\hbar$ is the reduced Planck constant, and $q^i$ is the $i$-th Cartesian component of the phonon crystal momentum and $N_\nu=[\exp(\hbar\omega_\nu/k_B \bar T)-1]^{-1}$ is the Bose-Einstein distribution of phonons with wavevector and mode index denoted by $\nu=(\bm{q},s)$ and energy $\hbar\omega_\nu$. 
Clearly,  Eq.~(\ref{eq:spec_momentum}) is positive definite because sum of positive contributions, and one can have $A^i\neq A^j$ for $i\neq j$ in anisotropic materials (e.g. for graphite $A^x=A^y\neq A^z$, as shown by Fig.~9 in Ref.\cite{Simoncelli2020}).

Second, we highlight that in the limit of a linear-isotropic band dispersion, energy and momentum are proportional, implying that $W_{0i}^j$ is positive definite and proportional to the identity. 
Considering that the specific heat $C$ and specific momentum $A^i$ are both positive definite, this implies that $\alpha^{ij}$ is also positive definite.

Third, it is straightforward to show that the coupling tensors $\alpha^{ij}$ and $\beta^{ij}$ are related as follows: 
\begin{equation}
\alpha^{ij}{=}\sqrt{C A^i \overline T }W_{0i}^j=\beta^{ij}\bar T
\end{equation}

Fourth, it follows from the symmetry of the LBTE collision operator\cite{Simoncelli2020} that $\gamma^{ij}$ is symmetric and positive definite.
The symmetries of the viscosity tensor are discussed in Sec.~\ref{sec:viscosity_decomposition}. The positive-definiteness and symmetry of the thermal conductivity tensor is discussed in Refs.~\cite{chaput_direct_2013,cepellotti_thermal_2016}.

For the materials in focus here, all the tensors $\alpha^{ij}$ are diagonal (off-diagonal componets are at least four orders of magnitude smaller than the diagonal componets), with the in-plane components differing from the in-plane ones (see Fig. 8 and 9 of Ref.~\cite{Simoncelli2020}).

In the case of an isotropic material, $A^i=A\forall i$, $\gamma^{ij}=\gamma\delta^{ij}$, and $\tau=A/\gamma$ becomes equivalent to Eq.~(15) of \cite{sendra_hydrodynamic_2022}.

We conclude by noting that we employed the approximation of considering the parameters entering the VHE, DPLE, and Fourier's law to be independent of frequency. This approximation is accurate in the MHz range considered here, as discussed in Ref.~\cite{chaput_direct_2013}.

\subsection{Graphite} 
\label{sub:graphite}
\begin{figure*}[htbp]
	\centering
	\includegraphics[width=0.85\textwidth]{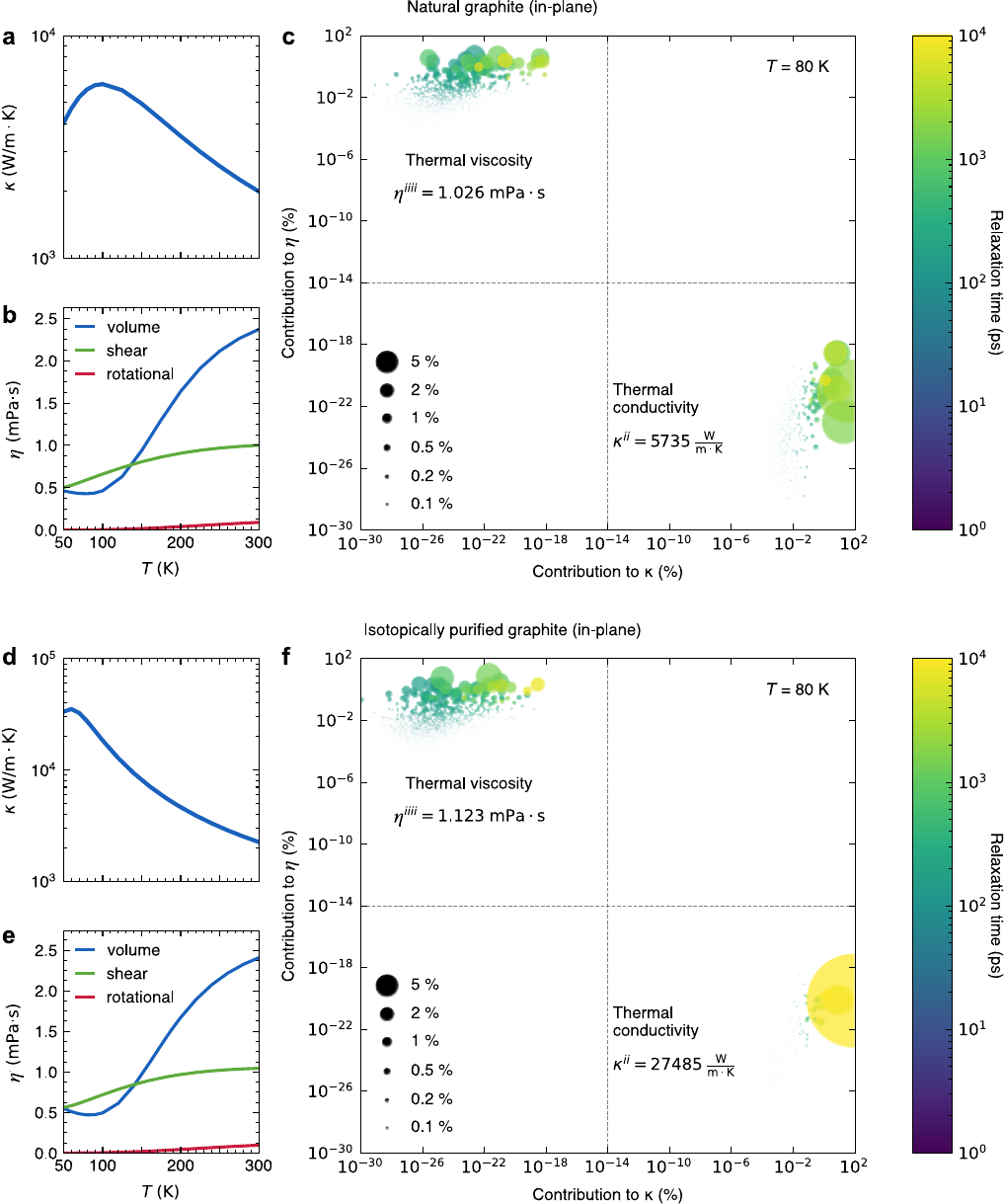}
	\caption{\textbf{Thermal conductivity and viscosity of graphite at various isotopic concentrations, and their decomposition in terms of relaxons.}
	The three plots in the upper row (\textbf{a, b, c}) refer to graphite at natural isotopic concentration (98.9 \% $^{12}$C, 1.1 \%$^{13}$C), the three in the bottom row (\textbf{d, e, f}) to isotopically purified graphite (99.9 \% $^{12}$C, 0.1 \%$^{13}$C).
	In each row, the in-plane bulk thermal conductivity as a function of temperature is shown in the upper left inset (\textbf{a, d}), while the bulk principal viscosity components as a function of temperature are shown in the lower left inset (\textbf{b, e}). 
	We note that the bulk thermal conductivity is significantly affected by isotopic impuriries; in contrast, the bulk principal viscosity components are not. 
	The right panels (\textbf{c, f}) shows the decomposition of conductivity and viscosity in terms of relaxon excitations. Each scatter point corresponds to a relaxon, with its color indicating the relaxation time and its area scaled proportionally to the sum of its relative contributions to both thermal conductivity and viscosity. Dashed lines emphasize how relaxons with odd parity determine the thermal conductivity and contribute negligibly to thermal viscosity, while relaxons with even parity determine thermal viscosity and have negligible impact on thermal conductivity
	\cite{Simoncelli2020}. 
	}
	\label{fig:relaxon_graphite}
\end{figure*}

All the parameters were computed from first principles, see Ref.~\cite{Simoncelli2020} for details. The parameters for graphite with natural-abundance isotopic-mass disorder (98.9 \% $^{12}$C, 1.1 \% $^{13}$C) are reported in Tab.~\ref{tab:param_graphite_nat}. The parameters for isotopically purified samples (99.9 \% $^{12}$C, 0.1 \% $^{13}$C) are reported in Tab.~\ref{tab:param_graphite}.
In Fig.~\ref{fig:relaxon_graphite} we analyze the influence of isotopic-mass disorder on thermal conductivity and viscosity, also resolving how each individual relaxon excitation contributes to them.
Tab.~\ref{tab:param_graphite_100} reports the VHE parameters for pure graphite (100 \% $^{12}$C).

\begin{table*}[htbp!]
  \caption{Parameters entering the viscous heat equations for graphite with natural isotope content. We report here only the in-plane components of the tensors needed to perform the calculations:
$\kappa^{ij}_{\rm P}=\kappa_{\rm P}\delta^{ij}$, $\kappa^{ij}_{\rm C}=\kappa_{\rm C}\delta^{ij}$, $K^{ij}_{S}=K_{S}\delta^{ij}$, ${D}_U^{ij}={D}_U\delta^{ij}$, $A=A^i\;\forall \;i$, $W^j_{0i}=W^j_{i0}= W\delta^{ij}$, where the indexes $i,j$ represent the in-plane directions $x,y$ only ($i,j=1,2$).}
  \label{tab:param_graphite_nat}
  \centering
\resizebox{\textwidth}{!}{%
  \begin{tabular}{cccccccccccccc}
  \hline
  \hline
  {T [K]} &
  {$\kappa_{\rm P}\;\big[\rm\frac{W}{m{\cdot}K}\big]$} & {$\kappa_{\rm C}\;\big[\rm\frac{W}{m{\cdot}K}\big]$} & ${K_S\;\big[\rm\frac{W}{m^2{\cdot}K}\big]}$ &
  $\eta_{\rm vol, bulk}\;[\rm Pa{\cdot}s]$ & $\eta_{\rm shr, bulk}\;[\rm Pa{\cdot}s]$ & $\eta_{\rm rot, bulk}\;[\rm Pa{\cdot}s]$ &
    $M_{\rm vol}\;[\rm\frac{ Pa{\cdot}s}{m}]$ & $M_{\rm shr}\;[\rm\frac{ Pa{\cdot}s}{m}]$& $M_{\rm rot}\;[\rm\frac{ Pa{\cdot}s}{m}]$ &
    ${D}_U\;[\rm ns^{-1}]$  &
    A $\big[\rm\frac{ pg}{\mu m^3}\big]$ &  $C$ $\big[\rm\frac{ pg}{\mu m{\cdot}ns^2{\cdot}K}\big]$ & W $\big[\rm \frac{\mu m}{ns}\big]$\\
  \hline
50 & 4.05937e+03 & 1.68603e-04 & 2.57075e+08 & 4.61815e-04 & 5.02357e-04 & 3.15700e-07 & 2.92722e+02 & 1.46453e+02 & 1.84400e-01 & 2.35602e-01 & 1.53433e-04 & 1.00900e-04 & 2.72761e+00 \\
60 & 4.71660e+03 & 2.57901e-04 & 3.95197e+08 & 4.43868e-04 & 5.28093e-04 & 5.92000e-07 & 4.76873e+02 & 2.38705e+02 & 5.39000e-01 & 3.43827e-01 & 2.29498e-04 & 1.41101e-04 & 3.00647e+00 \\
70 & 5.30227e+03 & 3.86737e-04 & 5.59073e+08 & 4.31686e-04 & 5.59864e-04 & 1.02525e-06 & 7.14658e+02 & 3.57984e+02 & 1.32370e+00 & 4.90888e-01 & 3.20256e-04 & 1.84524e-04 & 3.25249e+00 \\
80 & 5.73508e+03 & 5.99494e-04 & 7.45726e+08 & 4.28271e-04 & 5.93563e-04 & 1.67075e-06 & 1.00580e+03 & 5.04315e+02 & 2.87515e+00 & 6.99024e-01 & 4.25102e-04 & 2.30997e-04 & 3.45924e+00 \\
90 & 5.98509e+03 & 9.89803e-04 & 9.52360e+08 & 4.37016e-04 & 6.27481e-04 & 2.58790e-06 & 1.34836e+03 & 6.76963e+02 & 5.67725e+00 & 9.92275e-01 & 5.43532e-04 & 2.80534e-04 & 3.62371e+00 \\
100 & 6.05968e+03 & 1.70592e-03 & 1.17648e+09 & 4.62918e-04 & 6.60757e-04 & 3.83780e-06 & 1.73924e+03 & 8.74676e+02 & 1.03449e+01 &  1.39551e+00 & 6.75030e-04 & 3.33160e-04 & 3.74625e+00 \\
125 & 5.68808e+03 & 6.19470e-03 & 1.79932e+09 & 6.30614e-04 & 7.38393e-04 & 8.77595e-06 & 2.90233e+03 & 1.46805e+03 & 3.46960e+01 & 3.03975e+00 & 1.05685e-03 & 4.77779e-04 & 3.89427e+00 \\
150 & 4.95001e+03 & 1.68525e-02 & 2.48830e+09 & 9.37574e-04 & 8.05808e-04 & 1.66580e-05 & 4.28093e+03 & 2.18166e+03 & 8.46435e+01 & 5.82976e+00 & 1.50504e-03 & 6.38359e-04 & 3.88637e+00 \\
175 & 4.19048e+03 & 3.54534e-02 & 3.21691e+09 & 1.30108e-03 & 8.61970e-04 & 2.72463e-05 & 5.81999e+03 & 2.99059e+03 & 1.65497e+02 & 9.95230e+00 & 2.00676e-03 & 8.09822e-04 & 3.80167e+00 \\
200 & 3.53715e+03 & 6.22028e-02 & 3.96223e+09 & 1.63785e-03 & 9.07000e-04 & 3.96844e-05 & 7.47675e+03 & 3.87370e+03 & 2.77660e+02 & 1.54740e+01 & 2.54990e-03 & 9.86974e-04 & 3.68920e+00 \\
225 & 3.00949e+03 & 9.60903e-02 & 4.70645e+09 & 1.90990e-03 & 9.41864e-04 & 5.29272e-05 & 9.21972e+03 & 4.81387e+03 & 4.18293e+02 & 2.23622e+01 & 3.12432e-03 & 1.16581e-03 & 3.57259e+00 \\
250 & 2.59110e+03 & 1.35378e-01 & 5.43687e+09 & 2.11435e-03 & 9.68052e-04 & 6.60746e-05 & 1.10261e+04 & 5.79761e+03 & 5.83126e+02 & 3.05105e+01 & 3.72211e-03 & 1.34366e-03 & 3.46122e+00 \\
275 & 2.25923e+03 & 1.78035e-01 & 6.14489e+09 & 2.26296e-03 & 9.87217e-04 & 7.85060e-05 & 1.28794e+04 & 6.81445e+03 & 7.67646e+02 & 3.97658e+01 & 4.33732e-03 & 1.51881e-03 & 3.35792e+00 \\
300 & 1.99378e+03 & 2.22075e-01 & 6.82487e+09 & 2.36987e-03 & 1.00091e-03 & 8.98738e-05 & 1.47675e+04 & 7.85631e+03 & 9.67711e+02 & 4.99551e+01 & 4.96554e-03 & 1.69010e-03 & 3.26300e+00 \\
350 & 1.60318e+03 & 3.07855e-01 & 8.08784e+09 & 2.50315e-03 & 1.01696e-03 & 1.09003e-04 & 1.86142e+04 & 9.99151e+03 & 1.40093e+03 & 7.24549e+01 & 6.24865e-03 & 2.01775e-03 & 3.09612e+00 \\
\hline
  \end{tabular}
  }
\end{table*}

\begin{table*}[htbp!]
  \caption{Parameters entering the viscous heat equations for isotopically purified graphite (99.9 \% $^{12}$C, 0.1 \% $^{13}$C). We report here only the in-plane components of the tensors needed to perform the calculations: 
 $\kappa^{ij}_{\rm P}=\kappa_{\rm P}\delta^{ij}$, $\kappa^{ij}_{\rm C}=\kappa_{\rm C}\delta^{ij}$, $K^{ij}_{S}=K_{S}\delta^{ij}$, ${D}_{U}^{ij}={D}_{U}\delta^{ij}$, $A=A^i\;\forall \;i$, $W^j_{0i}=W^j_{i0}= W\delta^{ij}$, where the indexes $i,j$ represent the in-plane directions $x,y$ only ($i,j=1,2$ and $i\neq j$).}
  \label{tab:param_graphite}
  \centering
\resizebox{\textwidth}{!}{%
  \begin{tabular}{cccccccccccccc}
  \hline
  \hline
  {T [K]} & 
  {$\kappa_{\rm P}\;\big[\rm\frac{W}{m{\cdot}K}\big]$} & {$\kappa_{\rm C}\;\big[\rm\frac{W}{m{\cdot}K}\big]$} & ${K_S\;\big[\rm\frac{W}{m^2{\cdot}K}\big]}$ & 
  $\eta_{\rm vol, bulk}\;[\rm Pa{\cdot}s]$ & $\eta_{\rm shr, bulk}\;[\rm Pa{\cdot}s]$ & $\eta_{\rm rot, bulk}\;[\rm Pa{\cdot}s]$ &
    $M_{\rm vol}\;[\rm\frac{ Pa{\cdot}s}{m}]$ & $M_{\rm shr}\;[\rm\frac{ Pa{\cdot}s}{m}]$& $M_{\rm rot}\;[\rm\frac{ Pa{\cdot}s}{m}]$ &
    ${D}_{U}\;[\rm ns^{-1}]$  &
    A $\big[\rm\frac{ pg}{\mu m^3}\big]$ &  $C$ $\big[\rm\frac{ pg}{\mu m{\cdot}ns^2{\cdot}K}\big]$ & W $\big[\rm \frac{\mu m}{ns}\big]$\\
  \hline
50 & 3.34175e+04 & 8.17849e-05 & 2.57075e+08 & 5.50781e-04 & 5.61659e-04 & 4.44100e-07 & 2.92722e+02 & 1.46453e+02 & 1.84350e-01 & 2.41659e-02 & 1.53433e-04 & 1.00900e-04 & 2.72761e+00\\
60 & 3.52935e+04 & 1.44119e-04 & 3.95197e+08 & 5.10110e-04 & 5.83542e-04 & 7.91950e-07 & 4.76873e+02 & 2.38705e+02 & 5.39000e-01 & 4.34202e-02 & 2.29498e-04 & 1.41101e-04 & 3.00647e+00\\
70 & 3.24978e+04 & 2.40448e-04 & 5.59073e+08 & 4.83162e-04 & 6.13836e-04 & 1.32750e-06 & 7.14658e+02 & 3.57984e+02 & 1.32375e+00 & 8.50411e-02 & 3.20256e-04 & 1.84524e-04 & 3.25249e+00\\
80 & 2.74850e+04 & 4.06335e-04 & 7.45726e+08 & 4.70186e-04 & 6.47939e-04 & 2.11115e-06 & 1.00580e+03 & 5.04315e+02 & 2.87515e+00 & 1.68503e-01 & 4.25102e-04 & 2.30997e-04 & 3.45924e+00\\
90 & 2.25348e+04 & 7.19062e-04 & 9.52361e+08 & 4.72715e-04 & 6.83412e-04 & 3.20355e-06 & 1.34836e+03 & 6.76963e+02 & 5.67725e+00 & 3.18608e-01 & 5.43532e-04 & 2.80534e-04 & 3.62371e+00\\
100 & 1.84465e+04 & 1.30478e-03 & 1.17648e+09 & 4.94693e-04 & 7.18751e-04 & 4.66620e-06 & 1.73924e+03 & 8.74676e+02 & 1.03449e+01 & 5.63474e-01 & 6.75030e-04 & 3.33160e-04 & 3.74625e+00\\
120 & 1.27814e+04 & 3.97919e-03 & 1.66857e+09 & 6.12864e-04 & 7.85639e-04 & 8.93120e-06 & 2.65051e+03 & 1.33891e+03 & 2.80410e+01 & 1.45053e+00 & 9.74775e-04 & 4.47442e-04 & 3.88006e+00\\
140 & 9.33971e+03 & 9.87325e-03 & 2.20653e+09 & 8.30894e-04 & 8.44708e-04 & 1.52226e-05 & 3.70742e+03 & 1.88346e+03 & 6.11325e+01 & 3.02867e+00 & 1.31862e-03 & 5.72512e-04 & 3.90257e+00\\
160 & 7.14426e+03 & 2.01725e-02 & 2.77638e+09 & 1.11170e-03 & 8.94723e-04 & 2.35077e-05 & 4.87997e+03 & 2.49507e+03 & 1.13158e+02 & 5.43535e+00 & 1.70002e-03 & 7.05946e-04 & 3.85826e+00\\
180 & 5.66903e+03 & 3.55200e-02 & 3.36529e+09 & 1.40421e-03 & 9.35701e-04 & 3.34012e-05 & 6.14308e+03 & 3.16188e+03 & 1.85483e+02 & 8.74395e+00 & 2.11243e-03 & 8.44946e-04 & 3.78035e+00\\
185 & 5.37644e+03 & 4.01659e-02 & 3.51416e+09 & 1.47434e-03 & 9.44606e-04 & 3.60593e-05 & 6.47056e+03 & 3.33598e+03 & 2.06714e+02 & 9.71528e+00 & 2.21967e-03 & 8.80257e-04 & 3.75825e+00\\
190 & 5.10779e+03 & 4.51306e-02 & 3.66336e+09 & 1.54240e-03 & 9.53004e-04 & 3.87717e-05 & 6.80216e+03 & 3.51274e+03 & 2.29171e+02 & 1.07442e+01 & 2.32838e-03 & 9.15719e-04 & 3.73558e+00\\
195 & 4.86054e+03 & 5.04083e-02 & 3.81277e+09 & 1.60811e-03 & 9.60912e-04 & 4.15285e-05 & 7.13764e+03 & 3.69202e+03 & 2.52829e+02 & 1.18304e+01 & 2.43848e-03 & 9.51302e-04 & 3.71251e+00\\
200 & 4.63247e+03 & 5.59919e-02 & 3.96223e+09 & 1.67125e-03 & 9.68346e-04 & 4.43202e-05 & 7.47676e+03 & 3.87370e+03 & 2.77661e+02 & 1.29735e+01 & 2.54990e-03 & 9.86974e-04 & 3.68920e+00\\
220 & 3.87676e+03 & 8.11890e-02 & 4.55837e+09 & 1.89595e-03 & 9.93713e-04 & 5.56584e-05 & 8.86543e+03 & 4.62197e+03 & 3.88086e+02 & 1.80998e+01 & 3.00734e-03 & 1.13005e-03 & 3.59566e+00\\
240 & 3.30876e+03 & 1.10363e-01 & 5.14699e+09 & 2.07603e-03 & 1.01297e-03 & 6.68999e-05 & 1.02971e+04 & 5.39961e+03 & 5.14581e+02 & 2.40671e+01 & 3.48061e-03 & 1.27276e-03 & 3.50487e+00\\
260 & 2.87089e+03 & 1.42551e-01 & 5.72314e+09 & 2.21647e-03 & 1.02725e-03 & 7.76685e-05 & 1.17625e+04 & 6.20087e+03 & 6.54815e+02 & 3.07981e+01 & 3.96638e-03 & 1.41411e-03 & 3.41888e+00\\
280 & 2.52599e+03 & 1.76706e-01 & 6.28327e+09 & 2.32458e-03 & 1.03757e-03 & 8.77261e-05 & 1.32545e+04 & 7.02104e+03 & 8.06538e+02 & 3.82043e+01 & 4.46204e-03 & 1.55341e-03 & 3.33828e+00\\
300 & 2.24925e+03 & 2.11805e-01 & 6.82487e+09 & 2.40747e-03 & 1.04480e-03 & 9.69486e-05 & 1.47675e+04 & 7.85631e+03 & 9.67711e+02 & 4.61938e+01 & 4.96554e-03 & 1.69010e-03 & 3.26300e+00\\
350 & 1.75530e+03 & 2.98054e-01 & 8.08784e+09 & 2.54054e-03 & 1.05371e-03 & 1.16219e-04 & 1.86142e+04 & 9.99151e+03 & 1.40093e+03 & 6.81463e+01 & 6.24865e-03 & 2.01775e-03 & 3.09612e+00\\
400 & 1.43431e+03 & 3.74983e-01 & 9.21271e+09 & 2.61189e-03 & 1.05509e-03 & 1.30653e-04 & 2.25167e+04 & 1.21681e+04 & 1.86162e+03 & 9.20012e+01 & 7.55350e-03 & 2.32135e-03 & 2.95575e+00\\

 \hline
  \end{tabular}
  }
\end{table*}

\begin{table*}[htbp!]
  \caption{Parameters entering the viscous heat equations for pure graphite (100 \% $^{12}$C). We report here only the in-plane components of the tensors needed to perform the calculations: 
 $\kappa^{ij}_{\rm P}{=}\kappa_{\rm P}\delta^{ij}$, $\kappa^{ij}_{\rm C}{=}\kappa_{\rm C}\delta^{ij}$, $K^{ij}_{S}{=}K_{S}\delta^{ij}$, ${D}_{U}^{ij}{=}{D}_{U}\delta^{ij}$, $A{=}A^i\;\forall \;i$, $W^j_{0i}{=}W^j_{i0}{=} W\delta^{ij}$, where the indexes $i,j$ represent the in-plane directions $x,y$ only ($i,j=1,2$ and $i\neq j$).}
  \label{tab:param_graphite_100}
  \centering
\resizebox{\textwidth}{!}{%
  \begin{tabular}{cccccccccccccc}
  \hline
  \hline
  {T [K]} & 
  {$\kappa_{\rm P}\;\big[\rm\frac{W}{m{\cdot}K}\big]$} & {$\kappa_{\rm C}\;\big[\rm\frac{W}{m{\cdot}K}\big]$} & ${K_S\;\big[\rm\frac{W}{m^2{\cdot}K}\big]}$ & 
  $\eta_{\rm vol, bulk}\;[\rm Pa{\cdot}s]$ & $\eta_{\rm shr, bulk}\;[\rm Pa{\cdot}s]$ & $\eta_{\rm rot, bulk}\;[\rm Pa{\cdot}s]$ &
    $M_{\rm vol}\;[\rm\frac{ Pa{\cdot}s}{m}]$ & $M_{\rm shr}\;[\rm\frac{ Pa{\cdot}s}{m}]$& $M_{\rm rot}\;[\rm\frac{ Pa{\cdot}s}{m}]$ &
    ${D}_{U}\;[\rm ns^{-1}]$  &
    A $\big[\rm\frac{ pg}{\mu m^3}\big]$ &  $C$ $\big[\rm\frac{ pg}{\mu m{\cdot}ns^2{\cdot}K}\big]$ & W $\big[\rm \frac{\mu m}{ns}\big]$\\
  \hline

80 & 5.05800e+04 & 4.03690e-04 & 7.45726e+08 & 4.75473e-04 & 6.55777e-04 & 2.18415e-06 & 1.00580e+03 & 5.04315e+02 & 2.87515e+00 & 1.13119e-01 & 4.25102e-04 & 2.30997e-04 & 3.45924e+00 \\
150 & 8.76453e+03 & 1.42003e-02 & 2.48831e+09 & 1.85042e-03 & 8.99385e-04 & 4.72804e-04 & 4.28093e+03 & 2.18166e+03 & 8.46435e+01 & 3.94422e+00 & 1.50504e-03 & 6.38359e-04 & 3.88637e+00 \\
 \hline
 \end{tabular}
 }
\end{table*}

\subsection{Monoisotopic hexagonal boron nitride} 
\label{sub:monoisotopic_hexagonal_boron_nitride}
\begin{figure*}[htbp]
	\centering
	\includegraphics[width=0.9\textwidth]{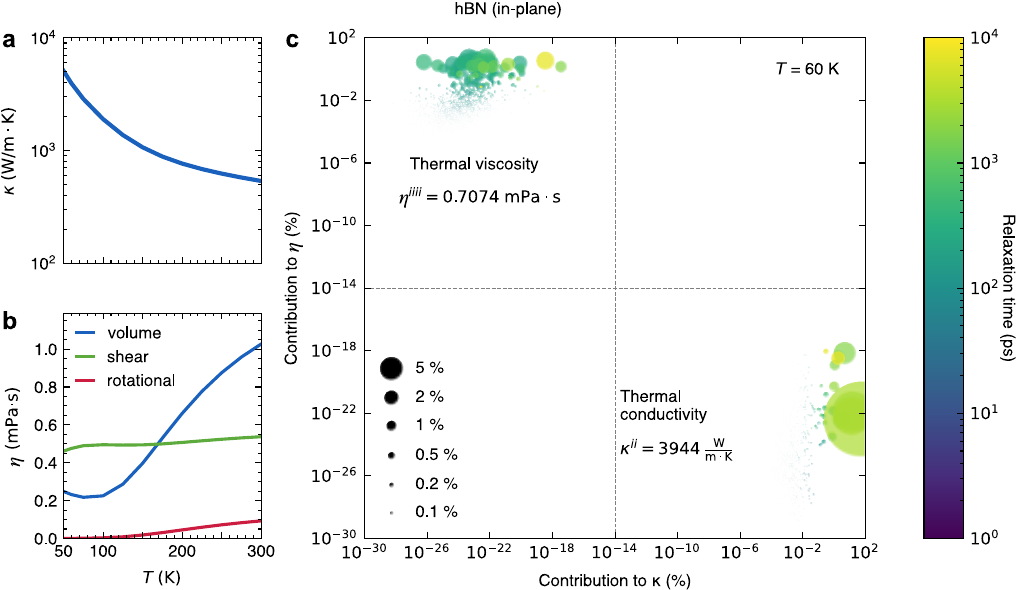}\\[-1.5mm]
	\caption{\textbf{Thermal conductivity and viscosity of h$^{11}$BN and their decomposition in terms of relaxons.}
	The in-plane bulk conductivity as a function of temperature is shown in the upper left inset, while the bulk principal viscosity components as a function of temperature are shown in the lower left inset. The left panel shows the decomposition of conductivity and viscosity in terms of relaxon excitations. Each scatter point corresponds to a relaxon, with its color indicating the relaxation time and its area scaled proportionally to the sum of its relative contributions to both thermal conductivity and viscosity. Dashed lines emphasize how relaxons with odd parity determine the thermal conductivity and contribute negligibly to thermal viscosity, while relaxons with even parity determine thermal viscosity and have negligible impact on thermal conductivity
	\cite{Simoncelli2020}.} 
	\label{fig:relaxon_hBN}
\end{figure*}
\begin{figure*}
    \centering
    \includegraphics[width=\textwidth]{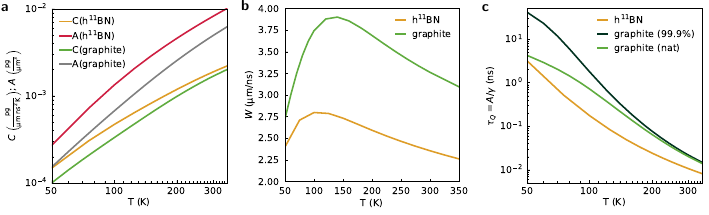}\\[-1.5mm]
    \caption{\textbf{Parameters entering the VHE for h$^{11}$BN and graphite at different isotopic concentrations.}
		The specific heat and specific momentum (\textbf{a}), and momentum relaxon velocity (\textbf{b}) are determined solely by the phonon band structure, therefore, within perturbation theory\cite{tamura_isotope_1983}, they are not influenced by isotopic-mass disorder.
		In contrast, the in-plane relaxation timescale due to Umklapp scattering (\textbf{c}) is influenced by isotipic concentration, and it progressively decreases from isotopically purified graphite (99.9 \% $^{12}$C, 0.1 \%$^{13}$C), natural graphite (98.9 \% $^{12}$C, 1.1 \%$^{13}$C) and h$^{11}$BN.}
    \label{fig:other_parameters}
\end{figure*}

First-principles calculations have been performed with Quantum ESPRESSO (QE) \cite{giannozzi2009quantum,giannozzi2017advanced} using the local-density approximation (LDA) for the exchange-correlation energy functional, since it has been shown to accurately describe the vibrational and thermal properties of hBN \cite{jiang_anisotropic_2018,PhysRevB.97.155435,yuan_modulating_2019}.
We have used norm-conserving and scalar-relativistic pseudopotentials 
 from the pseudo Dojo library \cite{van2018pseudodojo} 
with accuracy ``stringent'' and of type ONCVPSP v0.4.
The crystal structure of hBN has been taken from \cite{Kurakevych:sq3084} (Crystallographic Open Database~ \cite{COD_database} id 2016170).
Kinetic energy cutoffs of 90 and 360 Ry have been used for the wave functions and for the charge density; the Brillouin zone (BZ) is integrated with a Monkhorst-Pack mesh of $8{\times}8{\times}3$ points, with a (1,1,1) shift.
The equilibrium crystal structure has been computed performing a ``vc-relax'' calculation with QE, the resulting lattice parameters are $a=2.489089$ \AA ~and $c/a=2.604045$.
Second-order force constants have been computed on a $8{\times}8{\times}3$ mesh using density-functional perturbation theory~\cite{RevModPhys.73.515} and accounting for the non-analytic term correction due to the dielectric tensor and Born effective charges. 
Third-order force constants have been computed using the finite-difference method implemented in \textsc{ShengBTE}~\cite{li2014shengbte}, on a $4{\times}4{\times}2$ supercell, integrating the BZ with a $2{\times}2{\times}1$ Monkhorst-Pack mesh, and considering interactions up to 3.944 \AA~(corresponding to 6th nearest neighbor).
Then, anharmonic force constants have been converted from \textsc{ShengBTE} format to \texttt{mat3R} format using the program \texttt{d3\_import\_shengbte.x} in the \textsc{D3Q} package~\cite{paulatto2013anharmonic,paulatto2015first}, including interactions up to the second neighbouring cell for the re-centering of the third-order force constants (NFAR=2).
All the VHE parameters have been computed using a $49{\times}49{\times}3$ $\bm{q}$-mesh  and a Gaussian smearing of $6\;{\rm cm}^{-1}$, and are reported in Table~\ref{tab:param_hBN_11}. In Fig.~\ref{fig:relaxon_hBN} we analyze the temperature dependence of thermal conductivity and viscosity for h$^{11}$BN, also resolving how each individual relaxon excitation contributes to them. In Fig.~\ref{fig:other_parameters} we compare the specific heat $C$, specific momentum $A$, velocity $W$ and Umklapp dissipation timescale $\tau_Q=A/\gamma$ for the materials studied.

\begin{table*}[htbp!]
  \caption{Parameters entering the viscous heat equations for h$^{\rm 11}$BN. We report here only the in-plane components of the tensors needed to perform the calculations: 
 $\kappa^{ij}_{\rm P}=\kappa_{\rm P}\delta^{ij}$, $\kappa^{ij}_{\rm C}=\kappa_{\rm C}\delta^{ij}$, $K^{ij}_{S}=K_{S}\delta^{ij}$, ${D}_{U}^{ij}={D}_{U}\delta^{ij}$, $A=A^i\;\forall \;i$, $W^j_{0i}=W^j_{i0}= W\delta^{ij}$, where the indexes $i,j$ represent the in-plane directions $x,y$ only ($i,j=1,2$ and $i\neq j$).}
  \label{tab:param_hBN_11}
  \centering
\resizebox{\textwidth}{!}{%
  \begin{tabular}{cccccccccccccc}
  \hline
  \hline
  {T [K]} & 
  {$\kappa_{\rm P}\;\big[\rm\frac{W}{m{\cdot}K}\big]$} & {$\kappa_{\rm C}\;\big[\rm\frac{W}{m{\cdot}K}\big]$} & ${K_S\;\big[\rm\frac{W}{m^2{\cdot}K}\big]}$ & 
  $\eta_{\rm vol, bulk}\;[\rm Pa{\cdot}s]$ & $\eta_{\rm shr, bulk}\;[\rm Pa{\cdot}s]$ & $\eta_{\rm rot, bulk}\;[\rm Pa{\cdot}s]$ &
    $M_{\rm vol}\;[\rm\frac{ Pa{\cdot}s}{m}]$ & $M_{\rm shr}\;[\rm\frac{ Pa{\cdot}s}{m}]$& $M_{\rm rot}\;[\rm\frac{ Pa{\cdot}s}{m}]$ &
    ${D}_{U}\;[\rm ns^{-1}]$  &
    A $\big[\rm\frac{ pg}{\mu m^3}\big]$ &  $C$ $\big[\rm\frac{ pg}{\mu m{\cdot}ns^2{\cdot}K}\big]$ & W $\big[\rm \frac{\mu m}{ns}\big]$\\
  \hline
50 & 5.08496e+03 & 4.39299e-04 & 3.18233e+08 & 2.48335e-04 & 4.61274e-04 & 3.48800e-07 & 4.64880e+02 & 2.32754e+02 & 6.32750e-01 & 3.15987e-01 & 2.73574e-04 & 1.49113e-04 & 2.40299e+00 \\
60 & 3.94420e+03 & 6.91660e-04 & 4.77290e+08 & 2.31738e-04 & 4.75655e-04 & 5.97250e-07 & 7.67844e+02 & 3.84740e+02 & 1.65905e+00 & 7.48719e-01 & 4.26806e-04 & 2.08230e-04 & 2.56422e+00 \\
75 & 2.87382e+03 & 1.19669e-03 & 7.48233e+08 & 2.17434e-04 & 4.89862e-04 & 1.18350e-06 & 1.35937e+03 & 6.82172e+02 & 5.06755e+00 & 1.93310e+00 & 7.19449e-04 & 3.03986e-04 & 2.71119e+00 \\
100 & 1.89907e+03 & 2.58860e-03 & 1.24792e+09 & 2.25136e-04 & 4.95019e-04 & 3.36275e-06 & 2.61176e+03 & 1.31615e+03 & 2.08995e+01 &  5.71441e+00 & 1.33990e-03 & 4.71727e-04 & 2.79944e+00 \\
125 & 1.37164e+03 & 5.21646e-03 & 1.78806e+09 & 2.86497e-04 & 4.93230e-04 & 8.74825e-06 & 4.06757e+03 & 2.06353e+03 & 6.04160e+01 & 1.19136e+01 & 2.07076e-03 & 6.44647e-04 & 2.78843e+00 \\
150 & 1.06699e+03 & 9.66905e-03 & 2.36077e+09 & 3.97001e-04 & 4.94250e-04 & 1.82830e-05 & 5.64222e+03 & 2.88683e+03 & 1.33298e+02 & 2.02022e+01 & 2.86999e-03 & 8.22855e-04 & 2.73368e+00 \\
175 & 8.83008e+02 & 1.62740e-02 & 2.95342e+09 & 5.29753e-04 & 4.99399e-04 & 3.10292e-05 & 7.29517e+03 & 3.76699e+03 & 2.42027e+02 & 3.01303e+01 & 3.71435e-03 & 1.00492e-03 & 2.66425e+00 \\
200 & 7.65123e+02 & 2.51062e-02 & 3.55061e+09 & 6.60281e-04 & 5.07125e-04 & 4.51939e-05 & 9.00458e+03 & 4.69153e+03 & 3.83489e+02 & 4.13328e+01 & 4.58997e-03 & 1.18813e-03 & 2.59391e+00 \\
225 & 6.83803e+02 & 3.60584e-02 & 4.13896e+09 & 7.76465e-04 & 5.15676e-04 & 5.91933e-05 & 1.07565e+04 & 5.65070e+03 & 5.52098e+02 & 5.35343e+01 & 5.48778e-03 & 1.37008e-03 & 2.52756e+00 \\
250 & 6.23329e+02 & 4.88988e-02 & 4.70873e+09 & 8.75308e-04 & 5.23884e-04 & 7.20454e-05 & 1.25408e+04 & 6.63651e+03 & 7.41924e+02 & 6.65173e+01 & 6.40150e-03 & 1.54901e-03 & 2.46616e+00 \\
275 & 5.75413e+02 & 6.33176e-02 & 5.25353e+09 & 9.57987e-04 & 5.31173e-04 & 8.33178e-05 & 1.43497e+04 & 7.64246e+03 & 9.47691e+02 & 8.01021e+01 & 7.32669e-03 & 1.72367e-03 & 2.40940e+00 \\
300 & 5.35638e+02 & 7.89685e-02 & 5.76951e+09 & 1.02693e-03 & 5.37362e-04 & 9.29376e-05 & 1.61770e+04 & 8.66335e+03 & 1.16508e+03 & 9.41385e+01 & 8.26014e-03 & 1.89306e-03 & 2.35678e+00 \\
325 & 5.01567e+02 & 9.55019e-02 & 6.25468e+09 & 1.08456e-03 & 5.42481e-04 & 1.01021e-04 & 1.80179e+04 & 9.69506e+03 & 1.39072e+03 & 1.08503e+02 & 9.19946e-03 & 2.05633e-03 & 2.30792e+00 \\
350 & 4.71776e+02 & 1.12591e-01 & 6.70835e+09 & 1.13298e-03 & 5.46654e-04 & 1.07763e-04 & 1.98685e+04 & 1.07344e+04 & 1.62205e+03 & 1.23095e+02 & 1.01428e-02 & 2.21279e-03 & 2.26256e+00 \\
 \hline
  \end{tabular}
  }
\end{table*}

\newpage
\providecommand{\noopsort}[1]{}\providecommand{\singleletter}[1]{#1}%


\begin{thebibliography}{150}%
\makeatletter
\providecommand \@ifxundefined [1]{%
 \@ifx{#1\undefined}
}%
\providecommand \@ifnum [1]{%
 \ifnum #1\expandafter \@firstoftwo
 \else \expandafter \@secondoftwo
 \fi
}%
\providecommand \@ifx [1]{%
 \ifx #1\expandafter \@firstoftwo
 \else \expandafter \@secondoftwo
 \fi
}%
\providecommand \natexlab [1]{#1}%
\providecommand \enquote  [1]{``#1''}%
\providecommand \bibnamefont  [1]{#1}%
\providecommand \bibfnamefont [1]{#1}%
\providecommand \citenamefont [1]{#1}%
\providecommand \href@noop [0]{\@secondoftwo}%
\providecommand \href [0]{\begingroup \@sanitize@url \@href}%
\providecommand \@href[1]{\@@startlink{#1}\@@href}%
\providecommand \@@href[1]{\endgroup#1\@@endlink}%
\providecommand \@sanitize@url [0]{\catcode `\\12\catcode `\$12\catcode `\&12\catcode `\#12\catcode `\^12\catcode `\_12\catcode `\%12\relax}%
\providecommand \@@startlink[1]{}%
\providecommand \@@endlink[0]{}%
\providecommand \url  [0]{\begingroup\@sanitize@url \@url }%
\providecommand \@url [1]{\endgroup\@href {#1}{\urlprefix }}%
\providecommand \urlprefix  [0]{URL }%
\providecommand \Eprint [0]{\href }%
\providecommand \doibase [0]{https://doi.org/}%
\providecommand \selectlanguage [0]{\@gobble}%
\providecommand \bibinfo  [0]{\@secondoftwo}%
\providecommand \bibfield  [0]{\@secondoftwo}%
\providecommand \translation [1]{[#1]}%
\providecommand \BibitemOpen [0]{}%
\providecommand \bibitemStop [0]{}%
\providecommand \bibitemNoStop [0]{.\EOS\space}%
\providecommand \EOS [0]{\spacefactor3000\relax}%
\providecommand \BibitemShut  [1]{\csname bibitem#1\endcsname}%
\let\auto@bib@innerbib\@empty
\bibitem [{\citenamefont {Schmidt}\ \emph {et~al.}(2008)\citenamefont {Schmidt}, \citenamefont {Chen},\ and\ \citenamefont {Chen}}]{schmidt_pulse_2008}%
  \BibitemOpen
  \bibfield  {author} {\bibinfo {author} {\bibfnamefont {A.~J.}\ \bibnamefont {Schmidt}}, \bibinfo {author} {\bibfnamefont {X.}~\bibnamefont {Chen}},\ and\ \bibinfo {author} {\bibfnamefont {G.}~\bibnamefont {Chen}},\ }\bibfield  {title} {\bibinfo {title} {Pulse accumulation, radial heat conduction, and anisotropic thermal conductivity in pump-probe transient thermoreflectance},\ }\href {https://aip.scitation.org/doi/10.1063/1.3006335} {\bibfield  {journal} {\bibinfo  {journal} {Rev. Sci. Instrum.}\ }\textbf {\bibinfo {volume} {79}},\ \bibinfo {pages} {114902} (\bibinfo {year} {2008})}\BibitemShut {NoStop}%
\bibitem [{\citenamefont {Balandin}(2011)}]{balandin_thermal_2011}%
  \BibitemOpen
  \bibfield  {author} {\bibinfo {author} {\bibfnamefont {A.~A.}\ \bibnamefont {Balandin}},\ }\bibfield  {title} {\bibinfo {title} {Thermal properties of graphene and nanostructured carbon materials},\ }\href {https://www.nature.com/articles/nmat3064} {\bibfield  {journal} {\bibinfo  {journal} {Nat. Mater.}\ }\textbf {\bibinfo {volume} {10}},\ \bibinfo {pages} {569} (\bibinfo {year} {2011})}\BibitemShut {NoStop}%
\bibitem [{\citenamefont {Fugallo}\ \emph {et~al.}(2014)\citenamefont {Fugallo}, \citenamefont {Cepellotti}, \citenamefont {Paulatto}, \citenamefont {Lazzeri}, \citenamefont {Marzari},\ and\ \citenamefont {Mauri}}]{fugallo_thermal_2014}%
  \BibitemOpen
  \bibfield  {author} {\bibinfo {author} {\bibfnamefont {G.}~\bibnamefont {Fugallo}}, \bibinfo {author} {\bibfnamefont {A.}~\bibnamefont {Cepellotti}}, \bibinfo {author} {\bibfnamefont {L.}~\bibnamefont {Paulatto}}, \bibinfo {author} {\bibfnamefont {M.}~\bibnamefont {Lazzeri}}, \bibinfo {author} {\bibfnamefont {N.}~\bibnamefont {Marzari}},\ and\ \bibinfo {author} {\bibfnamefont {F.}~\bibnamefont {Mauri}},\ }\bibfield  {title} {\bibinfo {title} {Thermal {Conductivity} of {Graphene} and {Graphite}: {Collective} {Excitations} and {Mean} {Free} {Paths}},\ }\href {https://pubs.acs.org/doi/10.1021/nl502059f} {\bibfield  {journal} {\bibinfo  {journal} {Nano Lett.}\ }\textbf {\bibinfo {volume} {14}},\ \bibinfo {pages} {6109} (\bibinfo {year} {2014})}\BibitemShut {NoStop}%
\bibitem [{\citenamefont {Machida}\ \emph {et~al.}(2020)\citenamefont {Machida}, \citenamefont {Matsumoto}, \citenamefont {Isono},\ and\ \citenamefont {Behnia}}]{machida_phonon_2020}%
  \BibitemOpen
  \bibfield  {author} {\bibinfo {author} {\bibfnamefont {Y.}~\bibnamefont {Machida}}, \bibinfo {author} {\bibfnamefont {N.}~\bibnamefont {Matsumoto}}, \bibinfo {author} {\bibfnamefont {T.}~\bibnamefont {Isono}},\ and\ \bibinfo {author} {\bibfnamefont {K.}~\bibnamefont {Behnia}},\ }\bibfield  {title} {\bibinfo {title} {Phonon hydrodynamics and ultrahigh–room-temperature thermal conductivity in thin graphite},\ }\href {https://www.science.org/doi/10.1126/science.aaz8043} {\bibfield  {journal} {\bibinfo  {journal} {Science}\ }\textbf {\bibinfo {volume} {367}},\ \bibinfo {pages} {309} (\bibinfo {year} {2020})}\BibitemShut {NoStop}%
\bibitem [{\citenamefont {Jiang}\ \emph {et~al.}(2018)\citenamefont {Jiang}, \citenamefont {Qian}, \citenamefont {Yang},\ and\ \citenamefont {Lindsay}}]{jiang_anisotropic_2018}%
  \BibitemOpen
  \bibfield  {author} {\bibinfo {author} {\bibfnamefont {P.}~\bibnamefont {Jiang}}, \bibinfo {author} {\bibfnamefont {X.}~\bibnamefont {Qian}}, \bibinfo {author} {\bibfnamefont {R.}~\bibnamefont {Yang}},\ and\ \bibinfo {author} {\bibfnamefont {L.}~\bibnamefont {Lindsay}},\ }\bibfield  {title} {\bibinfo {title} {Anisotropic thermal transport in bulk hexagonal boron nitride},\ }\href {https://link.aps.org/doi/10.1103/PhysRevMaterials.2.064005} {\bibfield  {journal} {\bibinfo  {journal} {Phys. Rev. Materials}\ }\textbf {\bibinfo {volume} {2}},\ \bibinfo {pages} {064005} (\bibinfo {year} {2018})}\BibitemShut {NoStop}%
\bibitem [{\citenamefont {Yuan}\ \emph {et~al.}(2019)\citenamefont {Yuan}, \citenamefont {Li}, \citenamefont {Lindsay}, \citenamefont {Cherns}, \citenamefont {Pomeroy}, \citenamefont {Liu}, \citenamefont {Edgar},\ and\ \citenamefont {Kuball}}]{yuan_modulating_2019}%
  \BibitemOpen
  \bibfield  {author} {\bibinfo {author} {\bibfnamefont {C.}~\bibnamefont {Yuan}}, \bibinfo {author} {\bibfnamefont {J.}~\bibnamefont {Li}}, \bibinfo {author} {\bibfnamefont {L.}~\bibnamefont {Lindsay}}, \bibinfo {author} {\bibfnamefont {D.}~\bibnamefont {Cherns}}, \bibinfo {author} {\bibfnamefont {J.~W.}\ \bibnamefont {Pomeroy}}, \bibinfo {author} {\bibfnamefont {S.}~\bibnamefont {Liu}}, \bibinfo {author} {\bibfnamefont {J.~H.}\ \bibnamefont {Edgar}},\ and\ \bibinfo {author} {\bibfnamefont {M.}~\bibnamefont {Kuball}},\ }\bibfield  {title} {\bibinfo {title} {Modulating the thermal conductivity in hexagonal boron nitride via controlled boron isotope concentration},\ }\href {https://www.nature.com/articles/s42005-019-0145-5} {\bibfield  {journal} {\bibinfo  {journal} {Commun. Phys.}\ }\textbf {\bibinfo {volume} {2}} (\bibinfo {year} {2019})}\BibitemShut {NoStop}%
\bibitem [{\citenamefont {Qian}\ \emph {et~al.}(2021)\citenamefont {Qian}, \citenamefont {Zhou},\ and\ \citenamefont {Chen}}]{qian_phonon-engineered_2021}%
  \BibitemOpen
  \bibfield  {author} {\bibinfo {author} {\bibfnamefont {X.}~\bibnamefont {Qian}}, \bibinfo {author} {\bibfnamefont {J.}~\bibnamefont {Zhou}},\ and\ \bibinfo {author} {\bibfnamefont {G.}~\bibnamefont {Chen}},\ }\bibfield  {title} {\bibinfo {title} {Phonon-engineered extreme thermal conductivity materials},\ }\href {https://www.nature.com/articles/s41563-021-00918-3} {\bibfield  {journal} {\bibinfo  {journal} {Nat. Mater.}\ }\textbf {\bibinfo {volume} {20}},\ \bibinfo {pages} {1188} (\bibinfo {year} {2021})}\BibitemShut {NoStop}%
\bibitem [{\citenamefont {Moon}\ \emph {et~al.}(2023)\citenamefont {Moon}, \citenamefont {Kim}, \citenamefont {Park}, \citenamefont {Im}, \citenamefont {Kim}, \citenamefont {Hwang},\ and\ \citenamefont {Kim}}]{moon_hexagonal_2023}%
  \BibitemOpen
  \bibfield  {author} {\bibinfo {author} {\bibfnamefont {S.}~\bibnamefont {Moon}}, \bibinfo {author} {\bibfnamefont {J.}~\bibnamefont {Kim}}, \bibinfo {author} {\bibfnamefont {J.}~\bibnamefont {Park}}, \bibinfo {author} {\bibfnamefont {S.}~\bibnamefont {Im}}, \bibinfo {author} {\bibfnamefont {J.}~\bibnamefont {Kim}}, \bibinfo {author} {\bibfnamefont {I.}~\bibnamefont {Hwang}},\ and\ \bibinfo {author} {\bibfnamefont {J.~K.}\ \bibnamefont {Kim}},\ }\bibfield  {title} {\bibinfo {title} {Hexagonal {Boron} {Nitride} for {Next}-{Generation} {Photonics} and {Electronics}},\ }\href {https://onlinelibrary.wiley.com/doi/abs/10.1002/adma.202204161} {\bibfield  {journal} {\bibinfo  {journal} {Adv. Mater.}\ }\textbf {\bibinfo {volume} {35}},\ \bibinfo {pages} {2204161} (\bibinfo {year} {2023})}\BibitemShut {NoStop}%
\bibitem [{\citenamefont {Chen}(2021)}]{chen_non-fourier_2021}%
  \BibitemOpen
  \bibfield  {author} {\bibinfo {author} {\bibfnamefont {G.}~\bibnamefont {Chen}},\ }\bibfield  {title} {\bibinfo {title} {Non-{Fourier} phonon heat conduction at the microscale and nanoscale},\ }\href {https://www.nature.com/articles/s42254-021-00334-1} {\bibfield  {journal} {\bibinfo  {journal} {Nat. Rev. Phys.}\ }\textbf {\bibinfo {volume} {3}},\ \bibinfo {pages} {555} (\bibinfo {year} {2021})}\BibitemShut {NoStop}%
\bibitem [{\citenamefont {Huang}\ \emph {et~al.}(2023)\citenamefont {Huang}, \citenamefont {Guo}, \citenamefont {Wu}, \citenamefont {Masubuchi}, \citenamefont {Watanabe}, \citenamefont {Taniguchi}, \citenamefont {Zhang}, \citenamefont {Volz}, \citenamefont {Machida},\ and\ \citenamefont {Nomura}}]{huang_observation_2022}%
  \BibitemOpen
  \bibfield  {author} {\bibinfo {author} {\bibfnamefont {X.}~\bibnamefont {Huang}}, \bibinfo {author} {\bibfnamefont {Y.}~\bibnamefont {Guo}}, \bibinfo {author} {\bibfnamefont {Y.}~\bibnamefont {Wu}}, \bibinfo {author} {\bibfnamefont {S.}~\bibnamefont {Masubuchi}}, \bibinfo {author} {\bibfnamefont {K.}~\bibnamefont {Watanabe}}, \bibinfo {author} {\bibfnamefont {T.}~\bibnamefont {Taniguchi}}, \bibinfo {author} {\bibfnamefont {Z.}~\bibnamefont {Zhang}}, \bibinfo {author} {\bibfnamefont {S.}~\bibnamefont {Volz}}, \bibinfo {author} {\bibfnamefont {T.}~\bibnamefont {Machida}},\ and\ \bibinfo {author} {\bibfnamefont {M.}~\bibnamefont {Nomura}},\ }\bibfield  {title} {\bibinfo {title} {Observation of phonon {Poiseuille} flow in isotopically purified graphite ribbons},\ }\href {https://www.nature.com/articles/s41467-023-37380-5} {\bibfield  {journal} {\bibinfo  {journal} {Nat. Commun.}\ }\textbf {\bibinfo {volume} {14}},\ \bibinfo {pages} {2044} (\bibinfo {year} {2023})}\BibitemShut {NoStop}%
\bibitem [{\citenamefont {Huberman}\ \emph {et~al.}(2019)\citenamefont {Huberman}, \citenamefont {Duncan}, \citenamefont {Chen}, \citenamefont {Song}, \citenamefont {Chiloyan}, \citenamefont {Ding}, \citenamefont {Maznev}, \citenamefont {Chen},\ and\ \citenamefont {Nelson}}]{Huberman2019}%
  \BibitemOpen
  \bibfield  {author} {\bibinfo {author} {\bibfnamefont {S.}~\bibnamefont {Huberman}}, \bibinfo {author} {\bibfnamefont {R.~A.}\ \bibnamefont {Duncan}}, \bibinfo {author} {\bibfnamefont {K.}~\bibnamefont {Chen}}, \bibinfo {author} {\bibfnamefont {B.}~\bibnamefont {Song}}, \bibinfo {author} {\bibfnamefont {V.}~\bibnamefont {Chiloyan}}, \bibinfo {author} {\bibfnamefont {Z.}~\bibnamefont {Ding}}, \bibinfo {author} {\bibfnamefont {A.~A.}\ \bibnamefont {Maznev}}, \bibinfo {author} {\bibfnamefont {G.}~\bibnamefont {Chen}},\ and\ \bibinfo {author} {\bibfnamefont {K.~A.}\ \bibnamefont {Nelson}},\ }\bibfield  {title} {\bibinfo {title} {{Observation of second sound in graphite at temperatures above 100 K}},\ }\href@noop {} {\bibfield  {journal} {\bibinfo  {journal} {Science}\ }\textbf {\bibinfo {volume} {364}},\ \bibinfo {pages} {375} (\bibinfo {year} {2019})}\BibitemShut {NoStop}%
\bibitem [{\citenamefont {Jeong}\ \emph {et~al.}(2021)\citenamefont {Jeong}, \citenamefont {Li}, \citenamefont {Lee}, \citenamefont {Shi},\ and\ \citenamefont {Wang}}]{Jeong2021}%
  \BibitemOpen
  \bibfield  {author} {\bibinfo {author} {\bibfnamefont {J.}~\bibnamefont {Jeong}}, \bibinfo {author} {\bibfnamefont {X.}~\bibnamefont {Li}}, \bibinfo {author} {\bibfnamefont {S.}~\bibnamefont {Lee}}, \bibinfo {author} {\bibfnamefont {L.}~\bibnamefont {Shi}},\ and\ \bibinfo {author} {\bibfnamefont {Y.}~\bibnamefont {Wang}},\ }\bibfield  {title} {\bibinfo {title} {Transient hydrodynamic lattice cooling by picosecond laser irradiation of graphite},\ }\href {https://link.aps.org/doi/10.1103/PhysRevLett.127.085901} {\bibfield  {journal} {\bibinfo  {journal} {Phys. Rev. Lett.}\ }\textbf {\bibinfo {volume} {127}},\ \bibinfo {pages} {085901} (\bibinfo {year} {2021})}\BibitemShut {NoStop}%
\bibitem [{\citenamefont {Ding}\ \emph {et~al.}(2022)\citenamefont {Ding}, \citenamefont {Chen}, \citenamefont {Song}, \citenamefont {Shin}, \citenamefont {Maznev}, \citenamefont {Nelson},\ and\ \citenamefont {Chen}}]{Ding2022}%
  \BibitemOpen
  \bibfield  {author} {\bibinfo {author} {\bibfnamefont {Z.}~\bibnamefont {Ding}}, \bibinfo {author} {\bibfnamefont {K.}~\bibnamefont {Chen}}, \bibinfo {author} {\bibfnamefont {B.}~\bibnamefont {Song}}, \bibinfo {author} {\bibfnamefont {J.}~\bibnamefont {Shin}}, \bibinfo {author} {\bibfnamefont {A.~A.}\ \bibnamefont {Maznev}}, \bibinfo {author} {\bibfnamefont {K.~A.}\ \bibnamefont {Nelson}},\ and\ \bibinfo {author} {\bibfnamefont {G.}~\bibnamefont {Chen}},\ }\bibfield  {title} {\bibinfo {title} {Observation of second sound in graphite over 200 k},\ }\href {https://doi.org/10.1038/s41467-021-27907-z} {\bibfield  {journal} {\bibinfo  {journal} {Nat. Commun.}\ }\textbf {\bibinfo {volume} {13}},\ \bibinfo {pages} {285} (\bibinfo {year} {2022})}\BibitemShut {NoStop}%
\bibitem [{Note1()}]{Note1}%
  \BibitemOpen
  \bibinfo {note} {Specifically, Ref.~\cite {Ding2022} observed temperature oscillations at temperatures as high as 200 K in isotopically purified graphite, while the pioneering works \cite {Huberman2019,Jeong2021} observed temperature oscillations around 80-100~K in graphite at natural isotopic abundance}\BibitemShut {NoStop}%
\bibitem [{\citenamefont {Bocquet}\ and\ \citenamefont {Barrat}(2007)}]{bocquetFlowBoundaryConditions2007}%
  \BibitemOpen
  \bibfield  {author} {\bibinfo {author} {\bibfnamefont {L.}~\bibnamefont {Bocquet}}\ and\ \bibinfo {author} {\bibfnamefont {J.-L.}\ \bibnamefont {Barrat}},\ }\bibfield  {title} {\bibinfo {title} {Flow boundary conditions from nano- to micro-scales},\ }\href {https://doi.org/10.1039/b616490k} {\bibfield  {journal} {\bibinfo  {journal} {Soft Matter}\ }\textbf {\bibinfo {volume} {3}},\ \bibinfo {pages} {685} (\bibinfo {year} {2007})}\BibitemShut {NoStop}%
\bibitem [{\citenamefont {Peierls}(1955)}]{peierls1955quantum}%
  \BibitemOpen
  \bibfield  {author} {\bibinfo {author} {\bibfnamefont {R.~E.}\ \bibnamefont {Peierls}},\ }\href@noop {} {\emph {\bibinfo {title} {{Quantum theory of solids}}}}\ (\bibinfo  {publisher} {Oxford university press},\ \bibinfo {year} {1955})\BibitemShut {NoStop}%
\bibitem [{\citenamefont {Lindsay}\ \emph {et~al.}(2019)\citenamefont {Lindsay}, \citenamefont {Katre}, \citenamefont {Cepellotti},\ and\ \citenamefont {Mingo}}]{lindsay_perspective_2019}%
  \BibitemOpen
  \bibfield  {author} {\bibinfo {author} {\bibfnamefont {L.}~\bibnamefont {Lindsay}}, \bibinfo {author} {\bibfnamefont {A.}~\bibnamefont {Katre}}, \bibinfo {author} {\bibfnamefont {A.}~\bibnamefont {Cepellotti}},\ and\ \bibinfo {author} {\bibfnamefont {N.}~\bibnamefont {Mingo}},\ }\bibfield  {title} {\bibinfo {title} {Perspective on \textit{ab initio} phonon thermal transport},\ }\href {http://aip.scitation.org/doi/10.1063/1.5108651} {\bibfield  {journal} {\bibinfo  {journal} {J. Appl. Phys.}\ }\textbf {\bibinfo {volume} {126}},\ \bibinfo {pages} {050902} (\bibinfo {year} {2019})}\BibitemShut {NoStop}%
\bibitem [{\citenamefont {Chaput}(2013)}]{chaput_direct_2013}%
  \BibitemOpen
  \bibfield  {author} {\bibinfo {author} {\bibfnamefont {L.}~\bibnamefont {Chaput}},\ }\bibfield  {title} {\bibinfo {title} {Direct {Solution} to the {Linearized} {Phonon} {Boltzmann} {Equation}},\ }\href {https://link.aps.org/doi/10.1103/PhysRevLett.110.265506} {\bibfield  {journal} {\bibinfo  {journal} {Phys. Rev. Lett.}\ }\textbf {\bibinfo {volume} {110}},\ \bibinfo {pages} {265506} (\bibinfo {year} {2013})}\BibitemShut {NoStop}%
\bibitem [{\citenamefont {Fugallo}\ \emph {et~al.}(2013)\citenamefont {Fugallo}, \citenamefont {Lazzeri}, \citenamefont {Paulatto},\ and\ \citenamefont {Mauri}}]{fugallo_ab_2013}%
  \BibitemOpen
  \bibfield  {author} {\bibinfo {author} {\bibfnamefont {G.}~\bibnamefont {Fugallo}}, \bibinfo {author} {\bibfnamefont {M.}~\bibnamefont {Lazzeri}}, \bibinfo {author} {\bibfnamefont {L.}~\bibnamefont {Paulatto}},\ and\ \bibinfo {author} {\bibfnamefont {F.}~\bibnamefont {Mauri}},\ }\bibfield  {title} {\bibinfo {title} {\textit{{Ab} initio} variational approach for evaluating lattice thermal conductivity},\ }\href {https://link.aps.org/doi/10.1103/PhysRevB.88.045430} {\bibfield  {journal} {\bibinfo  {journal} {Phys. Rev. B}\ }\textbf {\bibinfo {volume} {88}},\ \bibinfo {pages} {045430} (\bibinfo {year} {2013})}\BibitemShut {NoStop}%
\bibitem [{\citenamefont {Lee}\ \emph {et~al.}(2015)\citenamefont {Lee}, \citenamefont {Broido}, \citenamefont {Esfarjani},\ and\ \citenamefont {Chen}}]{app_Lee2015}%
  \BibitemOpen
  \bibfield  {author} {\bibinfo {author} {\bibfnamefont {S.}~\bibnamefont {Lee}}, \bibinfo {author} {\bibfnamefont {D.}~\bibnamefont {Broido}}, \bibinfo {author} {\bibfnamefont {K.}~\bibnamefont {Esfarjani}},\ and\ \bibinfo {author} {\bibfnamefont {G.}~\bibnamefont {Chen}},\ }\bibfield  {title} {\bibinfo {title} {Hydrodynamic phonon transport in suspended graphene},\ }\href {https://doi.org/10.1038/ncomms7290} {\bibfield  {journal} {\bibinfo  {journal} {Nat. Commun.}\ }\textbf {\bibinfo {volume} {6}},\ \bibinfo {pages} {6290} (\bibinfo {year} {2015})}\BibitemShut {NoStop}%
\bibitem [{\citenamefont {Cepellotti}\ and\ \citenamefont {Marzari}(2016)}]{cepellotti_thermal_2016}%
  \BibitemOpen
  \bibfield  {author} {\bibinfo {author} {\bibfnamefont {A.}~\bibnamefont {Cepellotti}}\ and\ \bibinfo {author} {\bibfnamefont {N.}~\bibnamefont {Marzari}},\ }\bibfield  {title} {\bibinfo {title} {Thermal {Transport} in {Crystals} as a {Kinetic} {Theory} of {Relaxons}},\ }\href {https://link.aps.org/doi/10.1103/PhysRevX.6.041013} {\bibfield  {journal} {\bibinfo  {journal} {Phys. Rev. X}\ }\textbf {\bibinfo {volume} {6}},\ \bibinfo {pages} {041013} (\bibinfo {year} {2016})}\BibitemShut {NoStop}%
\bibitem [{\citenamefont {Ding}\ \emph {et~al.}(2018)\citenamefont {Ding}, \citenamefont {Zhou}, \citenamefont {Song}, \citenamefont {Chiloyan}, \citenamefont {Li}, \citenamefont {Liu},\ and\ \citenamefont {Chen}}]{app_Ding2018}%
  \BibitemOpen
  \bibfield  {author} {\bibinfo {author} {\bibfnamefont {Z.}~\bibnamefont {Ding}}, \bibinfo {author} {\bibfnamefont {J.}~\bibnamefont {Zhou}}, \bibinfo {author} {\bibfnamefont {B.}~\bibnamefont {Song}}, \bibinfo {author} {\bibfnamefont {V.}~\bibnamefont {Chiloyan}}, \bibinfo {author} {\bibfnamefont {M.}~\bibnamefont {Li}}, \bibinfo {author} {\bibfnamefont {T.-H.}\ \bibnamefont {Liu}},\ and\ \bibinfo {author} {\bibfnamefont {G.}~\bibnamefont {Chen}},\ }\bibfield  {title} {\bibinfo {title} {Phonon hydrodynamic heat conduction and knudsen minimum in graphite},\ }\href {https://doi.org/10.1021/acs.nanolett.7b04932} {\bibfield  {journal} {\bibinfo  {journal} {Nano Lett.}\ }\textbf {\bibinfo {volume} {18}},\ \bibinfo {pages} {638} (\bibinfo {year} {2018})}\BibitemShut {NoStop}%
\bibitem [{\citenamefont {Guo}\ \emph {et~al.}(2021{\natexlab{a}})\citenamefont {Guo}, \citenamefont {Zhang}, \citenamefont {Bescond}, \citenamefont {Xiong}, \citenamefont {Wang}, \citenamefont {Nomura},\ and\ \citenamefont {Volz}}]{guo_size_2021}%
  \BibitemOpen
  \bibfield  {author} {\bibinfo {author} {\bibfnamefont {Y.}~\bibnamefont {Guo}}, \bibinfo {author} {\bibfnamefont {Z.}~\bibnamefont {Zhang}}, \bibinfo {author} {\bibfnamefont {M.}~\bibnamefont {Bescond}}, \bibinfo {author} {\bibfnamefont {S.}~\bibnamefont {Xiong}}, \bibinfo {author} {\bibfnamefont {M.}~\bibnamefont {Wang}}, \bibinfo {author} {\bibfnamefont {M.}~\bibnamefont {Nomura}},\ and\ \bibinfo {author} {\bibfnamefont {S.}~\bibnamefont {Volz}},\ }\bibfield  {title} {\bibinfo {title} {Size effect on phonon hydrodynamics in graphite microstructures and nanostructures},\ }\href {https://link.aps.org/doi/10.1103/PhysRevB.104.075450} {\bibfield  {journal} {\bibinfo  {journal} {Phys. Rev. B}\ }\textbf {\bibinfo {volume} {104}},\ \bibinfo {pages} {075450} (\bibinfo {year} {2021}{\natexlab{a}})}\BibitemShut {NoStop}%
\bibitem [{\citenamefont {Li}\ \emph {et~al.}(2022)\citenamefont {Li}, \citenamefont {Lee}, \citenamefont {Ou}, \citenamefont {Lee},\ and\ \citenamefont {Shi}}]{li_reexamination_2022}%
  \BibitemOpen
  \bibfield  {author} {\bibinfo {author} {\bibfnamefont {X.}~\bibnamefont {Li}}, \bibinfo {author} {\bibfnamefont {H.}~\bibnamefont {Lee}}, \bibinfo {author} {\bibfnamefont {E.}~\bibnamefont {Ou}}, \bibinfo {author} {\bibfnamefont {S.}~\bibnamefont {Lee}},\ and\ \bibinfo {author} {\bibfnamefont {L.}~\bibnamefont {Shi}},\ }\bibfield  {title} {\bibinfo {title} {Reexamination of hydrodynamic phonon transport in thin graphite},\ }\href {https://aip.scitation.org/doi/10.1063/5.0078772} {\bibfield  {journal} {\bibinfo  {journal} {J. Appl. Phys.}\ }\textbf {\bibinfo {volume} {131}},\ \bibinfo {pages} {075104} (\bibinfo {year} {2022})}\BibitemShut {NoStop}%
\bibitem [{\citenamefont {Huang}\ \emph {et~al.}(2022)\citenamefont {Huang}, \citenamefont {Guo}, \citenamefont {Volz},\ and\ \citenamefont {Nomura}}]{huang_mapping_2022}%
  \BibitemOpen
  \bibfield  {author} {\bibinfo {author} {\bibfnamefont {X.}~\bibnamefont {Huang}}, \bibinfo {author} {\bibfnamefont {Y.}~\bibnamefont {Guo}}, \bibinfo {author} {\bibfnamefont {S.}~\bibnamefont {Volz}},\ and\ \bibinfo {author} {\bibfnamefont {M.}~\bibnamefont {Nomura}},\ }\bibfield  {title} {\bibinfo {title} {Mapping phonon hydrodynamic strength in micrometer-scale graphite structures},\ }\href {https://dx.doi.org/10.35848/1882-0786/ac8f82} {\bibfield  {journal} {\bibinfo  {journal} {Appl. Phys. Express}\ }\textbf {\bibinfo {volume} {15}},\ \bibinfo {pages} {105001} (\bibinfo {year} {2022})}\BibitemShut {NoStop}%
\bibitem [{\citenamefont {Cepellotti}\ \emph {et~al.}(2015)\citenamefont {Cepellotti}, \citenamefont {Fugallo}, \citenamefont {Paulatto}, \citenamefont {Lazzeri}, \citenamefont {Mauri},\ and\ \citenamefont {Marzari}}]{app_Cepellotti2015}%
  \BibitemOpen
  \bibfield  {author} {\bibinfo {author} {\bibfnamefont {A.}~\bibnamefont {Cepellotti}}, \bibinfo {author} {\bibfnamefont {G.}~\bibnamefont {Fugallo}}, \bibinfo {author} {\bibfnamefont {L.}~\bibnamefont {Paulatto}}, \bibinfo {author} {\bibfnamefont {M.}~\bibnamefont {Lazzeri}}, \bibinfo {author} {\bibfnamefont {F.}~\bibnamefont {Mauri}},\ and\ \bibinfo {author} {\bibfnamefont {N.}~\bibnamefont {Marzari}},\ }\bibfield  {title} {\bibinfo {title} {Phonon hydrodynamics in two-dimensional materials},\ }\href {https://doi.org/10.1038/ncomms7400} {\bibfield  {journal} {\bibinfo  {journal} {Nat. Commun.}\ }\textbf {\bibinfo {volume} {6}},\ \bibinfo {pages} {6400} (\bibinfo {year} {2015})}\BibitemShut {NoStop}%
\bibitem [{\citenamefont {Majee}\ and\ \citenamefont {Aksamija}(2018)}]{majee_dynamical_2018}%
  \BibitemOpen
  \bibfield  {author} {\bibinfo {author} {\bibfnamefont {A.~K.}\ \bibnamefont {Majee}}\ and\ \bibinfo {author} {\bibfnamefont {Z.}~\bibnamefont {Aksamija}},\ }\bibfield  {title} {\bibinfo {title} {Dynamical thermal conductivity of suspended graphene ribbons in the hydrodynamic regime},\ }\href {https://link.aps.org/doi/10.1103/PhysRevB.98.024303} {\bibfield  {journal} {\bibinfo  {journal} {Phys. Rev. B}\ }\textbf {\bibinfo {volume} {98}},\ \bibinfo {pages} {024303} (\bibinfo {year} {2018})}\BibitemShut {NoStop}%
\bibitem [{\citenamefont {Raya-Moreno}\ \emph {et~al.}(2022{\natexlab{a}})\citenamefont {Raya-Moreno}, \citenamefont {Carrete},\ and\ \citenamefont {Cartoixà}}]{raya-moreno_hydrodynamic_2022}%
  \BibitemOpen
  \bibfield  {author} {\bibinfo {author} {\bibfnamefont {M.}~\bibnamefont {Raya-Moreno}}, \bibinfo {author} {\bibfnamefont {J.}~\bibnamefont {Carrete}},\ and\ \bibinfo {author} {\bibfnamefont {X.}~\bibnamefont {Cartoixà}},\ }\bibfield  {title} {\bibinfo {title} {Hydrodynamic signatures in thermal transport in devices based on two-dimensional materials: {An} \textit{ab initio} study},\ }\href {https://link.aps.org/doi/10.1103/PhysRevB.106.014308} {\bibfield  {journal} {\bibinfo  {journal} {Phys. Rev. B}\ }\textbf {\bibinfo {volume} {106}},\ \bibinfo {pages} {014308} (\bibinfo {year} {2022}{\natexlab{a}})}\BibitemShut {NoStop}%
\bibitem [{\citenamefont {Guo}\ \emph {et~al.}(2021{\natexlab{b}})\citenamefont {Guo}, \citenamefont {Zhang}, \citenamefont {Nomura}, \citenamefont {Volz},\ and\ \citenamefont {Wang}}]{guo_phonon_2021}%
  \BibitemOpen
  \bibfield  {author} {\bibinfo {author} {\bibfnamefont {Y.}~\bibnamefont {Guo}}, \bibinfo {author} {\bibfnamefont {Z.}~\bibnamefont {Zhang}}, \bibinfo {author} {\bibfnamefont {M.}~\bibnamefont {Nomura}}, \bibinfo {author} {\bibfnamefont {S.}~\bibnamefont {Volz}},\ and\ \bibinfo {author} {\bibfnamefont {M.}~\bibnamefont {Wang}},\ }\bibfield  {title} {\bibinfo {title} {Phonon vortex dynamics in graphene ribbon by solving {Boltzmann} transport equation with ab initio scattering rates},\ }\href {https://linkinghub.elsevier.com/retrieve/pii/S0017931021000843} {\bibfield  {journal} {\bibinfo  {journal} {Int. J. Heat Mass Transf.}\ }\textbf {\bibinfo {volume} {169}},\ \bibinfo {pages} {120981} (\bibinfo {year} {2021}{\natexlab{b}})}\BibitemShut {NoStop}%
\bibitem [{\citenamefont {Zhang}\ \emph {et~al.}(2022)\citenamefont {Zhang}, \citenamefont {Huberman},\ and\ \citenamefont {Wu}}]{zhang_emergence_2022}%
  \BibitemOpen
  \bibfield  {author} {\bibinfo {author} {\bibfnamefont {C.}~\bibnamefont {Zhang}}, \bibinfo {author} {\bibfnamefont {S.}~\bibnamefont {Huberman}},\ and\ \bibinfo {author} {\bibfnamefont {L.}~\bibnamefont {Wu}},\ }\bibfield  {title} {\bibinfo {title} {On the emergence of heat waves in the transient thermal grating geometry},\ }\href {https://doi.org/10.1063/5.0102227} {\bibfield  {journal} {\bibinfo  {journal} {J. Appl. Phys.}\ }\textbf {\bibinfo {volume} {132}},\ \bibinfo {pages} {085103} (\bibinfo {year} {2022})}\BibitemShut {NoStop}%
\bibitem [{\citenamefont {Han}\ and\ \citenamefont {Ruan}(2023)}]{han_is_2023}%
  \BibitemOpen
  \bibfield  {author} {\bibinfo {author} {\bibfnamefont {Z.}~\bibnamefont {Han}}\ and\ \bibinfo {author} {\bibfnamefont {X.}~\bibnamefont {Ruan}},\ }\bibfield  {title} {\bibinfo {title} {Is {Thermal} {Conductivity} of {Graphene} {Divergent} and {Higher} {Than} {Diamond}?},\ }\href {http://arxiv.org/abs/2302.12216} {\bibfield  {journal} {\bibinfo  {journal} {arXiv: 2302.12216}\ } (\bibinfo {year} {2023})}\BibitemShut {NoStop}%
\bibitem [{\citenamefont {Cepellotti}\ and\ \citenamefont {Marzari}(2017{\natexlab{a}})}]{cepellotti_boltzmann_2017}%
  \BibitemOpen
  \bibfield  {author} {\bibinfo {author} {\bibfnamefont {A.}~\bibnamefont {Cepellotti}}\ and\ \bibinfo {author} {\bibfnamefont {N.}~\bibnamefont {Marzari}},\ }\bibfield  {title} {\bibinfo {title} {Boltzmann {Transport} in {Nanostructures} as a {Friction} {Effect}},\ }\href {https://doi.org/10.1021/acs.nanolett.7b01202} {\bibfield  {journal} {\bibinfo  {journal} {Nano Lett.}\ }\textbf {\bibinfo {volume} {17}},\ \bibinfo {pages} {4675} (\bibinfo {year} {2017}{\natexlab{a}})}\BibitemShut {NoStop}%
\bibitem [{\citenamefont {Machida}\ \emph {et~al.}(2018)\citenamefont {Machida}, \citenamefont {Subedi}, \citenamefont {Akiba}, \citenamefont {Miyake}, \citenamefont {Tokunaga}, \citenamefont {Akahama}, \citenamefont {Izawa},\ and\ \citenamefont {Behnia}}]{machida_observation_2018}%
  \BibitemOpen
  \bibfield  {author} {\bibinfo {author} {\bibfnamefont {Y.}~\bibnamefont {Machida}}, \bibinfo {author} {\bibfnamefont {A.}~\bibnamefont {Subedi}}, \bibinfo {author} {\bibfnamefont {K.}~\bibnamefont {Akiba}}, \bibinfo {author} {\bibfnamefont {A.}~\bibnamefont {Miyake}}, \bibinfo {author} {\bibfnamefont {M.}~\bibnamefont {Tokunaga}}, \bibinfo {author} {\bibfnamefont {Y.}~\bibnamefont {Akahama}}, \bibinfo {author} {\bibfnamefont {K.}~\bibnamefont {Izawa}},\ and\ \bibinfo {author} {\bibfnamefont {K.}~\bibnamefont {Behnia}},\ }\bibfield  {title} {\bibinfo {title} {Observation of {Poiseuille} flow of phonons in black phosphorus},\ }\href {https://www.science.org/doi/10.1126/sciadv.aat3374} {\bibfield  {journal} {\bibinfo  {journal} {Sci. Adv.}\ }\textbf {\bibinfo {volume} {4}},\ \bibinfo {pages} {eaat3374} (\bibinfo {year} {2018})}\BibitemShut {NoStop}%
\bibitem [{\citenamefont {Sendra}\ \emph {et~al.}(2022)\citenamefont {Sendra}, \citenamefont {Beardo}, \citenamefont {Bafaluy}, \citenamefont {Torres}, \citenamefont {Alvarez},\ and\ \citenamefont {Camacho}}]{sendra_hydrodynamic_2022}%
  \BibitemOpen
  \bibfield  {author} {\bibinfo {author} {\bibfnamefont {L.}~\bibnamefont {Sendra}}, \bibinfo {author} {\bibfnamefont {A.}~\bibnamefont {Beardo}}, \bibinfo {author} {\bibfnamefont {J.}~\bibnamefont {Bafaluy}}, \bibinfo {author} {\bibfnamefont {P.}~\bibnamefont {Torres}}, \bibinfo {author} {\bibfnamefont {F.~X.}\ \bibnamefont {Alvarez}},\ and\ \bibinfo {author} {\bibfnamefont {J.}~\bibnamefont {Camacho}},\ }\bibfield  {title} {\bibinfo {title} {Hydrodynamic heat transport in dielectric crystals in the collective limit and the drifting/driftless velocity conundrum},\ }\href {https://link.aps.org/doi/10.1103/PhysRevB.106.155301} {\bibfield  {journal} {\bibinfo  {journal} {Phys. Rev. B}\ }\textbf {\bibinfo {volume} {106}},\ \bibinfo {pages} {155301} (\bibinfo {year} {2022})}\BibitemShut {NoStop}%
\bibitem [{\citenamefont {Li}\ and\ \citenamefont {Lee}(2018)}]{li_role_2018}%
  \BibitemOpen
  \bibfield  {author} {\bibinfo {author} {\bibfnamefont {X.}~\bibnamefont {Li}}\ and\ \bibinfo {author} {\bibfnamefont {S.}~\bibnamefont {Lee}},\ }\bibfield  {title} {\bibinfo {title} {Role of hydrodynamic viscosity on phonon transport in suspended graphene},\ }\href {https://link.aps.org/doi/10.1103/PhysRevB.97.094309} {\bibfield  {journal} {\bibinfo  {journal} {Phys. Rev. B}\ }\textbf {\bibinfo {volume} {97}},\ \bibinfo {pages} {094309} (\bibinfo {year} {2018})}\BibitemShut {NoStop}%
\bibitem [{\citenamefont {Guo}\ \emph {et~al.}(2018)\citenamefont {Guo}, \citenamefont {Jou},\ and\ \citenamefont {Wang}}]{guo_nonequilibrium_2018}%
  \BibitemOpen
  \bibfield  {author} {\bibinfo {author} {\bibfnamefont {Y.}~\bibnamefont {Guo}}, \bibinfo {author} {\bibfnamefont {D.}~\bibnamefont {Jou}},\ and\ \bibinfo {author} {\bibfnamefont {M.}~\bibnamefont {Wang}},\ }\bibfield  {title} {\bibinfo {title} {Nonequilibrium thermodynamics of phonon hydrodynamic model for nanoscale heat transport},\ }\href {https://link.aps.org/doi/10.1103/PhysRevB.98.104304} {\bibfield  {journal} {\bibinfo  {journal} {Phys. Rev. B}\ }\textbf {\bibinfo {volume} {98}},\ \bibinfo {pages} {104304} (\bibinfo {year} {2018})}\BibitemShut {NoStop}%
\bibitem [{\citenamefont {Shang}\ \emph {et~al.}(2020)\citenamefont {Shang}, \citenamefont {Zhang}, \citenamefont {Guo},\ and\ \citenamefont {Lü}}]{shang_heat_2020}%
  \BibitemOpen
  \bibfield  {author} {\bibinfo {author} {\bibfnamefont {M.-Y.}\ \bibnamefont {Shang}}, \bibinfo {author} {\bibfnamefont {C.}~\bibnamefont {Zhang}}, \bibinfo {author} {\bibfnamefont {Z.}~\bibnamefont {Guo}},\ and\ \bibinfo {author} {\bibfnamefont {J.-T.}\ \bibnamefont {Lü}},\ }\bibfield  {title} {\bibinfo {title} {Heat vortex in hydrodynamic phonon transport of two-dimensional materials},\ }\href {http://www.nature.com/articles/s41598-020-65221-8} {\bibfield  {journal} {\bibinfo  {journal} {Sci. Rep.}\ }\textbf {\bibinfo {volume} {10}},\ \bibinfo {pages} {8272} (\bibinfo {year} {2020})}\BibitemShut {NoStop}%
\bibitem [{\citenamefont {Simoncelli}\ \emph {et~al.}(2020)\citenamefont {Simoncelli}, \citenamefont {Marzari},\ and\ \citenamefont {Cepellotti}}]{Simoncelli2020}%
  \BibitemOpen
  \bibfield  {author} {\bibinfo {author} {\bibfnamefont {M.}~\bibnamefont {Simoncelli}}, \bibinfo {author} {\bibfnamefont {N.}~\bibnamefont {Marzari}},\ and\ \bibinfo {author} {\bibfnamefont {A.}~\bibnamefont {Cepellotti}},\ }\bibfield  {title} {\bibinfo {title} {Generalization of fourier's law into viscous heat equations},\ }\href {https://link.aps.org/doi/10.1103/PhysRevX.10.011019} {\bibfield  {journal} {\bibinfo  {journal} {Phys. Rev. X}\ }\textbf {\bibinfo {volume} {10}},\ \bibinfo {pages} {011019} (\bibinfo {year} {2020})}\BibitemShut {NoStop}%
\bibitem [{\citenamefont {Sendra}\ \emph {et~al.}(2021)\citenamefont {Sendra}, \citenamefont {Beardo}, \citenamefont {Torres}, \citenamefont {Bafaluy}, \citenamefont {Alvarez},\ and\ \citenamefont {Camacho}}]{sendra_derivation_2021}%
  \BibitemOpen
  \bibfield  {author} {\bibinfo {author} {\bibfnamefont {L.}~\bibnamefont {Sendra}}, \bibinfo {author} {\bibfnamefont {A.}~\bibnamefont {Beardo}}, \bibinfo {author} {\bibfnamefont {P.}~\bibnamefont {Torres}}, \bibinfo {author} {\bibfnamefont {J.}~\bibnamefont {Bafaluy}}, \bibinfo {author} {\bibfnamefont {F.~X.}\ \bibnamefont {Alvarez}},\ and\ \bibinfo {author} {\bibfnamefont {J.}~\bibnamefont {Camacho}},\ }\bibfield  {title} {\bibinfo {title} {Derivation of a hydrodynamic heat equation from the phonon {Boltzmann} equation for general semiconductors},\ }\href {https://link.aps.org/doi/10.1103/PhysRevB.103.L140301} {\bibfield  {journal} {\bibinfo  {journal} {Phys. Rev. B}\ }\textbf {\bibinfo {volume} {103}},\ \bibinfo {pages} {L140301} (\bibinfo {year} {2021})}\BibitemShut {NoStop}%
\bibitem [{\citenamefont {Michalis}\ \emph {et~al.}(2010)\citenamefont {Michalis}, \citenamefont {Kalarakis}, \citenamefont {Skouras},\ and\ \citenamefont {Burganos}}]{michalis_rarefaction_2010}%
  \BibitemOpen
  \bibfield  {author} {\bibinfo {author} {\bibfnamefont {V.~K.}\ \bibnamefont {Michalis}}, \bibinfo {author} {\bibfnamefont {A.~N.}\ \bibnamefont {Kalarakis}}, \bibinfo {author} {\bibfnamefont {E.~D.}\ \bibnamefont {Skouras}},\ and\ \bibinfo {author} {\bibfnamefont {V.~N.}\ \bibnamefont {Burganos}},\ }\bibfield  {title} {\bibinfo {title} {Rarefaction effects on gas viscosity in the {Knudsen} transition regime},\ }\href {https://doi.org/10.1007/s10404-010-0606-3} {\bibfield  {journal} {\bibinfo  {journal} {Microfluidics and Nanofluidics}\ }\textbf {\bibinfo {volume} {9}},\ \bibinfo {pages} {847} (\bibinfo {year} {2010})}\BibitemShut {NoStop}%
\bibitem [{Note2()}]{Note2}%
  \BibitemOpen
  \bibinfo {note} {We note that the VHE have a mathematical form analogous to the damped and diffusion-extended Navier-Stokes equations, used to describe rarefied fluids flowing in porous media in isothermal and laminar conditions, with VHE's temperature mapping to fluid's pressure and VHE's drift velocity mapping to fluid velocity. To see this, we highlight how the linear damping term $-\gamma ^{ij}u^j(\protect \bm {r},t)$ appearing in the second VHE~(\ref {viscous_heat_U}) is analogous to the dissipative term appearing in the linearized ``damped'' Navier-Stokes equations \cite {balasubramanian_darcys_1987,dardis_lattice_1998,bresch_existence_2003,cai_weak_2008,zhang_uniqueness_2011} used, e.g., to describe a fluid flowing through a porous medium. Then, the diffusive term appearing in the first VHE~(\ref {viscous_heat_T}) $-\kappa _D^{ij}\protect \tfrac {\partial ^2 T(\protect \bm {r},t)}{\partial r^i\partial r^j}$ is analogous to the self-diffusion term appearing for pressure in the extended Navier-Stokes equations \cite {brenner_navierstokes_2005,sambasivam_numerical_2014,schwarz_openfoam_2023} used to describe the flow of an isothermal, compressible, and rarefied gas \cite {maurer_second-order_2003,dongari_pressure-driven_2009}.}\BibitemShut {Stop}%
\bibitem [{\citenamefont {Joseph}\ and\ \citenamefont {Preziosi}(1989)}]{joseph_heat_1989}%
  \BibitemOpen
  \bibfield  {author} {\bibinfo {author} {\bibfnamefont {D.~D.}\ \bibnamefont {Joseph}}\ and\ \bibinfo {author} {\bibfnamefont {L.}~\bibnamefont {Preziosi}},\ }\bibfield  {title} {\bibinfo {title} {Heat waves},\ }\href {https://link.aps.org/doi/10.1103/RevModPhys.61.41} {\bibfield  {journal} {\bibinfo  {journal} {Rev. Mod. Phys.}\ }\textbf {\bibinfo {volume} {61}},\ \bibinfo {pages} {41} (\bibinfo {year} {1989})}\BibitemShut {NoStop}%
\bibitem [{\citenamefont {Tzou}(1995)}]{tzou_unified_1995}%
  \BibitemOpen
  \bibfield  {author} {\bibinfo {author} {\bibfnamefont {D.~Y.}\ \bibnamefont {Tzou}},\ }\bibfield  {title} {\bibinfo {title} {A {Unified} {Field} {Approach} for {Heat} {Conduction} {From} {Macro}- to {Micro}-{Scales}},\ }\href@noop {} {\bibfield  {journal} {\bibinfo  {journal} {J Heat Transf}\ }\textbf {\bibinfo {volume} {117}},\ \bibinfo {pages} {8} (\bibinfo {year} {1995})}\BibitemShut {NoStop}%
\bibitem [{Note3()}]{Note3}%
  \BibitemOpen
  \bibinfo {note} {The DPLE encompasses Cattaneo's second-sound equation \cite {cattaneo1958form} as a special case, as discussed in SM~\ref {sec:DPLE_derivation} and \cite {tzou_unified_1995}}\BibitemShut {NoStop}%
\bibitem [{\citenamefont {Guyer}\ and\ \citenamefont {Krumhansl}(1966)}]{guyer_solution_1966}%
  \BibitemOpen
  \bibfield  {author} {\bibinfo {author} {\bibfnamefont {R.~A.}\ \bibnamefont {Guyer}}\ and\ \bibinfo {author} {\bibfnamefont {J.~A.}\ \bibnamefont {Krumhansl}},\ }\bibfield  {title} {\bibinfo {title} {Solution of the {Linearized} {Phonon} {Boltzmann} {Equation}},\ }\href {https://link.aps.org/doi/10.1103/PhysRev.148.766} {\bibfield  {journal} {\bibinfo  {journal} {Phys. Rev.}\ }\textbf {\bibinfo {volume} {148}},\ \bibinfo {pages} {766} (\bibinfo {year} {1966})}\BibitemShut {NoStop}%
\bibitem [{Note4()}]{Note4}%
  \BibitemOpen
  \bibinfo {note} {We investigate monoisotopic h$^{11}$BN because Refs.~\cite {yuan_modulating_2019,SimoncelliPhD} suggest that heat hydrodynamics in h$^{11}$BN is stronger than in hexagonal boron nitride with natural isotopic-mass disorder (19.9\% $^{10}$B and 80.1\% $^{11}$B)}\BibitemShut {NoStop}%
\bibitem [{\citenamefont {Aharon-Steinberg}\ \emph {et~al.}(2022)\citenamefont {Aharon-Steinberg}, \citenamefont {Völkl}, \citenamefont {Kaplan}, \citenamefont {Pariari}, \citenamefont {Roy}, \citenamefont {Holder}, \citenamefont {Wolf}, \citenamefont {Meltzer}, \citenamefont {Myasoedov}, \citenamefont {Huber}, \citenamefont {Yan}, \citenamefont {Falkovich}, \citenamefont {Levitov}, \citenamefont {Hücker},\ and\ \citenamefont {Zeldov}}]{e-vortexes}%
  \BibitemOpen
  \bibfield  {author} {\bibinfo {author} {\bibfnamefont {A.}~\bibnamefont {Aharon-Steinberg}}, \bibinfo {author} {\bibfnamefont {T.}~\bibnamefont {Völkl}}, \bibinfo {author} {\bibfnamefont {A.}~\bibnamefont {Kaplan}}, \bibinfo {author} {\bibfnamefont {A.~K.}\ \bibnamefont {Pariari}}, \bibinfo {author} {\bibfnamefont {I.}~\bibnamefont {Roy}}, \bibinfo {author} {\bibfnamefont {T.}~\bibnamefont {Holder}}, \bibinfo {author} {\bibfnamefont {Y.}~\bibnamefont {Wolf}}, \bibinfo {author} {\bibfnamefont {A.~Y.}\ \bibnamefont {Meltzer}}, \bibinfo {author} {\bibfnamefont {Y.}~\bibnamefont {Myasoedov}}, \bibinfo {author} {\bibfnamefont {M.~E.}\ \bibnamefont {Huber}}, \bibinfo {author} {\bibfnamefont {B.}~\bibnamefont {Yan}}, \bibinfo {author} {\bibfnamefont {G.}~\bibnamefont {Falkovich}}, \bibinfo {author} {\bibfnamefont {L.~S.}\ \bibnamefont {Levitov}}, \bibinfo {author} {\bibfnamefont {M.}~\bibnamefont {Hücker}},\ and\ \bibinfo {author} {\bibfnamefont {E.}~\bibnamefont {Zeldov}},\ }\bibfield  {title} {\bibinfo {title} {Direct observation of vortices in an electron fluid},\ }\href {https://doi.org/10.1038/s41586-022-04794-y} {\bibfield  {journal} {\bibinfo  {journal} {Nature}\ }\textbf {\bibinfo {volume} {607}},\ \bibinfo {pages} {74} (\bibinfo {year} {2022})}\BibitemShut {NoStop}%
\bibitem [{\citenamefont {Menges}\ \emph {et~al.}(2016)\citenamefont {Menges}, \citenamefont {Riel}, \citenamefont {Stemmer},\ and\ \citenamefont {Gotsmann}}]{menges_nanoscale_2016}%
  \BibitemOpen
  \bibfield  {author} {\bibinfo {author} {\bibfnamefont {F.}~\bibnamefont {Menges}}, \bibinfo {author} {\bibfnamefont {H.}~\bibnamefont {Riel}}, \bibinfo {author} {\bibfnamefont {A.}~\bibnamefont {Stemmer}},\ and\ \bibinfo {author} {\bibfnamefont {B.}~\bibnamefont {Gotsmann}},\ }\bibfield  {title} {\bibinfo {title} {Nanoscale thermometry by scanning thermal microscopy},\ }\href {https://aip.scitation.org/doi/full/10.1063/1.4955449} {\bibfield  {journal} {\bibinfo  {journal} {Rev. Sci. Instrum.}\ }\textbf {\bibinfo {volume} {87}},\ \bibinfo {pages} {074902} (\bibinfo {year} {2016})}\BibitemShut {NoStop}%
\bibitem [{\citenamefont {Cheng}\ \emph {et~al.}(2022)\citenamefont {Cheng}, \citenamefont {Ji},\ and\ \citenamefont {Cahill}}]{cheng_battery_2022}%
  \BibitemOpen
  \bibfield  {author} {\bibinfo {author} {\bibfnamefont {Z.}~\bibnamefont {Cheng}}, \bibinfo {author} {\bibfnamefont {X.}~\bibnamefont {Ji}},\ and\ \bibinfo {author} {\bibfnamefont {D.~G.}\ \bibnamefont {Cahill}},\ }\bibfield  {title} {\bibinfo {title} {Battery absorbs heat during charging uncovered by ultra-sensitive thermometry},\ }\href {https://www.sciencedirect.com/science/article/pii/S0378775321012544} {\bibfield  {journal} {\bibinfo  {journal} {J. Power Sources}\ }\textbf {\bibinfo {volume} {518}},\ \bibinfo {pages} {230762} (\bibinfo {year} {2022})}\BibitemShut {NoStop}%
\bibitem [{\citenamefont {Cahill}\ \emph {et~al.}(2014)\citenamefont {Cahill}, \citenamefont {Braun}, \citenamefont {Chen}, \citenamefont {Clarke}, \citenamefont {Fan}, \citenamefont {Goodson}, \citenamefont {Keblinski}, \citenamefont {King}, \citenamefont {Mahan}, \citenamefont {Majumdar}, \citenamefont {Maris}, \citenamefont {Phillpot}, \citenamefont {Pop},\ and\ \citenamefont {Shi}}]{cahill_nanoscale_2014}%
  \BibitemOpen
  \bibfield  {author} {\bibinfo {author} {\bibfnamefont {D.~G.}\ \bibnamefont {Cahill}}, \bibinfo {author} {\bibfnamefont {P.~V.}\ \bibnamefont {Braun}}, \bibinfo {author} {\bibfnamefont {G.}~\bibnamefont {Chen}}, \bibinfo {author} {\bibfnamefont {D.~R.}\ \bibnamefont {Clarke}}, \bibinfo {author} {\bibfnamefont {S.}~\bibnamefont {Fan}}, \bibinfo {author} {\bibfnamefont {K.~E.}\ \bibnamefont {Goodson}}, \bibinfo {author} {\bibfnamefont {P.}~\bibnamefont {Keblinski}}, \bibinfo {author} {\bibfnamefont {W.~P.}\ \bibnamefont {King}}, \bibinfo {author} {\bibfnamefont {G.~D.}\ \bibnamefont {Mahan}}, \bibinfo {author} {\bibfnamefont {A.}~\bibnamefont {Majumdar}}, \bibinfo {author} {\bibfnamefont {H.~J.}\ \bibnamefont {Maris}}, \bibinfo {author} {\bibfnamefont {S.~R.}\ \bibnamefont {Phillpot}}, \bibinfo {author} {\bibfnamefont {E.}~\bibnamefont {Pop}},\ and\ \bibinfo {author} {\bibfnamefont {L.}~\bibnamefont {Shi}},\ }\bibfield  {title} {\bibinfo {title} {Nanoscale thermal transport. {II}. 2003–2012},\ }\href {http://aip.scitation.org/doi/10.1063/1.4832615} {\bibfield  {journal} {\bibinfo  {journal} {Appl. Phys. Rev.}\ }\textbf {\bibinfo {volume} {1}},\ \bibinfo {pages} {011305} (\bibinfo {year} {2014})}\BibitemShut {NoStop}%
\bibitem [{\citenamefont {Braun}\ \emph {et~al.}(2022)\citenamefont {Braun}, \citenamefont {Furrer}, \citenamefont {Butti}, \citenamefont {Thodkar}, \citenamefont {Shorubalko}, \citenamefont {Zardo}, \citenamefont {Calame},\ and\ \citenamefont {Perrin}}]{braun_spatially_2022}%
  \BibitemOpen
  \bibfield  {author} {\bibinfo {author} {\bibfnamefont {O.}~\bibnamefont {Braun}}, \bibinfo {author} {\bibfnamefont {R.}~\bibnamefont {Furrer}}, \bibinfo {author} {\bibfnamefont {P.}~\bibnamefont {Butti}}, \bibinfo {author} {\bibfnamefont {K.}~\bibnamefont {Thodkar}}, \bibinfo {author} {\bibfnamefont {I.}~\bibnamefont {Shorubalko}}, \bibinfo {author} {\bibfnamefont {I.}~\bibnamefont {Zardo}}, \bibinfo {author} {\bibfnamefont {M.}~\bibnamefont {Calame}},\ and\ \bibinfo {author} {\bibfnamefont {M.~L.}\ \bibnamefont {Perrin}},\ }\bibfield  {title} {\bibinfo {title} {Spatially mapping thermal transport in graphene by an opto-thermal method},\ }\href {https://www.nature.com/articles/s41699-021-00277-2} {\bibfield  {journal} {\bibinfo  {journal} {NPJ 2D Mater. Appl.}\ }\textbf {\bibinfo {volume} {6}},\ \bibinfo {pages} {1} (\bibinfo {year} {2022})}\BibitemShut {NoStop}%
\bibitem [{\citenamefont {Ziabari}\ \emph {et~al.}(2018)\citenamefont {Ziabari}, \citenamefont {Torres}, \citenamefont {Vermeersch}, \citenamefont {Xuan}, \citenamefont {Cartoixà}, \citenamefont {Torelló}, \citenamefont {Bahk}, \citenamefont {Koh}, \citenamefont {Parsa}, \citenamefont {Ye}, \citenamefont {Alvarez},\ and\ \citenamefont {Shakouri}}]{ziabari_full-field_2018}%
  \BibitemOpen
  \bibfield  {author} {\bibinfo {author} {\bibfnamefont {A.}~\bibnamefont {Ziabari}}, \bibinfo {author} {\bibfnamefont {P.}~\bibnamefont {Torres}}, \bibinfo {author} {\bibfnamefont {B.}~\bibnamefont {Vermeersch}}, \bibinfo {author} {\bibfnamefont {Y.}~\bibnamefont {Xuan}}, \bibinfo {author} {\bibfnamefont {X.}~\bibnamefont {Cartoixà}}, \bibinfo {author} {\bibfnamefont {A.}~\bibnamefont {Torelló}}, \bibinfo {author} {\bibfnamefont {J.-H.}\ \bibnamefont {Bahk}}, \bibinfo {author} {\bibfnamefont {Y.~R.}\ \bibnamefont {Koh}}, \bibinfo {author} {\bibfnamefont {M.}~\bibnamefont {Parsa}}, \bibinfo {author} {\bibfnamefont {P.~D.}\ \bibnamefont {Ye}}, \bibinfo {author} {\bibfnamefont {F.~X.}\ \bibnamefont {Alvarez}},\ and\ \bibinfo {author} {\bibfnamefont {A.}~\bibnamefont {Shakouri}},\ }\bibfield  {title} {\bibinfo {title} {Full-field thermal imaging of quasiballistic crosstalk reduction in nanoscale devices},\ }\href {http://www.nature.com/articles/s41467-017-02652-4} {\bibfield  {journal} {\bibinfo  {journal} {Nat. Commun.}\ }\textbf {\bibinfo {volume} {9}},\ \bibinfo {pages} {255} (\bibinfo {year} {2018})}\BibitemShut {NoStop}%
\bibitem [{\citenamefont {Reihani}\ \emph {et~al.}(2022)\citenamefont {Reihani}, \citenamefont {Luan}, \citenamefont {Yan}, \citenamefont {Lim}, \citenamefont {Meyhofer},\ and\ \citenamefont {Reddy}}]{reihaniQuantitativeMappingUnmodulated2022}%
  \BibitemOpen
  \bibfield  {author} {\bibinfo {author} {\bibfnamefont {A.}~\bibnamefont {Reihani}}, \bibinfo {author} {\bibfnamefont {Y.}~\bibnamefont {Luan}}, \bibinfo {author} {\bibfnamefont {S.}~\bibnamefont {Yan}}, \bibinfo {author} {\bibfnamefont {J.~W.}\ \bibnamefont {Lim}}, \bibinfo {author} {\bibfnamefont {E.}~\bibnamefont {Meyhofer}},\ and\ \bibinfo {author} {\bibfnamefont {P.}~\bibnamefont {Reddy}},\ }\bibfield  {title} {\bibinfo {title} {Quantitative {Mapping} of {Unmodulated} {Temperature} {Fields} with {Nanometer} {Resolution}},\ }\href {https://doi.org/10.1021/acsnano.1c08513} {\bibfield  {journal} {\bibinfo  {journal} {ACS Nano}\ }\textbf {\bibinfo {volume} {16}},\ \bibinfo {pages} {939} (\bibinfo {year} {2022})}\BibitemShut {NoStop}%
\bibitem [{\citenamefont {Goblot}\ \emph {et~al.}(2024)\citenamefont {Goblot}, \citenamefont {Wu}, \citenamefont {Lucente}, \citenamefont {Zhu}, \citenamefont {Losero}, \citenamefont {Jobert}, \citenamefont {Concha}, \citenamefont {Marzari}, \citenamefont {Simoncelli},\ and\ \citenamefont {Galland}}]{goblotImagingHeatTransport2024}%
  \BibitemOpen
  \bibfield  {author} {\bibinfo {author} {\bibfnamefont {V.}~\bibnamefont {Goblot}}, \bibinfo {author} {\bibfnamefont {K.}~\bibnamefont {Wu}}, \bibinfo {author} {\bibfnamefont {E.~D.}\ \bibnamefont {Lucente}}, \bibinfo {author} {\bibfnamefont {Y.}~\bibnamefont {Zhu}}, \bibinfo {author} {\bibfnamefont {E.}~\bibnamefont {Losero}}, \bibinfo {author} {\bibfnamefont {Q.}~\bibnamefont {Jobert}}, \bibinfo {author} {\bibfnamefont {C.~J.}\ \bibnamefont {Concha}}, \bibinfo {author} {\bibfnamefont {N.}~\bibnamefont {Marzari}}, \bibinfo {author} {\bibfnamefont {M.}~\bibnamefont {Simoncelli}},\ and\ \bibinfo {author} {\bibfnamefont {C.}~\bibnamefont {Galland}},\ }\href {http://arxiv.org/abs/2411.04065} {\bibinfo {title} {Imaging heat transport in suspended diamond nanostructures with integrated spin defect thermometers}} (\bibinfo {year} {2024}),\ \bibinfo {note} {arXiv:2411.04065}\BibitemShut {NoStop}%
\bibitem [{Note5()}]{Note5}%
  \BibitemOpen
  \bibinfo {note} {We recall that the curl of a gradient is zero, so Fourier's heat flux $\protect \bm {Q}^{\delta }={-} \kappa _D\nabla T$ is irrotational.}\BibitemShut {Stop}%
\bibitem [{\citenamefont {Cepellotti}\ and\ \citenamefont {Marzari}(2017{\natexlab{b}})}]{cepellotti_transport_2017}%
  \BibitemOpen
  \bibfield  {author} {\bibinfo {author} {\bibfnamefont {A.}~\bibnamefont {Cepellotti}}\ and\ \bibinfo {author} {\bibfnamefont {N.}~\bibnamefont {Marzari}},\ }\bibfield  {title} {\bibinfo {title} {Transport waves as crystal excitations},\ }\href {https://link.aps.org/doi/10.1103/PhysRevMaterials.1.045406} {\bibfield  {journal} {\bibinfo  {journal} {Phys. Rev. Materials}\ }\textbf {\bibinfo {volume} {1}},\ \bibinfo {pages} {045406} (\bibinfo {year} {2017}{\natexlab{b}})}\BibitemShut {NoStop}%
\bibitem [{\citenamefont {Zhang}\ and\ \citenamefont {Guo}(2021)}]{zhang_transient_2021}%
  \BibitemOpen
  \bibfield  {author} {\bibinfo {author} {\bibfnamefont {C.}~\bibnamefont {Zhang}}\ and\ \bibinfo {author} {\bibfnamefont {Z.}~\bibnamefont {Guo}},\ }\bibfield  {title} {\bibinfo {title} {A transient heat conduction phenomenon to distinguish the hydrodynamic and (quasi) ballistic phonon transport},\ }\href {https://www.sciencedirect.com/science/article/pii/S0017931021009522} {\bibfield  {journal} {\bibinfo  {journal} {Int. J. Heat Mass Transf.}\ }\textbf {\bibinfo {volume} {181}},\ \bibinfo {pages} {121847} (\bibinfo {year} {2021})}\BibitemShut {NoStop}%
\bibitem [{\citenamefont {Xu}\ and\ \citenamefont {Wang}(2002)}]{xu_thermal_2002}%
  \BibitemOpen
  \bibfield  {author} {\bibinfo {author} {\bibfnamefont {M.}~\bibnamefont {Xu}}\ and\ \bibinfo {author} {\bibfnamefont {L.}~\bibnamefont {Wang}},\ }\bibfield  {title} {\bibinfo {title} {Thermal oscillation and resonance in dual-phase-lagging heat conduction},\ }\href {https://www.sciencedirect.com/science/article/pii/S0017931001001995} {\bibfield  {journal} {\bibinfo  {journal} {Int. J. Heat Mass Transf.}\ }\textbf {\bibinfo {volume} {45}},\ \bibinfo {pages} {1055} (\bibinfo {year} {2002})}\BibitemShut {NoStop}%
\bibitem [{\citenamefont {Ordóñez-Miranda}\ and\ \citenamefont {Alvarado-Gil}(2010)}]{ordonez-miranda_exact_2010}%
  \BibitemOpen
  \bibfield  {author} {\bibinfo {author} {\bibfnamefont {J.}~\bibnamefont {Ordóñez-Miranda}}\ and\ \bibinfo {author} {\bibfnamefont {J.~J.}\ \bibnamefont {Alvarado-Gil}},\ }\bibfield  {title} {\bibinfo {title} {Exact solution of the dual-phase-lag heat conduction model for a one-dimensional system excited with a periodic heat source},\ }\href {https://www.sciencedirect.com/science/article/pii/S009364131000039X} {\bibfield  {journal} {\bibinfo  {journal} {Mech. Res. Commun.}\ }\textbf {\bibinfo {volume} {37}},\ \bibinfo {pages} {276} (\bibinfo {year} {2010})}\BibitemShut {NoStop}%
\bibitem [{\citenamefont {Kang}\ \emph {et~al.}(2017)\citenamefont {Kang}, \citenamefont {Zhu}, \citenamefont {Gui},\ and\ \citenamefont {Wang}}]{kang_method_2017}%
  \BibitemOpen
  \bibfield  {author} {\bibinfo {author} {\bibfnamefont {Z.}~\bibnamefont {Kang}}, \bibinfo {author} {\bibfnamefont {P.}~\bibnamefont {Zhu}}, \bibinfo {author} {\bibfnamefont {D.}~\bibnamefont {Gui}},\ and\ \bibinfo {author} {\bibfnamefont {L.}~\bibnamefont {Wang}},\ }\bibfield  {title} {\bibinfo {title} {A method for predicting thermal waves in dual-phase-lag heat conduction},\ }\href {https://www.sciencedirect.com/science/article/pii/S0017931017310980} {\bibfield  {journal} {\bibinfo  {journal} {Int. J. Heat Mass Transf.}\ }\textbf {\bibinfo {volume} {115}},\ \bibinfo {pages} {250} (\bibinfo {year} {2017})}\BibitemShut {NoStop}%
\bibitem [{\citenamefont {Gandolfi}\ \emph {et~al.}(2019)\citenamefont {Gandolfi}, \citenamefont {Benetti}, \citenamefont {Glorieux}, \citenamefont {Giannetti},\ and\ \citenamefont {Banfi}}]{gandolfi_accessing_2019}%
  \BibitemOpen
  \bibfield  {author} {\bibinfo {author} {\bibfnamefont {M.}~\bibnamefont {Gandolfi}}, \bibinfo {author} {\bibfnamefont {G.}~\bibnamefont {Benetti}}, \bibinfo {author} {\bibfnamefont {C.}~\bibnamefont {Glorieux}}, \bibinfo {author} {\bibfnamefont {C.}~\bibnamefont {Giannetti}},\ and\ \bibinfo {author} {\bibfnamefont {F.}~\bibnamefont {Banfi}},\ }\bibfield  {title} {\bibinfo {title} {Accessing temperature waves: {A} dispersion relation perspective},\ }\href {https://www.sciencedirect.com/science/article/pii/S0017931019322999} {\bibfield  {journal} {\bibinfo  {journal} {Int. J. Heat Mass Transf.}\ }\textbf {\bibinfo {volume} {143}},\ \bibinfo {pages} {118553} (\bibinfo {year} {2019})}\BibitemShut {NoStop}%
\bibitem [{\citenamefont {Xu}(2021)}]{xu_thermal_2021}%
  \BibitemOpen
  \bibfield  {author} {\bibinfo {author} {\bibfnamefont {M.}~\bibnamefont {Xu}},\ }\bibfield  {title} {\bibinfo {title} {Thermal oscillations, second sound and thermal resonance in phonon hydrodynamics},\ }\href {https://royalsocietypublishing.org/doi/10.1098/rspa.2020.0913} {\bibfield  {journal} {\bibinfo  {journal} {Proc. Math. Phys. Eng.}\ }\textbf {\bibinfo {volume} {477}},\ \bibinfo {pages} {20200913} (\bibinfo {year} {2021})}\BibitemShut {NoStop}%
\bibitem [{\citenamefont {Mazza}\ \emph {et~al.}(2021)\citenamefont {Mazza}, \citenamefont {Gandolfi}, \citenamefont {Capone}, \citenamefont {Banfi},\ and\ \citenamefont {Giannetti}}]{mazza_thermal_2021}%
  \BibitemOpen
  \bibfield  {author} {\bibinfo {author} {\bibfnamefont {G.}~\bibnamefont {Mazza}}, \bibinfo {author} {\bibfnamefont {M.}~\bibnamefont {Gandolfi}}, \bibinfo {author} {\bibfnamefont {M.}~\bibnamefont {Capone}}, \bibinfo {author} {\bibfnamefont {F.}~\bibnamefont {Banfi}},\ and\ \bibinfo {author} {\bibfnamefont {C.}~\bibnamefont {Giannetti}},\ }\bibfield  {title} {\bibinfo {title} {Thermal dynamics and electronic temperature waves in layered correlated materials},\ }\href {https://www.nature.com/articles/s41467-021-27081-2} {\bibfield  {journal} {\bibinfo  {journal} {Nat. Commun.}\ }\textbf {\bibinfo {volume} {12}},\ \bibinfo {pages} {6904} (\bibinfo {year} {2021})}\BibitemShut {NoStop}%
\bibitem [{Note6()}]{Note6}%
  \BibitemOpen
  \bibinfo {note} {In this time-dependent simulation we choose an equilibrium temperature of 80 K to match the temperature at which transient hydrodynamic heat propagation has been observed in recent experiments \cite {Huberman2019,Jeong2021}.}\BibitemShut {Stop}%
\bibitem [{foo({\natexlab{a}})}]{footnote_param}%
  \BibitemOpen
  \href@noop {} {\emph {\bibinfo {title} {{\normalfont To ensure that the perturbation created causes variations within 10\% of the equilibrium temperature, we used the following parameters: $\mathcal{H}=0.013\tfrac{W}{\mu m^3}$, $t_{\rm heat}=0.4ns$, $x_c=5 \mu m$, $\sigma_x=2\mu m$, $\sigma_y=2.8\mu m$.}}}}\BibitemShut {Stop}%
\bibitem [{\citenamefont {Skinner}(2015)}]{skinner_university_nodate}%
  \BibitemOpen
  \bibfield  {author} {\bibinfo {author} {\bibfnamefont {D.}~\bibnamefont {Skinner}},\ }\href {https://www.damtp.cam.ac.uk/user/dbs26/1Bmethods.html} {\emph {\bibinfo {title} {{Mathematical Methods, University of Cambridge}}}}\ (\bibinfo {year} {2015})\BibitemShut {NoStop}%
\bibitem [{foo({\natexlab{b}})}]{footnote_legend}%
  \BibitemOpen
  \href@noop {} {\emph {\bibinfo {title} {{\normalfont We report the length of the longest side, the shortest side is 0.8 times shorter.}}}}\BibitemShut {Stop}%
\bibitem [{foo({\natexlab{c}})}]{footnote_fig4}%
  \BibitemOpen
  \href@noop {} {}\bibinfo {note} {{\normalfont Refs.~\cite{Huberman2019,Ding2022} quantified the hydrodynamic strength as the dip depth of the TTG signal, see e.g. Fig. 1a in Ref.~\cite{Ding2022}.}}\BibitemShut {Stop}%
\bibitem [{\citenamefont {Barletta}\ and\ \citenamefont {Zanchini}(1996)}]{barletta_hyperbolic_1996}%
  \BibitemOpen
  \bibfield  {author} {\bibinfo {author} {\bibfnamefont {A.}~\bibnamefont {Barletta}}\ and\ \bibinfo {author} {\bibfnamefont {E.}~\bibnamefont {Zanchini}},\ }\bibfield  {title} {\bibinfo {title} {Hyperbolic heat conduction and thermal resonances in a cylindrical solid carrying a steady-periodic electric field},\ }\href {https://www.sciencedirect.com/science/article/pii/0017931095002022} {\bibfield  {journal} {\bibinfo  {journal} {Int. J. Heat Mass Transf.}\ }\textbf {\bibinfo {volume} {39}},\ \bibinfo {pages} {1307} (\bibinfo {year} {1996})}\BibitemShut {NoStop}%
\bibitem [{Note7()}]{Note7}%
  \BibitemOpen
  \bibinfo {note} {All the simulations shown in Fig.~\ref {fig:rect_freq} were performed accounting for grain-boundary scattering as in Ref. \cite {Simoncelli2020}, and considering a grain size of 20 $\mu $m. This value was estimated considering the largest grains in Fig.~S3 of Ref.~\cite {Huberman2019}, Fig.~S1 of Ref. \cite {Jeong2021}, Fig.~S11 of Ref.~\cite {Ding2022}, and Fig. 1e of Ref.~\cite {yuan_modulating_2019}, which are all in broad agreement with 20 $\mu $m.}\BibitemShut {Stop}%
\bibitem [{Note8()}]{Note8}%
  \BibitemOpen
  \bibinfo {note} {Experiments \cite {Ding2022} for isotopically purified graphite are available only at temperatures higher than 100 K, preventing us from comparing VHE and DPLE in the low-temperature limit, where Fig.~\ref {fig:freq_VHE_vs_DPLE}\protect \textbf {a,b} show that viscous effects are largest.}\BibitemShut {Stop}%
\bibitem [{\citenamefont {Raya-Moreno}\ \emph {et~al.}(2022{\natexlab{b}})\citenamefont {Raya-Moreno}, \citenamefont {Cartoixà},\ and\ \citenamefont {Carrete}}]{raya-moreno_bte-barna_2022}%
  \BibitemOpen
  \bibfield  {author} {\bibinfo {author} {\bibfnamefont {M.}~\bibnamefont {Raya-Moreno}}, \bibinfo {author} {\bibfnamefont {X.}~\bibnamefont {Cartoixà}},\ and\ \bibinfo {author} {\bibfnamefont {J.}~\bibnamefont {Carrete}},\ }\bibfield  {title} {\bibinfo {title} {{BTE}-{Barna}: {An} extension of {almaBTE} for thermal simulation of devices based on {2D} materials},\ }\href {https://doi.org/10.1016/j.cpc.2022.108504} {\bibfield  {journal} {\bibinfo  {journal} {Computer Physics Communications}\ }\textbf {\bibinfo {volume} {281}},\ \bibinfo {pages} {108504} (\bibinfo {year} {2022}{\natexlab{b}})}\BibitemShut {NoStop}%
\bibitem [{\citenamefont {Nataf}\ \emph {et~al.}(2024)\citenamefont {Nataf}, \citenamefont {Volz}, \citenamefont {Ordonez-Miranda}, \citenamefont {Íñiguez González}, \citenamefont {Rurali},\ and\ \citenamefont {Dkhil}}]{nataf_using_2024}%
  \BibitemOpen
  \bibfield  {author} {\bibinfo {author} {\bibfnamefont {G.~F.}\ \bibnamefont {Nataf}}, \bibinfo {author} {\bibfnamefont {S.}~\bibnamefont {Volz}}, \bibinfo {author} {\bibfnamefont {J.}~\bibnamefont {Ordonez-Miranda}}, \bibinfo {author} {\bibfnamefont {J.}~\bibnamefont {Íñiguez González}}, \bibinfo {author} {\bibfnamefont {R.}~\bibnamefont {Rurali}},\ and\ \bibinfo {author} {\bibfnamefont {B.}~\bibnamefont {Dkhil}},\ }\bibfield  {title} {\bibinfo {title} {Using oxides to compute with heat},\ }\href {https://doi.org/10.1038/s41578-024-00690-1} {\bibfield  {journal} {\bibinfo  {journal} {Nature Reviews Materials}\ ,\ \bibinfo {pages} {1}} (\bibinfo {year} {2024})},\ \bibinfo {note} {publisher: Nature Publishing Group}\BibitemShut {NoStop}%
\bibitem [{\citenamefont {Torres}\ \emph {et~al.}(2023)\citenamefont {Torres}, \citenamefont {Basaran},\ and\ \citenamefont {Schuller}}]{torres_thermal_2023}%
  \BibitemOpen
  \bibfield  {author} {\bibinfo {author} {\bibfnamefont {F.}~\bibnamefont {Torres}}, \bibinfo {author} {\bibfnamefont {A.~C.}\ \bibnamefont {Basaran}},\ and\ \bibinfo {author} {\bibfnamefont {I.~K.}\ \bibnamefont {Schuller}},\ }\bibfield  {title} {\bibinfo {title} {Thermal {Management} in {Neuromorphic} {Materials}, {Devices}, and {Networks}},\ }\href {https://doi.org/10.1002/adma.202205098} {\bibfield  {journal} {\bibinfo  {journal} {Advanced Materials}\ }\textbf {\bibinfo {volume} {35}},\ \bibinfo {pages} {2205098} (\bibinfo {year} {2023})}\BibitemShut {NoStop}%
\bibitem [{\citenamefont {Palm}\ \emph {et~al.}(2024)\citenamefont {Palm}, \citenamefont {Ding}, \citenamefont {Huxter}, \citenamefont {Taniguchi}, \citenamefont {Watanabe},\ and\ \citenamefont {Degen}}]{palm_observation_2024}%
  \BibitemOpen
  \bibfield  {author} {\bibinfo {author} {\bibfnamefont {M.~L.}\ \bibnamefont {Palm}}, \bibinfo {author} {\bibfnamefont {C.}~\bibnamefont {Ding}}, \bibinfo {author} {\bibfnamefont {W.~S.}\ \bibnamefont {Huxter}}, \bibinfo {author} {\bibfnamefont {T.}~\bibnamefont {Taniguchi}}, \bibinfo {author} {\bibfnamefont {K.}~\bibnamefont {Watanabe}},\ and\ \bibinfo {author} {\bibfnamefont {C.~L.}\ \bibnamefont {Degen}},\ }\bibfield  {title} {\bibinfo {title} {Observation of current whirlpools in graphene at room temperature},\ }\href {https://doi.org/10.1126/science.adj2167} {\bibfield  {journal} {\bibinfo  {journal} {Science}\ }\textbf {\bibinfo {volume} {384}},\ \bibinfo {pages} {465} (\bibinfo {year} {2024})},\ \bibinfo {note} {publisher: American Association for the Advancement of Science}\BibitemShut {NoStop}%
\bibitem [{\citenamefont {Coulter}\ \emph {et~al.}(2018)\citenamefont {Coulter}, \citenamefont {Sundararaman},\ and\ \citenamefont {Narang}}]{coulter_microscopic_2018}%
  \BibitemOpen
  \bibfield  {author} {\bibinfo {author} {\bibfnamefont {J.}~\bibnamefont {Coulter}}, \bibinfo {author} {\bibfnamefont {R.}~\bibnamefont {Sundararaman}},\ and\ \bibinfo {author} {\bibfnamefont {P.}~\bibnamefont {Narang}},\ }\bibfield  {title} {\bibinfo {title} {{Microscopic origins of hydrodynamic transport in the type-{II} {Weyl} semimetal WP$_2$}},\ }\href {https://link.aps.org/doi/10.1103/PhysRevB.98.115130} {\bibfield  {journal} {\bibinfo  {journal} {Phys. Rev. B}\ }\textbf {\bibinfo {volume} {98}},\ \bibinfo {pages} {115130} (\bibinfo {year} {2018})}\BibitemShut {NoStop}%
\bibitem [{\citenamefont {Vool}\ \emph {et~al.}(2021)\citenamefont {Vool}, \citenamefont {Hamo}, \citenamefont {Varnavides}, \citenamefont {Wang}, \citenamefont {Zhou}, \citenamefont {Kumar}, \citenamefont {Dovzhenko}, \citenamefont {Qiu}, \citenamefont {Garcia}, \citenamefont {Pierce}, \citenamefont {Gooth}, \citenamefont {Anikeeva}, \citenamefont {Felser}, \citenamefont {Narang},\ and\ \citenamefont {Yacoby}}]{vool_imaging_2021}%
  \BibitemOpen
  \bibfield  {author} {\bibinfo {author} {\bibfnamefont {U.}~\bibnamefont {Vool}}, \bibinfo {author} {\bibfnamefont {A.}~\bibnamefont {Hamo}}, \bibinfo {author} {\bibfnamefont {G.}~\bibnamefont {Varnavides}}, \bibinfo {author} {\bibfnamefont {Y.}~\bibnamefont {Wang}}, \bibinfo {author} {\bibfnamefont {T.~X.}\ \bibnamefont {Zhou}}, \bibinfo {author} {\bibfnamefont {N.}~\bibnamefont {Kumar}}, \bibinfo {author} {\bibfnamefont {Y.}~\bibnamefont {Dovzhenko}}, \bibinfo {author} {\bibfnamefont {Z.}~\bibnamefont {Qiu}}, \bibinfo {author} {\bibfnamefont {C.~A.~C.}\ \bibnamefont {Garcia}}, \bibinfo {author} {\bibfnamefont {A.~T.}\ \bibnamefont {Pierce}}, \bibinfo {author} {\bibfnamefont {J.}~\bibnamefont {Gooth}}, \bibinfo {author} {\bibfnamefont {P.}~\bibnamefont {Anikeeva}}, \bibinfo {author} {\bibfnamefont {C.}~\bibnamefont {Felser}}, \bibinfo {author} {\bibfnamefont {P.}~\bibnamefont {Narang}},\ and\ \bibinfo {author} {\bibfnamefont {A.}~\bibnamefont {Yacoby}},\ }\bibfield  {title} {\bibinfo {title} {Imaging phonon-mediated hydrodynamic flow in {WTe2}},\ }\href {https://www.nature.com/articles/s41567-021-01341-w} {\bibfield  {journal} {\bibinfo  {journal} {Nat. Phys}\ }\textbf {\bibinfo {volume} {17}},\ \bibinfo {pages} {1216} (\bibinfo {year} {2021})}\BibitemShut {NoStop}%
\bibitem [{\citenamefont {Yang}\ \emph {et~al.}(2021)\citenamefont {Yang}, \citenamefont {Yao}, \citenamefont {Plisson}, \citenamefont {Mozaffari}, \citenamefont {Scheifers}, \citenamefont {Savvidou}, \citenamefont {Choi}, \citenamefont {McCandless}, \citenamefont {Padlewski}, \citenamefont {Putzke}, \citenamefont {Moll}, \citenamefont {Chan}, \citenamefont {Balicas}, \citenamefont {Burch},\ and\ \citenamefont {Tafti}}]{yang_evidence_2021}%
  \BibitemOpen
  \bibfield  {author} {\bibinfo {author} {\bibfnamefont {H.-Y.}\ \bibnamefont {Yang}}, \bibinfo {author} {\bibfnamefont {X.}~\bibnamefont {Yao}}, \bibinfo {author} {\bibfnamefont {V.}~\bibnamefont {Plisson}}, \bibinfo {author} {\bibfnamefont {S.}~\bibnamefont {Mozaffari}}, \bibinfo {author} {\bibfnamefont {J.~P.}\ \bibnamefont {Scheifers}}, \bibinfo {author} {\bibfnamefont {A.~F.}\ \bibnamefont {Savvidou}}, \bibinfo {author} {\bibfnamefont {E.~S.}\ \bibnamefont {Choi}}, \bibinfo {author} {\bibfnamefont {G.~T.}\ \bibnamefont {McCandless}}, \bibinfo {author} {\bibfnamefont {M.~F.}\ \bibnamefont {Padlewski}}, \bibinfo {author} {\bibfnamefont {C.}~\bibnamefont {Putzke}}, \bibinfo {author} {\bibfnamefont {P.~J.~W.}\ \bibnamefont {Moll}}, \bibinfo {author} {\bibfnamefont {J.~Y.}\ \bibnamefont {Chan}}, \bibinfo {author} {\bibfnamefont {L.}~\bibnamefont {Balicas}}, \bibinfo {author} {\bibfnamefont {K.~S.}\ \bibnamefont {Burch}},\ and\ \bibinfo {author} {\bibfnamefont {F.}~\bibnamefont {Tafti}},\ }\bibfield  {title} {\bibinfo {title} {Evidence of a coupled electron-phonon liquid in {NbGe2}},\ }\href {https://www.nature.com/articles/s41467-021-25547-x} {\bibfield  {journal} {\bibinfo  {journal} {Nat. Commun.}\ }\textbf {\bibinfo {volume} {12}},\ \bibinfo {pages} {5292} (\bibinfo {year} {2021})}\BibitemShut {NoStop}%
\bibitem [{\citenamefont {Jaoui}\ \emph {et~al.}(2022)\citenamefont {Jaoui}, \citenamefont {Gourgout}, \citenamefont {Seyfarth}, \citenamefont {Subedi}, \citenamefont {Lorenz}, \citenamefont {Fauqué},\ and\ \citenamefont {Behnia}}]{jaoui_formation_2022}%
  \BibitemOpen
  \bibfield  {author} {\bibinfo {author} {\bibfnamefont {A.}~\bibnamefont {Jaoui}}, \bibinfo {author} {\bibfnamefont {A.}~\bibnamefont {Gourgout}}, \bibinfo {author} {\bibfnamefont {G.}~\bibnamefont {Seyfarth}}, \bibinfo {author} {\bibfnamefont {A.}~\bibnamefont {Subedi}}, \bibinfo {author} {\bibfnamefont {T.}~\bibnamefont {Lorenz}}, \bibinfo {author} {\bibfnamefont {B.}~\bibnamefont {Fauqué}},\ and\ \bibinfo {author} {\bibfnamefont {K.}~\bibnamefont {Behnia}},\ }\bibfield  {title} {\bibinfo {title} {Formation of an {Electron}-{Phonon} {Bifluid} in {Bulk} {Antimony}},\ }\href {https://link.aps.org/doi/10.1103/PhysRevX.12.031023} {\bibfield  {journal} {\bibinfo  {journal} {Phys. Rev. X}\ }\textbf {\bibinfo {volume} {12}},\ \bibinfo {pages} {031023} (\bibinfo {year} {2022})}\BibitemShut {NoStop}%
\bibitem [{\citenamefont {Huang}\ and\ \citenamefont {Lucas}(2021)}]{huang_electron-phonon_2021}%
  \BibitemOpen
  \bibfield  {author} {\bibinfo {author} {\bibfnamefont {X.}~\bibnamefont {Huang}}\ and\ \bibinfo {author} {\bibfnamefont {A.}~\bibnamefont {Lucas}},\ }\bibfield  {title} {\bibinfo {title} {Electron-phonon hydrodynamics},\ }\href {https://link.aps.org/doi/10.1103/PhysRevB.103.155128} {\bibfield  {journal} {\bibinfo  {journal} {Phys. Rev. B}\ }\textbf {\bibinfo {volume} {103}},\ \bibinfo {pages} {155128} (\bibinfo {year} {2021})}\BibitemShut {NoStop}%
\bibitem [{\citenamefont {Protik}\ \emph {et~al.}(2022)\citenamefont {Protik}, \citenamefont {Li}, \citenamefont {Pruneda}, \citenamefont {Broido},\ and\ \citenamefont {Ordejón}}]{protik_elphbolt_2022}%
  \BibitemOpen
  \bibfield  {author} {\bibinfo {author} {\bibfnamefont {N.~H.}\ \bibnamefont {Protik}}, \bibinfo {author} {\bibfnamefont {C.}~\bibnamefont {Li}}, \bibinfo {author} {\bibfnamefont {M.}~\bibnamefont {Pruneda}}, \bibinfo {author} {\bibfnamefont {D.}~\bibnamefont {Broido}},\ and\ \bibinfo {author} {\bibfnamefont {P.}~\bibnamefont {Ordejón}},\ }\bibfield  {title} {\bibinfo {title} {The elphbolt ab initio solver for the coupled electron-phonon {Boltzmann} transport equations},\ }\href {https://www.nature.com/articles/s41524-022-00710-0} {\bibfield  {journal} {\bibinfo  {journal} {NPJ Comput. Mater}\ }\textbf {\bibinfo {volume} {8}},\ \bibinfo {pages} {28} (\bibinfo {year} {2022})}\BibitemShut {NoStop}%
\bibitem [{\citenamefont {Levchenko}\ and\ \citenamefont {Schmalian}(2020)}]{levchenko_transport_2020}%
  \BibitemOpen
  \bibfield  {author} {\bibinfo {author} {\bibfnamefont {A.}~\bibnamefont {Levchenko}}\ and\ \bibinfo {author} {\bibfnamefont {J.}~\bibnamefont {Schmalian}},\ }\bibfield  {title} {\bibinfo {title} {Transport properties of strongly coupled electron–phonon liquids},\ }\href {https://linkinghub.elsevier.com/retrieve/pii/S0003491620301524} {\bibfield  {journal} {\bibinfo  {journal} {Ann. Phys.}\ }\textbf {\bibinfo {volume} {419}},\ \bibinfo {pages} {168218} (\bibinfo {year} {2020})}\BibitemShut {NoStop}%
\bibitem [{\citenamefont {Coulter}\ \emph {et~al.}(2025)\citenamefont {Coulter}, \citenamefont {Rajkov},\ and\ \citenamefont {Simoncelli}}]{coulterCoupledElectronphononHydrodynamics2025a}%
  \BibitemOpen
  \bibfield  {author} {\bibinfo {author} {\bibfnamefont {J.}~\bibnamefont {Coulter}}, \bibinfo {author} {\bibfnamefont {B.}~\bibnamefont {Rajkov}},\ and\ \bibinfo {author} {\bibfnamefont {M.}~\bibnamefont {Simoncelli}},\ }\href {https://doi.org/10.48550/arXiv.2503.07560} {\bibinfo {title} {Coupled electron-phonon hydrodynamics and viscous thermoelectric equations}} (\bibinfo {year} {2025}),\ \bibinfo {note} {arXiv:2503.07560 [cond-mat]}\BibitemShut {NoStop}%
\bibitem [{\citenamefont {Wei}\ \emph {et~al.}(2022)\citenamefont {Wei}, \citenamefont {Santos}, \citenamefont {Lusero}, \citenamefont {Bauer}, \citenamefont {Ben~Youssef},\ and\ \citenamefont {van Wees}}]{wei_giant_2022}%
  \BibitemOpen
  \bibfield  {author} {\bibinfo {author} {\bibfnamefont {X.-Y.}\ \bibnamefont {Wei}}, \bibinfo {author} {\bibfnamefont {O.~A.}\ \bibnamefont {Santos}}, \bibinfo {author} {\bibfnamefont {C.~H.~S.}\ \bibnamefont {Lusero}}, \bibinfo {author} {\bibfnamefont {G.~E.~W.}\ \bibnamefont {Bauer}}, \bibinfo {author} {\bibfnamefont {J.}~\bibnamefont {Ben~Youssef}},\ and\ \bibinfo {author} {\bibfnamefont {B.~J.}\ \bibnamefont {van Wees}},\ }\bibfield  {title} {\bibinfo {title} {Giant magnon spin conductivity in ultrathin yttrium iron garnet films},\ }\href {https://www.nature.com/articles/s41563-022-01369-0} {\bibfield  {journal} {\bibinfo  {journal} {Nat. Mater.}\ ,\ \bibinfo {pages} {1352}} (\bibinfo {year} {2022})}\BibitemShut {NoStop}%
\bibitem [{\citenamefont {Rosales}\ \emph {et~al.}(2023)\citenamefont {Rosales}, \citenamefont {Albarrac\'{\i}n}, \citenamefont {Pujol},\ and\ \citenamefont {Jaubert}}]{PhysRevLett.130.106703}%
  \BibitemOpen
  \bibfield  {author} {\bibinfo {author} {\bibfnamefont {H.~D.}\ \bibnamefont {Rosales}}, \bibinfo {author} {\bibfnamefont {F.~A.~G.}\ \bibnamefont {Albarrac\'{\i}n}}, \bibinfo {author} {\bibfnamefont {P.}~\bibnamefont {Pujol}},\ and\ \bibinfo {author} {\bibfnamefont {L.~D.~C.}\ \bibnamefont {Jaubert}},\ }\bibfield  {title} {\bibinfo {title} {Skyrmion fluid and bimeron glass protected by a chiral spin liquid on a kagome lattice},\ }\href {https://link.aps.org/doi/10.1103/PhysRevLett.130.106703} {\bibfield  {journal} {\bibinfo  {journal} {Phys. Rev. Lett.}\ }\textbf {\bibinfo {volume} {130}},\ \bibinfo {pages} {106703} (\bibinfo {year} {2023})}\BibitemShut {NoStop}%
\bibitem [{\citenamefont {Raya-Moreno}(2022)}]{raya-moreno_heat_nodate}%
  \BibitemOpen
  \bibfield  {author} {\bibinfo {author} {\bibfnamefont {M.}~\bibnamefont {Raya-Moreno}},\ }\emph {\bibinfo {title} {Heat transport in binary semiconductor polytypes and devices based on {2D} materials: an ab initio study}},\ \href {https://www.tdx.cat/handle/10803/675596#page=1} {Ph.D. thesis},\ \bibinfo  {school} {Universitat Autònoma de Barcelona} (\bibinfo {year} {2022})\BibitemShut {NoStop}%
\bibitem [{\citenamefont {Sabatti}\ \emph {et~al.}(2016)\citenamefont {Sabatti}, \citenamefont {Goodnick},\ and\ \citenamefont {Saraniti}}]{sabattiSimulationPhononTransport2016a}%
  \BibitemOpen
  \bibfield  {author} {\bibinfo {author} {\bibfnamefont {F.~F.~M.}\ \bibnamefont {Sabatti}}, \bibinfo {author} {\bibfnamefont {S.~M.}\ \bibnamefont {Goodnick}},\ and\ \bibinfo {author} {\bibfnamefont {M.}~\bibnamefont {Saraniti}},\ }\bibfield  {title} {\bibinfo {title} {Simulation of {Phonon} {Transport} in {Semiconductors} {Using} a {Population}-{Dependent} {Many}-{Body} {Cellular} {Monte} {Carlo} {Approach}},\ }\bibfield  {journal} {\bibinfo  {journal} {Journal of Heat Transfer}\ }\textbf {\bibinfo {volume} {139}},\ \href {https://doi.org/10.1115/1.4035042} {10.1115/1.4035042} (\bibinfo {year} {2016})\BibitemShut {NoStop}%
\bibitem [{Note9()}]{Note9}%
  \BibitemOpen
  \bibinfo {note} {Implemented in \protect \texttt {BTE-Barna} following Sec.~4.2.2.1 of Ref.~\cite {raya-moreno_heat_nodate}}\BibitemShut {NoStop}%
\bibitem [{\citenamefont {Laurent}\ \emph {et~al.}(2011)\citenamefont {Laurent}, \citenamefont {Drezet}, \citenamefont {Sellier}, \citenamefont {Chevrier},\ and\ \citenamefont {Huant}}]{laurentLargeVariationBoundaryCondition2011}%
  \BibitemOpen
  \bibfield  {author} {\bibinfo {author} {\bibfnamefont {J.}~\bibnamefont {Laurent}}, \bibinfo {author} {\bibfnamefont {A.}~\bibnamefont {Drezet}}, \bibinfo {author} {\bibfnamefont {H.}~\bibnamefont {Sellier}}, \bibinfo {author} {\bibfnamefont {J.}~\bibnamefont {Chevrier}},\ and\ \bibinfo {author} {\bibfnamefont {S.}~\bibnamefont {Huant}},\ }\bibfield  {title} {\bibinfo {title} {Large {Variation} in the {Boundary}-{Condition} {Slippage} for a {Rarefied} {Gas} {Flowing} between {Two} {Surfaces}},\ }\href {https://doi.org/10.1103/PhysRevLett.107.164501} {\bibfield  {journal} {\bibinfo  {journal} {Physical Review Letters}\ }\textbf {\bibinfo {volume} {107}},\ \bibinfo {pages} {164501} (\bibinfo {year} {2011})},\ \bibinfo {note} {publisher: American Physical Society}\BibitemShut {NoStop}%
\bibitem [{\citenamefont {Gurcan}\ \emph {et~al.}(2013)\citenamefont {Gurcan}, \citenamefont {Diamond}, \citenamefont {Garbet}, \citenamefont {Berionni}, \citenamefont {Dif-Pradalier}, \citenamefont {Hennequin}, \citenamefont {Morel}, \citenamefont {Kosuga},\ and\ \citenamefont {Vermare}}]{gurcan_transport_2013}%
  \BibitemOpen
  \bibfield  {author} {\bibinfo {author} {\bibfnamefont {O.~D.}\ \bibnamefont {Gurcan}}, \bibinfo {author} {\bibfnamefont {P.~H.}\ \bibnamefont {Diamond}}, \bibinfo {author} {\bibfnamefont {X.}~\bibnamefont {Garbet}}, \bibinfo {author} {\bibfnamefont {V.}~\bibnamefont {Berionni}}, \bibinfo {author} {\bibfnamefont {G.}~\bibnamefont {Dif-Pradalier}}, \bibinfo {author} {\bibfnamefont {P.}~\bibnamefont {Hennequin}}, \bibinfo {author} {\bibfnamefont {P.}~\bibnamefont {Morel}}, \bibinfo {author} {\bibfnamefont {Y.}~\bibnamefont {Kosuga}},\ and\ \bibinfo {author} {\bibfnamefont {L.}~\bibnamefont {Vermare}},\ }\bibfield  {title} {\bibinfo {title} {Transport of radial heat flux and second sound in fusion plasmas},\ }\href {https://doi.org/10.1063/1.4792161} {\bibfield  {journal} {\bibinfo  {journal} {Physics of Plasmas}\ }\textbf {\bibinfo {volume} {20}},\ \bibinfo {pages} {022307} (\bibinfo {year} {2013})}\BibitemShut {NoStop}%
\bibitem [{\citenamefont {Allen}(2018)}]{allen_analysis_2018}%
  \BibitemOpen
  \bibfield  {author} {\bibinfo {author} {\bibfnamefont {P.~B.}\ \bibnamefont {Allen}},\ }\bibfield  {title} {\bibinfo {title} {Analysis of nonlocal phonon thermal conductivity simulations showing the ballistic to diffusive crossover},\ }\href {https://doi.org/10.1103/PhysRevB.97.134307} {\bibfield  {journal} {\bibinfo  {journal} {Physical Review B}\ }\textbf {\bibinfo {volume} {97}},\ \bibinfo {pages} {134307} (\bibinfo {year} {2018})}\BibitemShut {NoStop}%
\bibitem [{\citenamefont {Shang}\ \emph {et~al.}(2022)\citenamefont {Shang}, \citenamefont {Mao}, \citenamefont {Yang}, \citenamefont {Li},\ and\ \citenamefont {Lü}}]{shang_unified_2022}%
  \BibitemOpen
  \bibfield  {author} {\bibinfo {author} {\bibfnamefont {M.-Y.}\ \bibnamefont {Shang}}, \bibinfo {author} {\bibfnamefont {W.-H.}\ \bibnamefont {Mao}}, \bibinfo {author} {\bibfnamefont {N.}~\bibnamefont {Yang}}, \bibinfo {author} {\bibfnamefont {B.}~\bibnamefont {Li}},\ and\ \bibinfo {author} {\bibfnamefont {J.-T.}\ \bibnamefont {Lü}},\ }\bibfield  {title} {\bibinfo {title} {Unified theory of second sound in two-dimensional materials},\ }\href {https://doi.org/10.1103/PhysRevB.105.165423} {\bibfield  {journal} {\bibinfo  {journal} {Physical Review B}\ }\textbf {\bibinfo {volume} {105}},\ \bibinfo {pages} {165423} (\bibinfo {year} {2022})}\BibitemShut {NoStop}%
\bibitem [{\citenamefont {Yadav}\ \emph {et~al.}(2025)\citenamefont {Yadav}, \citenamefont {Goutham}, \citenamefont {Meshram}, \citenamefont {Pathak}, \citenamefont {Bansal},\ and\ \citenamefont {Agrawal}}]{yadavDerivationGeneralizedHeat2025a}%
  \BibitemOpen
  \bibfield  {author} {\bibinfo {author} {\bibfnamefont {U.}~\bibnamefont {Yadav}}, \bibinfo {author} {\bibfnamefont {C.~N.~S.}\ \bibnamefont {Goutham}}, \bibinfo {author} {\bibfnamefont {K.}~\bibnamefont {Meshram}}, \bibinfo {author} {\bibfnamefont {A.}~\bibnamefont {Pathak}}, \bibinfo {author} {\bibfnamefont {D.}~\bibnamefont {Bansal}},\ and\ \bibinfo {author} {\bibfnamefont {A.}~\bibnamefont {Agrawal}},\ }\bibfield  {title} {\bibinfo {title} {Derivation of the generalized heat transport equation and comparison with existing models},\ }\href {https://doi.org/10.1103/xvbk-75bf} {\bibfield  {journal} {\bibinfo  {journal} {Physical Review Applied}\ }\textbf {\bibinfo {volume} {24}},\ \bibinfo {pages} {034024} (\bibinfo {year} {2025})}\BibitemShut {NoStop}%
\bibitem [{\citenamefont {Spohn}(2006)}]{spohn_phonon_2006}%
  \BibitemOpen
  \bibfield  {author} {\bibinfo {author} {\bibfnamefont {H.}~\bibnamefont {Spohn}},\ }\bibfield  {title} {\bibinfo {title} {The {Phonon} {Boltzmann} {Equation}, {Properties} and {Link} to {Weakly} {Anharmonic} {Lattice} {Dynamics}},\ }\href {https://doi.org/10.1007/s10955-005-8088-5} {\bibfield  {journal} {\bibinfo  {journal} {Journal of Statistical Physics}\ }\textbf {\bibinfo {volume} {124}},\ \bibinfo {pages} {1041} (\bibinfo {year} {2006})}\BibitemShut {NoStop}%
\bibitem [{Note10()}]{Note10}%
  \BibitemOpen
  \bibinfo {note} {One can also directly see that $\protect \big <\theta ^{0}_{\nu }\protect \big | v^i_\nu \protect \big |\theta ^{0}_{\nu }\protect \big >\propto \DOTSB \sum@ \slimits@ _s\DOTSI \intop \ilimits@ _{\protect \mathfrak {B}}\omega _{\protect \bm {q}s}^2 v^i_{\protect \bm {q}s} d^3q =0$, because obtained by integrating an odd-parity integrand over the symmetric Brillouin zone $\protect \mathfrak {B}$.}\BibitemShut {Stop}%
\bibitem [{Note11()}]{Note11}%
  \BibitemOpen
  \bibinfo {note} {The relaxon \cite {cepellotti_thermal_2016} and variational \cite {fugallo_ab_2013} methods always provide results that are numerically compatible within 0.1\%. For example, in natural graphite at 80 K the bulk thermal conductivity obtained from the variational method \cite {fugallo_ab_2013} at 80 K is 5735.08 W/mK while with the relaxon formalism it is 5735.03 W/mK; in isotopically purified graphite (99.9\% $^{12}$C, 0.1 \%$^{13}$C) the bulk thermal conductivity at 80 K obtained from the variational method is 27484.96 W/mK while with the relaxon formalism it is 27485.09 W/mK; for h$^{11}$BN the variational method at 60 K yields 3944.20 W/mK, while the result of the relaxon method is 3944.23 W/mK.}\BibitemShut {Stop}%
\bibitem [{\citenamefont {Allen}\ and\ \citenamefont {Perebeinos}(2018)}]{allen_temperature_2018}%
  \BibitemOpen
  \bibfield  {author} {\bibinfo {author} {\bibfnamefont {P.~B.}\ \bibnamefont {Allen}}\ and\ \bibinfo {author} {\bibfnamefont {V.}~\bibnamefont {Perebeinos}},\ }\bibfield  {title} {\bibinfo {title} {Temperature in a {Peierls}-{Boltzmann} treatment of nonlocal phonon heat transport},\ }\href {https://doi.org/10.1103/PhysRevB.98.085427} {\bibfield  {journal} {\bibinfo  {journal} {Physical Review B}\ }\textbf {\bibinfo {volume} {98}},\ \bibinfo {pages} {085427} (\bibinfo {year} {2018})}\BibitemShut {NoStop}%
\bibitem [{\citenamefont {Ziman}(1960)}]{ziman1960electrons}%
  \BibitemOpen
  \bibfield  {author} {\bibinfo {author} {\bibfnamefont {J.~M.}\ \bibnamefont {Ziman}},\ }\href@noop {} {\emph {\bibinfo {title} {Electrons and phonons: the theory of transport phenomena in solids}}}\ (\bibinfo  {publisher} {Oxford university press},\ \bibinfo {year} {1960})\BibitemShut {NoStop}%
\bibitem [{\citenamefont {Winkler}(2003)}]{winklerSpinOrbitCouplingEffects2003}%
  \BibitemOpen
  \bibfield  {author} {\bibinfo {author} {\bibfnamefont {R.}~\bibnamefont {Winkler}},\ }\href@noop {} {\emph {\bibinfo {title} {Spin-{Orbit} {Coupling} {Effects} in {Two}-{Dimensional} {Electron} and {Hole} {Systems}}}}\ (\bibinfo  {publisher} {Springer Berlin / Heidelberg},\ \bibinfo {year} {2003})\BibitemShut {NoStop}%
\bibitem [{\citenamefont {Bravyi}\ \emph {et~al.}(2011)\citenamefont {Bravyi}, \citenamefont {DiVincenzo},\ and\ \citenamefont {Loss}}]{bravyiSchriefferWolffTransformation2011}%
  \BibitemOpen
  \bibfield  {author} {\bibinfo {author} {\bibfnamefont {S.}~\bibnamefont {Bravyi}}, \bibinfo {author} {\bibfnamefont {D.~P.}\ \bibnamefont {DiVincenzo}},\ and\ \bibinfo {author} {\bibfnamefont {D.}~\bibnamefont {Loss}},\ }\bibfield  {title} {\bibinfo {title} {Schrieffer–{Wolff} transformation for quantum many-body systems},\ }\href {https://doi.org/10.1016/j.aop.2011.06.004} {\bibfield  {journal} {\bibinfo  {journal} {Annals of Physics}\ }\textbf {\bibinfo {volume} {326}},\ \bibinfo {pages} {2793} (\bibinfo {year} {2011})}\BibitemShut {NoStop}%
\bibitem [{\citenamefont {Horn}\ and\ \citenamefont {Johnson}(1991)}]{hornTopicsMatrixAnalysis1991}%
  \BibitemOpen
  \bibfield  {author} {\bibinfo {author} {\bibfnamefont {R.~A.}\ \bibnamefont {Horn}}\ and\ \bibinfo {author} {\bibfnamefont {C.~R.}\ \bibnamefont {Johnson}},\ }\href {https://doi.org/10.1017/CBO9780511840371} {\emph {\bibinfo {title} {Topics in {Matrix} {Analysis}}}}\ (\bibinfo  {publisher} {Cambridge University Press},\ \bibinfo {year} {1991})\BibitemShut {NoStop}%
\bibitem [{Note12()}]{Note12}%
  \BibitemOpen
  \bibinfo {note} {Specifically, Eq.~(0.7.3.1) of Ref.~\cite {hornTopicsMatrixAnalysis1991} shows that for a partitioned nonsingular matrix A, \begin {equation}A=\left [\begin {array}{ll} A_{11} & A_{12} \\ A_{21} & A_{22} \end {array}\right ] \end {equation} its inverse can be written in partitioned form as: {\relax \protect \fontsize {5}{6}\protect \selectfont \begin {equation} A^{-1}{=}\left [\begin {array}{lc} \left (A_{11}-A_{12} A_{22}^{-1} A_{21}\right )^{-1} & A_{11}^{-1} A_{12}\left (A_{21} A_{11}^{-1} A_{12}-A_{22}\right )^{-1} \\ A_{22}^{-1} A_{21}\left (A_{12} A_{22}^{-1} A_{21}-A_{11}\right )^{-1} & \left (A_{22}-A_{21} A_{11}^{-1} A_{12}\right )^{-1} \end {array}\right ] \end {equation}}}\BibitemShut {NoStop}%
\bibitem [{\citenamefont {Gurzhi}(1968)}]{gurzhiHYDRODYNAMICEFFECTSSOLIDS1968}%
  \BibitemOpen
  \bibfield  {author} {\bibinfo {author} {\bibfnamefont {R.~N.}\ \bibnamefont {Gurzhi}},\ }\bibfield  {title} {\bibinfo {title} {{HYDRODYNAMIC} {EFFECTS} {IN} {SOLIDS} {AT} {LOW} {TEMPERATURE}},\ }\href@noop {} {\bibfield  {journal} {\bibinfo  {journal} {Soviet Physics Uspekhi}\ }\textbf {\bibinfo {volume} {11}},\ \bibinfo {pages} {255} (\bibinfo {year} {1968})}\BibitemShut {NoStop}%
\bibitem [{\citenamefont {Bandurin}\ \emph {et~al.}(2016)\citenamefont {Bandurin}, \citenamefont {Torre}, \citenamefont {Kumar}, \citenamefont {Ben~Shalom}, \citenamefont {Tomadin}, \citenamefont {Principi}, \citenamefont {Auton}, \citenamefont {Khestanova}, \citenamefont {Novoselov}, \citenamefont {Grigorieva}, \citenamefont {Ponomarenko}, \citenamefont {Geim},\ and\ \citenamefont {Polini}}]{bandurin_negative_2016}%
  \BibitemOpen
  \bibfield  {author} {\bibinfo {author} {\bibfnamefont {D.~A.}\ \bibnamefont {Bandurin}}, \bibinfo {author} {\bibfnamefont {I.}~\bibnamefont {Torre}}, \bibinfo {author} {\bibfnamefont {R.~K.}\ \bibnamefont {Kumar}}, \bibinfo {author} {\bibfnamefont {M.}~\bibnamefont {Ben~Shalom}}, \bibinfo {author} {\bibfnamefont {A.}~\bibnamefont {Tomadin}}, \bibinfo {author} {\bibfnamefont {A.}~\bibnamefont {Principi}}, \bibinfo {author} {\bibfnamefont {G.~H.}\ \bibnamefont {Auton}}, \bibinfo {author} {\bibfnamefont {E.}~\bibnamefont {Khestanova}}, \bibinfo {author} {\bibfnamefont {K.~S.}\ \bibnamefont {Novoselov}}, \bibinfo {author} {\bibfnamefont {I.~V.}\ \bibnamefont {Grigorieva}}, \bibinfo {author} {\bibfnamefont {L.~A.}\ \bibnamefont {Ponomarenko}}, \bibinfo {author} {\bibfnamefont {A.~K.}\ \bibnamefont {Geim}},\ and\ \bibinfo {author} {\bibfnamefont {M.}~\bibnamefont {Polini}},\ }\bibfield  {title} {\bibinfo {title} {Negative local resistance caused by viscous electron backflow in graphene},\ }\href {https://www.science.org/doi/10.1126/science.aad0201} {\bibfield  {journal} {\bibinfo  {journal} {Science}\ }\textbf {\bibinfo {volume} {351}},\ \bibinfo {pages} {1055} (\bibinfo {year} {2016})}\BibitemShut {NoStop}%
\bibitem [{\citenamefont {Kovács}\ \emph {et~al.}(2022)\citenamefont {Kovács}, \citenamefont {Fehér},\ and\ \citenamefont {Sobolev}}]{kovacs_two-temperature_2022}%
  \BibitemOpen
  \bibfield  {author} {\bibinfo {author} {\bibfnamefont {R.}~\bibnamefont {Kovács}}, \bibinfo {author} {\bibfnamefont {A.}~\bibnamefont {Fehér}},\ and\ \bibinfo {author} {\bibfnamefont {S.}~\bibnamefont {Sobolev}},\ }\bibfield  {title} {\bibinfo {title} {On the two-temperature description of heterogeneous materials},\ }\href {https://doi.org/10.1016/j.ijheatmasstransfer.2022.123021} {\bibfield  {journal} {\bibinfo  {journal} {International Journal of Heat and Mass Transfer}\ }\textbf {\bibinfo {volume} {194}},\ \bibinfo {pages} {123021} (\bibinfo {year} {2022})}\BibitemShut {NoStop}%
\bibitem [{\citenamefont {Sambasivam}\ \emph {et~al.}(2014)\citenamefont {Sambasivam}, \citenamefont {Chakraborty},\ and\ \citenamefont {Durst}}]{sambasivam_numerical_2014}%
  \BibitemOpen
  \bibfield  {author} {\bibinfo {author} {\bibfnamefont {R.}~\bibnamefont {Sambasivam}}, \bibinfo {author} {\bibfnamefont {S.}~\bibnamefont {Chakraborty}},\ and\ \bibinfo {author} {\bibfnamefont {F.}~\bibnamefont {Durst}},\ }\bibfield  {title} {\bibinfo {title} {Numerical predictions of backward-facing step flows in microchannels using extended {Navier}–{Stokes} equations},\ }\href {http://link.springer.com/10.1007/s10404-013-1254-1} {\bibfield  {journal} {\bibinfo  {journal} {Microfluid. Nanofluid.}\ }\textbf {\bibinfo {volume} {16}},\ \bibinfo {pages} {757} (\bibinfo {year} {2014})}\BibitemShut {NoStop}%
\bibitem [{Note13()}]{Note13}%
  \BibitemOpen
  \bibinfo {note} {The volume viscosity is also called the bulk viscosity, here we use the term 'bulk viscosity' for the value of the viscosity tensor in the bulk of the crystal, without accounting for finite size effects.}\BibitemShut {Stop}%
\bibitem [{\citenamefont {Landau}\ and\ \citenamefont {Lifshitz}(1987)}]{landau_fluid}%
  \BibitemOpen
  \bibfield  {author} {\bibinfo {author} {\bibfnamefont {L.}~\bibnamefont {Landau}}\ and\ \bibinfo {author} {\bibfnamefont {E.}~\bibnamefont {Lifshitz}},\ }\href {https://doi.org/10.1016/C2013-0-03799-1} {\emph {\bibinfo {title} {Fluid Mechanics}}}\ (\bibinfo  {publisher} {Pergamon Press},\ \bibinfo {year} {1987})\BibitemShut {NoStop}%
\bibitem [{\citenamefont {De~Groot}\ and\ \citenamefont {Mazur}(2013)}]{groot_thermodynamics}%
  \BibitemOpen
  \bibfield  {author} {\bibinfo {author} {\bibfnamefont {S.}~\bibnamefont {De~Groot}}\ and\ \bibinfo {author} {\bibfnamefont {P.}~\bibnamefont {Mazur}},\ }\href {https://books.google.co.uk/books?id=mfFyG9jfaMYC} {\emph {\bibinfo {title} {Non-Equilibrium Thermodynamics}}},\ Dover Books on Physics\ (\bibinfo  {publisher} {Dover Publications},\ \bibinfo {year} {2013})\BibitemShut {NoStop}%
\bibitem [{\citenamefont {Cook}\ and\ \citenamefont {Lucas}(2019)}]{el_rot_viscosity}%
  \BibitemOpen
  \bibfield  {author} {\bibinfo {author} {\bibfnamefont {C.~Q.}\ \bibnamefont {Cook}}\ and\ \bibinfo {author} {\bibfnamefont {A.}~\bibnamefont {Lucas}},\ }\bibfield  {title} {\bibinfo {title} {Electron hydrodynamics with a polygonal fermi surface},\ }\href {https://doi.org/10.1103/PhysRevB.99.235148} {\bibfield  {journal} {\bibinfo  {journal} {Phys. Rev. B}\ }\textbf {\bibinfo {volume} {99}},\ \bibinfo {pages} {235148} (\bibinfo {year} {2019})}\BibitemShut {NoStop}%
\bibitem [{Note14()}]{Note14}%
  \BibitemOpen
  \bibinfo {note} {Strictly speaking, in proximity of a perfectly thermalized boundary (temperature fixed at a certain value and $\protect \bm {u}=\protect \bm {0}$), the temperature gradient emerging from the VHE is weakly space dependent, due to the coupling between the temperature gradient and the drift velocity in Eq.~(\ref {viscous_heat_U}). In practice, in a rectangular geometry of graphite with gradient applied around 70 K, the temperature gradient and drift velocity reach a practically space-independent value within about 1 $\mu m$ from the boundaries. Analytical details on the mechanisms determining this lengthscale can be found in Appendix G of Ref.~\cite {Simoncelli2020}, and numerical simulations in Fig. 4 of the same reference.}\BibitemShut {Stop}%
\bibitem [{Note15()}]{Note15}%
  \BibitemOpen
  \bibinfo {note} {For a well-behaved vector field, the circulation computed on a closed streamline $\protect \vec {\protect \mathcal {S}}$ is nonzero: $\protect \mathcal {C}=\DOTSI \ointop \ilimits@ _{\protect \mathcal {S}}\protect \bm {Q}^D{\cdot } d\protect \vec {\protect \mathcal {S}}\protect \neq 0$; therefore, one sees from Stokes' theorem that this implies nonzero vorticity: $\protect \mathcal {C}=\DOTSI \intop \ilimits@ _{\protect \mathcal {A}}(\nabla {\times } \protect \bm {Q}^D){\cdot }\protect \bm {m} d{\protect \mathcal {A}}\protect \neq 0$, where $\protect \bm {m}$ is the unit vector outward-normal to the surface $\protect \mathcal {A}$ enclosed by $\protect \vec {\protect \mathcal {S}}$.}\BibitemShut {Stop}%
\bibitem [{Note16()}]{Note16}%
  \BibitemOpen
  \bibinfo {note} {Specifically, in the region where temperature inversion is most pronounced in Fig.~\ref {fig:1_vortex}, one has $\beta \protect \frac {\partial T({\protect \bm {r}}, t)}{\partial r^y} {\sim } \eta ^{yyyy} \protect \frac {\partial ^2 u^y({\protect \bm {r}}, t)}{\partial r_y^2}$, a condition that permits temperature gradient and drift velocity to be aligned, thus temperature-gradient heat flux $\protect \bm {Q}^\delta =-\kappa _D\nabla T$ and drifting heat flux $\protect \bm {Q}^D=\alpha \protect \bm {u}$ to have opposite directions.}\BibitemShut {Stop}%
\bibitem [{\citenamefont {Hunter}(2007)}]{Hunter_2007}%
  \BibitemOpen
  \bibfield  {author} {\bibinfo {author} {\bibfnamefont {J.~D.}\ \bibnamefont {Hunter}},\ }\bibfield  {title} {\bibinfo {title} {Matplotlib: A 2d graphics environment},\ }\href {https://doi.org/10.1109/MCSE.2007.55} {\bibfield  {journal} {\bibinfo  {journal} {Computing in Science \& Engineering}\ }\textbf {\bibinfo {volume} {9}},\ \bibinfo {pages} {90} (\bibinfo {year} {2007})}\BibitemShut {NoStop}%
\bibitem [{\citenamefont {{Matplotlib~developers}}(2025)}]{streamlines_curve_level_streamfunction}%
  \BibitemOpen
  \bibfield  {author} {\bibinfo {author} {\bibnamefont {{Matplotlib~developers}}},\ }\href {https://github.com/matplotlib/matplotlib/issues/27261} {\bibinfo {title} {{Matplotlib} repository}} (\bibinfo {year} {2025}),\ \bibinfo {note} {\url{https://github.com/matplotlib/matplotlib/issues/27261}}\BibitemShut {NoStop}%
\bibitem [{\citenamefont {Péraud}\ and\ \citenamefont {Hadjiconstantinou}(2011)}]{peraud_efficient_2011}%
  \BibitemOpen
  \bibfield  {author} {\bibinfo {author} {\bibfnamefont {J.-P.~M.}\ \bibnamefont {Péraud}}\ and\ \bibinfo {author} {\bibfnamefont {N.~G.}\ \bibnamefont {Hadjiconstantinou}},\ }\bibfield  {title} {\bibinfo {title} {Efficient simulation of multidimensional phonon transport using energy-based variance-reduced {Monte} {Carlo} formulations},\ }\href {https://doi.org/10.1103/PhysRevB.84.205331} {\bibfield  {journal} {\bibinfo  {journal} {Physical Review B}\ }\textbf {\bibinfo {volume} {84}},\ \bibinfo {pages} {205331} (\bibinfo {year} {2011})}\BibitemShut {NoStop}%
\bibitem [{Note17()}]{Note17}%
  \BibitemOpen
  \bibinfo {note} {The VHE solution in Fig.~\ref {fig:1_vortex} was computed in less than two minutes on a laptop, while the corresponding LBTE solution in Fig.~\ref {fig:BTE_revision} took about 72 hours on one node (128 cores) of the Kelvin2 supercomputer from the Northern Ireland High Performance Computing center}\BibitemShut {NoStop}%
\bibitem [{\citenamefont {Paulatto}\ \emph {et~al.}(2013)\citenamefont {Paulatto}, \citenamefont {Mauri},\ and\ \citenamefont {Lazzeri}}]{paulatto2013anharmonic}%
  \BibitemOpen
  \bibfield  {author} {\bibinfo {author} {\bibfnamefont {L.}~\bibnamefont {Paulatto}}, \bibinfo {author} {\bibfnamefont {F.}~\bibnamefont {Mauri}},\ and\ \bibinfo {author} {\bibfnamefont {M.}~\bibnamefont {Lazzeri}},\ }\bibfield  {title} {\bibinfo {title} {{Anharmonic properties from a generalized third-order ab initio approach: Theory and applications to graphite and graphene}},\ }\href@noop {} {\bibfield  {journal} {\bibinfo  {journal} {Phys. Rev. B}\ }\textbf {\bibinfo {volume} {87}},\ \bibinfo {pages} {214303} (\bibinfo {year} {2013})}\BibitemShut {NoStop}%
\bibitem [{\citenamefont {Paulatto}\ \emph {et~al.}(2015)\citenamefont {Paulatto}, \citenamefont {Errea}, \citenamefont {Calandra},\ and\ \citenamefont {Mauri}}]{paulatto2015first}%
  \BibitemOpen
  \bibfield  {author} {\bibinfo {author} {\bibfnamefont {L.}~\bibnamefont {Paulatto}}, \bibinfo {author} {\bibfnamefont {I.}~\bibnamefont {Errea}}, \bibinfo {author} {\bibfnamefont {M.}~\bibnamefont {Calandra}},\ and\ \bibinfo {author} {\bibfnamefont {F.}~\bibnamefont {Mauri}},\ }\bibfield  {title} {\bibinfo {title} {First-principles calculations of phonon frequencies, lifetimes, and spectral functions from weak to strong anharmonicity: The example of palladium hydrides},\ }\href@noop {} {\bibfield  {journal} {\bibinfo  {journal} {Phys. Rev. B}\ }\textbf {\bibinfo {volume} {91}},\ \bibinfo {pages} {054304} (\bibinfo {year} {2015})}\BibitemShut {NoStop}%
\bibitem [{\citenamefont {Togo}(2023)}]{togo_first-principles_2023}%
  \BibitemOpen
  \bibfield  {author} {\bibinfo {author} {\bibfnamefont {A.}~\bibnamefont {Togo}},\ }\bibfield  {title} {\bibinfo {title} {First-principles {Phonon} {Calculations} with {Phonopy} and {Phono3py}},\ }\href {https://doi.org/10.7566/JPSJ.92.012001} {\bibfield  {journal} {\bibinfo  {journal} {Journal of the Physical Society of Japan}\ }\textbf {\bibinfo {volume} {92}},\ \bibinfo {pages} {012001} (\bibinfo {year} {2023})}\BibitemShut {NoStop}%
\bibitem [{Pho()}]{Phonopy_parser}%
  \BibitemOpen
  \href@noop {} {}\bibinfo {note} {\url{https://phonopy.github.io/phonopy/qe.html}}\BibitemShut {NoStop}%
\bibitem [{the(2023)}]{thermal2module}%
  \BibitemOpen
  \href@noop {} {\bibinfo {title} {Thermal2 module}} (\bibinfo {year} {2023}),\ \bibinfo {note} {\url{https://github.com/anharmonic/d3q/tree/main/thermal2}}\BibitemShut {NoStop}%
\bibitem [{\citenamefont {Li}\ \emph {et~al.}(2014)\citenamefont {Li}, \citenamefont {Carrete}, \citenamefont {Katcho},\ and\ \citenamefont {Mingo}}]{li2014shengbte}%
  \BibitemOpen
  \bibfield  {author} {\bibinfo {author} {\bibfnamefont {W.}~\bibnamefont {Li}}, \bibinfo {author} {\bibfnamefont {J.}~\bibnamefont {Carrete}}, \bibinfo {author} {\bibfnamefont {N.~A.}\ \bibnamefont {Katcho}},\ and\ \bibinfo {author} {\bibfnamefont {N.}~\bibnamefont {Mingo}},\ }\bibfield  {title} {\bibinfo {title} {{ShengBTE: A solver of the Boltzmann transport equation for phonons}},\ }\href@noop {} {\bibfield  {journal} {\bibinfo  {journal} {Comput. Phys. Commun.}\ }\textbf {\bibinfo {volume} {185}},\ \bibinfo {pages} {1747 } (\bibinfo {year} {2014})}\BibitemShut {NoStop}%
\bibitem [{\citenamefont {Tamura}(1983)}]{tamura_isotope_1983}%
  \BibitemOpen
  \bibfield  {author} {\bibinfo {author} {\bibfnamefont {S.-i.}\ \bibnamefont {Tamura}},\ }\bibfield  {title} {\bibinfo {title} {Isotope scattering of dispersive phonons in {Ge}},\ }\href {https://doi.org/10.1103/PhysRevB.27.858} {\bibfield  {journal} {\bibinfo  {journal} {Physical Review B}\ }\textbf {\bibinfo {volume} {27}},\ \bibinfo {pages} {858} (\bibinfo {year} {1983})}\BibitemShut {NoStop}%
\bibitem [{\citenamefont {openfoam.org}(2024)}]{greenshields_openfoam_2024}%
  \BibitemOpen
  \bibfield  {author} {\bibinfo {author} {\bibnamefont {openfoam.org}},\ }\href {https://doc.cfd.direct/openfoam/user-guide-v12/index/} {\bibinfo {title} {{OpenFOAM} v12 {User} {Guide}}} (\bibinfo {year} {2024}),\ \bibinfo {note} {\url{https://doc.cfd.direct/openfoam/user-guide-v12/boundaries}}\BibitemShut {NoStop}%
\bibitem [{\citenamefont {Tu}(2008)}]{Bump}%
  \BibitemOpen
  \bibfield  {author} {\bibinfo {author} {\bibfnamefont {L.~W.}\ \bibnamefont {Tu}},\ }\href@noop {} {\emph {\bibinfo {title} {An Introduction to Manifolds}}}\ (\bibinfo  {publisher} {Springer Science + Business Media, LLC},\ \bibinfo {year} {2008})\ pp.\ \bibinfo {pages} {127--130}\BibitemShut {NoStop}%
\bibitem [{\citenamefont {Phillips}(2019)}]{Sigmoid}%
  \BibitemOpen
  \bibfield  {author} {\bibinfo {author} {\bibfnamefont {F.}~\bibnamefont {Phillips}},\ }\href@noop {} {\bibinfo {title} {{SmootherStep: An improved sigmoidal interpolation function}}},\ \bibinfo {howpublished} {\url{https://resources.wolframcloud.com/FunctionRepository/resources/SmootherStep/}} (\bibinfo {year} {2019})\BibitemShut {NoStop}%
\bibitem [{\citenamefont {Carruthers}(1961)}]{carruthers_theory_1961}%
  \BibitemOpen
  \bibfield  {author} {\bibinfo {author} {\bibfnamefont {P.}~\bibnamefont {Carruthers}},\ }\bibfield  {title} {\bibinfo {title} {Theory of {Thermal} {Conductivity} of {Solids} at {Low} {Temperatures}},\ }\href {https://doi.org/10.1103/RevModPhys.33.92} {\bibfield  {journal} {\bibinfo  {journal} {Reviews of Modern Physics}\ }\textbf {\bibinfo {volume} {33}},\ \bibinfo {pages} {92} (\bibinfo {year} {1961})}\BibitemShut {NoStop}%
\bibitem [{\citenamefont {{Simoncelli, M. and Marzari, N. and Mauri, F.}}(2019)}]{simoncelli2019unified}%
  \BibitemOpen
  \bibfield  {author} {\bibinfo {author} {\bibnamefont {{Simoncelli, M. and Marzari, N. and Mauri, F.}}},\ }\bibfield  {title} {\bibinfo {title} {{Unified theory of thermal transport in crystals and glasses}},\ }\href@noop {} {\bibfield  {journal} {\bibinfo  {journal} {Nat. Phys.}\ }\textbf {\bibinfo {volume} {15}},\ \bibinfo {pages} {809} (\bibinfo {year} {2019})}\BibitemShut {NoStop}%
\bibitem [{\citenamefont {Simoncelli}\ \emph {et~al.}(2022)\citenamefont {Simoncelli}, \citenamefont {Marzari},\ and\ \citenamefont {Mauri}}]{simoncelli2021Wigner}%
  \BibitemOpen
  \bibfield  {author} {\bibinfo {author} {\bibfnamefont {M.}~\bibnamefont {Simoncelli}}, \bibinfo {author} {\bibfnamefont {N.}~\bibnamefont {Marzari}},\ and\ \bibinfo {author} {\bibfnamefont {F.}~\bibnamefont {Mauri}},\ }\bibfield  {title} {\bibinfo {title} {Wigner formulation of thermal transport in solids},\ }\href {https://link.aps.org/doi/10.1103/PhysRevX.12.041011} {\bibfield  {journal} {\bibinfo  {journal} {Phys. Rev. X}\ }\textbf {\bibinfo {volume} {12}},\ \bibinfo {pages} {041011} (\bibinfo {year} {2022})}\BibitemShut {NoStop}%
\bibitem [{\citenamefont {Caldarelli}\ \emph {et~al.}(2022)\citenamefont {Caldarelli}, \citenamefont {Simoncelli}, \citenamefont {Marzari}, \citenamefont {Mauri},\ and\ \citenamefont {Benfatto}}]{caldarelli_many-body_2022}%
  \BibitemOpen
  \bibfield  {author} {\bibinfo {author} {\bibfnamefont {G.}~\bibnamefont {Caldarelli}}, \bibinfo {author} {\bibfnamefont {M.}~\bibnamefont {Simoncelli}}, \bibinfo {author} {\bibfnamefont {N.}~\bibnamefont {Marzari}}, \bibinfo {author} {\bibfnamefont {F.}~\bibnamefont {Mauri}},\ and\ \bibinfo {author} {\bibfnamefont {L.}~\bibnamefont {Benfatto}},\ }\bibfield  {title} {\bibinfo {title} {Many-body {Green}'s function approach to lattice thermal transport},\ }\href {https://link.aps.org/doi/10.1103/PhysRevB.106.024312} {\bibfield  {journal} {\bibinfo  {journal} {Phys. Rev. B}\ }\textbf {\bibinfo {volume} {106}},\ \bibinfo {pages} {024312} (\bibinfo {year} {2022})}\BibitemShut {NoStop}%
\bibitem [{\citenamefont {Di~Lucente}\ \emph {et~al.}(2023)\citenamefont {Di~Lucente}, \citenamefont {Simoncelli},\ and\ \citenamefont {Marzari}}]{di_lucente_crossover_2023}%
  \BibitemOpen
  \bibfield  {author} {\bibinfo {author} {\bibfnamefont {E.}~\bibnamefont {Di~Lucente}}, \bibinfo {author} {\bibfnamefont {M.}~\bibnamefont {Simoncelli}},\ and\ \bibinfo {author} {\bibfnamefont {N.}~\bibnamefont {Marzari}},\ }\bibfield  {title} {\bibinfo {title} {Crossover from boltzmann to wigner thermal transport in thermoelectric skutterudites},\ }\href {https://doi.org/10.1103/PhysRevResearch.5.033125} {\bibfield  {journal} {\bibinfo  {journal} {Phys. Rev. Res.}\ }\textbf {\bibinfo {volume} {5}},\ \bibinfo {pages} {033125} (\bibinfo {year} {2023})}\BibitemShut {NoStop}%
\bibitem [{\citenamefont {Giannozzi}\ \emph {et~al.}(2009)\citenamefont {Giannozzi}, \citenamefont {Baroni}, \citenamefont {Bonini}, \citenamefont {Calandra}, \citenamefont {Car}, \citenamefont {Cavazzoni}, \citenamefont {Ceresoli}, \citenamefont {Chiarotti}, \citenamefont {Cococcioni}, \citenamefont {Dabo} \emph {et~al.}}]{giannozzi2009quantum}%
  \BibitemOpen
  \bibfield  {author} {\bibinfo {author} {\bibfnamefont {P.}~\bibnamefont {Giannozzi}}, \bibinfo {author} {\bibfnamefont {S.}~\bibnamefont {Baroni}}, \bibinfo {author} {\bibfnamefont {N.}~\bibnamefont {Bonini}}, \bibinfo {author} {\bibfnamefont {M.}~\bibnamefont {Calandra}}, \bibinfo {author} {\bibfnamefont {R.}~\bibnamefont {Car}}, \bibinfo {author} {\bibfnamefont {C.}~\bibnamefont {Cavazzoni}}, \bibinfo {author} {\bibfnamefont {D.}~\bibnamefont {Ceresoli}}, \bibinfo {author} {\bibfnamefont {G.~L.}\ \bibnamefont {Chiarotti}}, \bibinfo {author} {\bibfnamefont {M.}~\bibnamefont {Cococcioni}}, \bibinfo {author} {\bibfnamefont {I.}~\bibnamefont {Dabo}}, \emph {et~al.},\ }\bibfield  {title} {\bibinfo {title} {{QUANTUM ESPRESSO: a modular and open-source software project for quantum simulations of materials}},\ }\href@noop {} {\bibfield  {journal} {\bibinfo  {journal} {J. Phys. Condens. Matter}\ }\textbf {\bibinfo {volume} {21}},\ \bibinfo {pages} {395502} (\bibinfo {year} {2009})}\BibitemShut {NoStop}%
\bibitem [{\citenamefont {Giannozzi}\ \emph {et~al.}(2017)\citenamefont {Giannozzi}, \citenamefont {Andreussi}, \citenamefont {Brumme}, \citenamefont {Bunau}, \citenamefont {Nardelli}, \citenamefont {Calandra}, \citenamefont {Car}, \citenamefont {Cavazzoni}, \citenamefont {Ceresoli}, \citenamefont {Cococcioni}, \citenamefont {Colonna}, \citenamefont {Carnimeo}, \citenamefont {Corso}, \citenamefont {de~Gironcoli}, \citenamefont {Delugas}, \citenamefont {Jr}, \citenamefont {Ferretti}, \citenamefont {Floris}, \citenamefont {Fratesi}, \citenamefont {Fugallo}, \citenamefont {Gebauer}, \citenamefont {Gerstmann}, \citenamefont {Giustino}, \citenamefont {Gorni}, \citenamefont {Jia}, \citenamefont {Kawamura}, \citenamefont {Ko}, \citenamefont {Kokalj}, \citenamefont {K{\"u}{\c c}{\"u}kbenli}, \citenamefont {Lazzeri}, \citenamefont {Marsili}, \citenamefont {Marzari}, \citenamefont {Mauri}, \citenamefont {Nguyen}, \citenamefont {Nguyen}, \citenamefont {de-la Roza}, \citenamefont {Paulatto}, \citenamefont {Ponc{\'e}}, \citenamefont {Rocca}, \citenamefont {Sabatini}, \citenamefont {Santra}, \citenamefont {Schlipf}, \citenamefont {Seitsonen}, \citenamefont {Smogunov}, \citenamefont {Timrov}, \citenamefont {Thonhauser}, \citenamefont {Umari}, \citenamefont {Vast}, \citenamefont {Wu},\ and\ \citenamefont {Baroni}}]{giannozzi2017advanced}%
  \BibitemOpen
  \bibfield  {author} {\bibinfo {author} {\bibfnamefont {P.}~\bibnamefont {Giannozzi}}, \bibinfo {author} {\bibfnamefont {O.}~\bibnamefont {Andreussi}}, \bibinfo {author} {\bibfnamefont {T.}~\bibnamefont {Brumme}}, \bibinfo {author} {\bibfnamefont {O.}~\bibnamefont {Bunau}}, \bibinfo {author} {\bibfnamefont {M.~B.}\ \bibnamefont {Nardelli}}, \bibinfo {author} {\bibfnamefont {M.}~\bibnamefont {Calandra}}, \bibinfo {author} {\bibfnamefont {R.}~\bibnamefont {Car}}, \bibinfo {author} {\bibfnamefont {C.}~\bibnamefont {Cavazzoni}}, \bibinfo {author} {\bibfnamefont {D.}~\bibnamefont {Ceresoli}}, \bibinfo {author} {\bibfnamefont {M.}~\bibnamefont {Cococcioni}}, \bibinfo {author} {\bibfnamefont {N.}~\bibnamefont {Colonna}}, \bibinfo {author} {\bibfnamefont {I.}~\bibnamefont {Carnimeo}}, \bibinfo {author} {\bibfnamefont {A.~D.}\ \bibnamefont {Corso}}, \bibinfo {author} {\bibfnamefont {S.}~\bibnamefont {de~Gironcoli}}, \bibinfo {author} {\bibfnamefont {P.}~\bibnamefont {Delugas}}, \bibinfo {author} {\bibfnamefont {R.~A.~D.}\ \bibnamefont {Jr}}, \bibinfo {author} {\bibfnamefont {A.}~\bibnamefont {Ferretti}}, \bibinfo {author} {\bibfnamefont {A.}~\bibnamefont {Floris}}, \bibinfo {author} {\bibfnamefont {G.}~\bibnamefont {Fratesi}}, \bibinfo {author} {\bibfnamefont {G.}~\bibnamefont {Fugallo}}, \bibinfo {author} {\bibfnamefont {R.}~\bibnamefont {Gebauer}}, \bibinfo {author} {\bibfnamefont {U.}~\bibnamefont {Gerstmann}}, \bibinfo {author} {\bibfnamefont {F.}~\bibnamefont {Giustino}}, \bibinfo {author} {\bibfnamefont {T.}~\bibnamefont {Gorni}}, \bibinfo {author} {\bibfnamefont {J.}~\bibnamefont {Jia}}, \bibinfo {author} {\bibfnamefont {M.}~\bibnamefont {Kawamura}}, \bibinfo {author} {\bibfnamefont {H.-Y.}\ \bibnamefont {Ko}}, \bibinfo {author} {\bibfnamefont {A.}~\bibnamefont {Kokalj}}, \bibinfo {author} {\bibfnamefont {E.}~\bibnamefont {K{\"u}{\c c}{\"u}kbenli}}, \bibinfo {author} {\bibfnamefont {M.}~\bibnamefont {Lazzeri}}, \bibinfo {author} {\bibfnamefont {M.}~\bibnamefont {Marsili}}, \bibinfo {author} {\bibfnamefont {N.}~\bibnamefont {Marzari}}, \bibinfo {author} {\bibfnamefont {F.}~\bibnamefont {Mauri}}, \bibinfo {author} {\bibfnamefont {N.~L.}\ \bibnamefont {Nguyen}}, \bibinfo {author} {\bibfnamefont {H.-V.}\ \bibnamefont {Nguyen}}, \bibinfo {author} {\bibfnamefont {A.~O.}\ \bibnamefont {de-la Roza}}, \bibinfo {author} {\bibfnamefont {L.}~\bibnamefont {Paulatto}}, \bibinfo {author} {\bibfnamefont {S.}~\bibnamefont {Ponc{\'e}}}, \bibinfo {author} {\bibfnamefont {D.}~\bibnamefont {Rocca}}, \bibinfo {author} {\bibfnamefont {R.}~\bibnamefont {Sabatini}}, \bibinfo {author} {\bibfnamefont {B.}~\bibnamefont {Santra}}, \bibinfo {author} {\bibfnamefont {M.}~\bibnamefont {Schlipf}}, \bibinfo {author} {\bibfnamefont {A.~P.}\ \bibnamefont {Seitsonen}}, \bibinfo {author} {\bibfnamefont {A.}~\bibnamefont {Smogunov}}, \bibinfo {author} {\bibfnamefont {I.}~\bibnamefont {Timrov}}, \bibinfo {author} {\bibfnamefont {T.}~\bibnamefont {Thonhauser}}, \bibinfo {author} {\bibfnamefont {P.}~\bibnamefont {Umari}}, \bibinfo {author} {\bibfnamefont {N.}~\bibnamefont {Vast}}, \bibinfo {author} {\bibfnamefont {X.}~\bibnamefont {Wu}},\ and\ \bibinfo {author} {\bibfnamefont {S.}~\bibnamefont {Baroni}},\ }\bibfield  {title} {\bibinfo {title} {{Advanced capabilities for materials modelling with Quantum ESPRESSO}},\ }\href@noop {} {\bibfield  {journal} {\bibinfo  {journal} {J. Phys. Condens. Matter}\ }\textbf {\bibinfo {volume} {29}},\ \bibinfo {pages} {465901} (\bibinfo {year} {2017})}\BibitemShut {NoStop}%
\bibitem [{\citenamefont {Cusc\'o}\ \emph {et~al.}(2018)\citenamefont {Cusc\'o}, \citenamefont {Art\'us}, \citenamefont {Edgar}, \citenamefont {Liu}, \citenamefont {Cassabois},\ and\ \citenamefont {Gil}}]{PhysRevB.97.155435}%
  \BibitemOpen
  \bibfield  {author} {\bibinfo {author} {\bibfnamefont {R.}~\bibnamefont {Cusc\'o}}, \bibinfo {author} {\bibfnamefont {L.}~\bibnamefont {Art\'us}}, \bibinfo {author} {\bibfnamefont {J.~H.}\ \bibnamefont {Edgar}}, \bibinfo {author} {\bibfnamefont {S.}~\bibnamefont {Liu}}, \bibinfo {author} {\bibfnamefont {G.}~\bibnamefont {Cassabois}},\ and\ \bibinfo {author} {\bibfnamefont {B.}~\bibnamefont {Gil}},\ }\bibfield  {title} {\bibinfo {title} {Isotopic effects on phonon anharmonicity in layered van der waals crystals: Isotopically pure hexagonal boron nitride},\ }\href@noop {} {\bibfield  {journal} {\bibinfo  {journal} {Phys. Rev. B}\ }\textbf {\bibinfo {volume} {97}},\ \bibinfo {pages} {155435} (\bibinfo {year} {2018})}\BibitemShut {NoStop}%
\bibitem [{\citenamefont {Van~Setten}\ \emph {et~al.}(2018)\citenamefont {Van~Setten}, \citenamefont {Giantomassi}, \citenamefont {Bousquet}, \citenamefont {Verstraete}, \citenamefont {Hamann}, \citenamefont {Gonze},\ and\ \citenamefont {Rignanese}}]{van2018pseudodojo}%
  \BibitemOpen
  \bibfield  {author} {\bibinfo {author} {\bibfnamefont {M.}~\bibnamefont {Van~Setten}}, \bibinfo {author} {\bibfnamefont {M.}~\bibnamefont {Giantomassi}}, \bibinfo {author} {\bibfnamefont {E.}~\bibnamefont {Bousquet}}, \bibinfo {author} {\bibfnamefont {M.~J.}\ \bibnamefont {Verstraete}}, \bibinfo {author} {\bibfnamefont {D.~R.}\ \bibnamefont {Hamann}}, \bibinfo {author} {\bibfnamefont {X.}~\bibnamefont {Gonze}},\ and\ \bibinfo {author} {\bibfnamefont {G.-M.}\ \bibnamefont {Rignanese}},\ }\bibfield  {title} {\bibinfo {title} {The pseudodojo: Training and grading a 85 element optimized norm-conserving pseudopotential table},\ }\href@noop {} {\bibfield  {journal} {\bibinfo  {journal} {Comput. Phys. Commun.}\ }\textbf {\bibinfo {volume} {226}},\ \bibinfo {pages} {39} (\bibinfo {year} {2018})}\BibitemShut {NoStop}%
\bibitem [{\citenamefont {Kurakevych}\ and\ \citenamefont {Solozhenko}(2007)}]{Kurakevych:sq3084}%
  \BibitemOpen
  \bibfield  {author} {\bibinfo {author} {\bibfnamefont {O.~O.}\ \bibnamefont {Kurakevych}}\ and\ \bibinfo {author} {\bibfnamefont {V.~L.}\ \bibnamefont {Solozhenko}},\ }\bibfield  {title} {\bibinfo {title} {{Rhombohedral boron subnitride, B${\sb 13}$N${\sb 2}$, by X-ray powder diffraction}},\ }\href@noop {} {\bibfield  {journal} {\bibinfo  {journal} {Acta Crystallogr., Sect. C}\ }\textbf {\bibinfo {volume} {63}},\ \bibinfo {pages} {i80} (\bibinfo {year} {2007})}\BibitemShut {NoStop}%
\bibitem [{\citenamefont {Gra{\v{z}}ulis}\ \emph {et~al.}(2009)\citenamefont {Gra{\v{z}}ulis}, \citenamefont {Chateigner}, \citenamefont {Downs}, \citenamefont {Yokochi}, \citenamefont {Quir{\'{o}}s}, \citenamefont {Lutterotti}, \citenamefont {Manakova}, \citenamefont {Butkus}, \citenamefont {Moeck},\ and\ \citenamefont {Le~Bail}}]{COD_database}%
  \BibitemOpen
  \bibfield  {author} {\bibinfo {author} {\bibfnamefont {S.}~\bibnamefont {Gra{\v{z}}ulis}}, \bibinfo {author} {\bibfnamefont {D.}~\bibnamefont {Chateigner}}, \bibinfo {author} {\bibfnamefont {R.~T.}\ \bibnamefont {Downs}}, \bibinfo {author} {\bibfnamefont {A.~F.~T.}\ \bibnamefont {Yokochi}}, \bibinfo {author} {\bibfnamefont {M.}~\bibnamefont {Quir{\'{o}}s}}, \bibinfo {author} {\bibfnamefont {L.}~\bibnamefont {Lutterotti}}, \bibinfo {author} {\bibfnamefont {E.}~\bibnamefont {Manakova}}, \bibinfo {author} {\bibfnamefont {J.}~\bibnamefont {Butkus}}, \bibinfo {author} {\bibfnamefont {P.}~\bibnamefont {Moeck}},\ and\ \bibinfo {author} {\bibfnamefont {A.}~\bibnamefont {Le~Bail}},\ }\bibfield  {title} {\bibinfo {title} {{Crystallography Open Database {--} an open-access collection of crystal structures}},\ }\href@noop {} {\bibfield  {journal} {\bibinfo  {journal} {J. Appl. Crystallogr.}\ }\textbf {\bibinfo {volume} {42}},\ \bibinfo {pages} {726} (\bibinfo {year} {2009})}\BibitemShut {NoStop}%
\bibitem [{\citenamefont {Baroni}\ \emph {et~al.}(2001)\citenamefont {Baroni}, \citenamefont {de~Gironcoli}, \citenamefont {Dal~Corso},\ and\ \citenamefont {Giannozzi}}]{RevModPhys.73.515}%
  \BibitemOpen
  \bibfield  {author} {\bibinfo {author} {\bibfnamefont {S.}~\bibnamefont {Baroni}}, \bibinfo {author} {\bibfnamefont {S.}~\bibnamefont {de~Gironcoli}}, \bibinfo {author} {\bibfnamefont {A.}~\bibnamefont {Dal~Corso}},\ and\ \bibinfo {author} {\bibfnamefont {P.}~\bibnamefont {Giannozzi}},\ }\bibfield  {title} {\bibinfo {title} {{Phonons and related crystal properties from density-functional perturbation theory}},\ }\href@noop {} {\bibfield  {journal} {\bibinfo  {journal} {Rev. Mod. Phys.}\ }\textbf {\bibinfo {volume} {73}},\ \bibinfo {pages} {515} (\bibinfo {year} {2001})}\BibitemShut {NoStop}%
\bibitem [{\citenamefont {Balasubramanian}\ \emph {et~al.}(1987)\citenamefont {Balasubramanian}, \citenamefont {Hayot},\ and\ \citenamefont {Saam}}]{balasubramanian_darcys_1987}%
  \BibitemOpen
  \bibfield  {author} {\bibinfo {author} {\bibfnamefont {K.}~\bibnamefont {Balasubramanian}}, \bibinfo {author} {\bibfnamefont {F.}~\bibnamefont {Hayot}},\ and\ \bibinfo {author} {\bibfnamefont {W.~F.}\ \bibnamefont {Saam}},\ }\bibfield  {title} {\bibinfo {title} {Darcy's law from lattice-gas hydrodynamics},\ }\href {https://link.aps.org/doi/10.1103/PhysRevA.36.2248} {\bibfield  {journal} {\bibinfo  {journal} {Phys. Rev. A}\ }\textbf {\bibinfo {volume} {36}},\ \bibinfo {pages} {2248} (\bibinfo {year} {1987})}\BibitemShut {NoStop}%
\bibitem [{\citenamefont {Dardis}\ and\ \citenamefont {McCloskey}(1998)}]{dardis_lattice_1998}%
  \BibitemOpen
  \bibfield  {author} {\bibinfo {author} {\bibfnamefont {O.}~\bibnamefont {Dardis}}\ and\ \bibinfo {author} {\bibfnamefont {J.}~\bibnamefont {McCloskey}},\ }\bibfield  {title} {\bibinfo {title} {Lattice {Boltzmann} scheme with real numbered solid density for the simulation of flow in porous media},\ }\href {https://link.aps.org/doi/10.1103/PhysRevE.57.4834} {\bibfield  {journal} {\bibinfo  {journal} {Phys. Rev. E}\ }\textbf {\bibinfo {volume} {57}},\ \bibinfo {pages} {4834} (\bibinfo {year} {1998})}\BibitemShut {NoStop}%
\bibitem [{\citenamefont {Bresch}\ and\ \citenamefont {Desjardins}(2003)}]{bresch_existence_2003}%
  \BibitemOpen
  \bibfield  {author} {\bibinfo {author} {\bibfnamefont {D.}~\bibnamefont {Bresch}}\ and\ \bibinfo {author} {\bibfnamefont {B.}~\bibnamefont {Desjardins}},\ }\bibfield  {title} {\bibinfo {title} {Existence of {Global} {Weak} {Solutions} for a {2D} {Viscous} {Shallow} {Water} {Equations} and {Convergence} to the {Quasi}-{Geostrophic} {Model}},\ }\href {https://doi.org/10.1007/s00220-003-0859-8} {\bibfield  {journal} {\bibinfo  {journal} {Commun. Math. Phys}\ }\textbf {\bibinfo {volume} {238}},\ \bibinfo {pages} {211} (\bibinfo {year} {2003})}\BibitemShut {NoStop}%
\bibitem [{\citenamefont {Cai}\ and\ \citenamefont {Jiu}(2008)}]{cai_weak_2008}%
  \BibitemOpen
  \bibfield  {author} {\bibinfo {author} {\bibfnamefont {X.}~\bibnamefont {Cai}}\ and\ \bibinfo {author} {\bibfnamefont {Q.}~\bibnamefont {Jiu}},\ }\bibfield  {title} {\bibinfo {title} {Weak and strong solutions for the incompressible {Navier}–{Stokes} equations with damping},\ }\href {https://www.sciencedirect.com/science/article/pii/S0022247X08000553} {\bibfield  {journal} {\bibinfo  {journal} {J. Math. Anal.}\ }\textbf {\bibinfo {volume} {343}},\ \bibinfo {pages} {799} (\bibinfo {year} {2008})}\BibitemShut {NoStop}%
\bibitem [{\citenamefont {Zhang}\ \emph {et~al.}(2011)\citenamefont {Zhang}, \citenamefont {Wu},\ and\ \citenamefont {Lu}}]{zhang_uniqueness_2011}%
  \BibitemOpen
  \bibfield  {author} {\bibinfo {author} {\bibfnamefont {Z.}~\bibnamefont {Zhang}}, \bibinfo {author} {\bibfnamefont {X.}~\bibnamefont {Wu}},\ and\ \bibinfo {author} {\bibfnamefont {M.}~\bibnamefont {Lu}},\ }\bibfield  {title} {\bibinfo {title} {On the uniqueness of strong solution to the incompressible {Navier}--{Stokes} equations with damping},\ }\href@noop {} {\bibfield  {journal} {\bibinfo  {journal} {J. Math. Anal.}\ }\textbf {\bibinfo {volume} {377}},\ \bibinfo {pages} {414} (\bibinfo {year} {2011})}\BibitemShut {NoStop}%
\bibitem [{\citenamefont {Brenner}(2005)}]{brenner_navierstokes_2005}%
  \BibitemOpen
  \bibfield  {author} {\bibinfo {author} {\bibfnamefont {H.}~\bibnamefont {Brenner}},\ }\bibfield  {title} {\bibinfo {title} {Navier–{Stokes} revisited},\ }\href {https://www.sciencedirect.com/science/article/pii/S037843710401324X} {\bibfield  {journal} {\bibinfo  {journal} {Physica A}\ }\textbf {\bibinfo {volume} {349}},\ \bibinfo {pages} {60} (\bibinfo {year} {2005})}\BibitemShut {NoStop}%
\bibitem [{\citenamefont {Schwarz}\ \emph {et~al.}(2023)\citenamefont {Schwarz}, \citenamefont {Axelsson}, \citenamefont {Anheuer}, \citenamefont {Richter}, \citenamefont {Adam}, \citenamefont {Heinrich},\ and\ \citenamefont {Schwarze}}]{schwarz_openfoam_2023}%
  \BibitemOpen
  \bibfield  {author} {\bibinfo {author} {\bibfnamefont {J.}~\bibnamefont {Schwarz}}, \bibinfo {author} {\bibfnamefont {K.}~\bibnamefont {Axelsson}}, \bibinfo {author} {\bibfnamefont {D.}~\bibnamefont {Anheuer}}, \bibinfo {author} {\bibfnamefont {M.}~\bibnamefont {Richter}}, \bibinfo {author} {\bibfnamefont {J.}~\bibnamefont {Adam}}, \bibinfo {author} {\bibfnamefont {M.}~\bibnamefont {Heinrich}},\ and\ \bibinfo {author} {\bibfnamefont {R.}~\bibnamefont {Schwarze}},\ }\bibfield  {title} {\bibinfo {title} {An {OpenFOAM} solver for the extended {Navier}–{Stokes} equations},\ }\href {https://www.sciencedirect.com/science/article/pii/S2352711023000742} {\bibfield  {journal} {\bibinfo  {journal} {SoftwareX}\ }\textbf {\bibinfo {volume} {22}},\ \bibinfo {pages} {101378} (\bibinfo {year} {2023})}\BibitemShut {NoStop}%
\bibitem [{\citenamefont {Maurer}\ \emph {et~al.}(2003)\citenamefont {Maurer}, \citenamefont {Tabeling}, \citenamefont {Joseph},\ and\ \citenamefont {Willaime}}]{maurer_second-order_2003}%
  \BibitemOpen
  \bibfield  {author} {\bibinfo {author} {\bibfnamefont {J.}~\bibnamefont {Maurer}}, \bibinfo {author} {\bibfnamefont {P.}~\bibnamefont {Tabeling}}, \bibinfo {author} {\bibfnamefont {P.}~\bibnamefont {Joseph}},\ and\ \bibinfo {author} {\bibfnamefont {H.}~\bibnamefont {Willaime}},\ }\bibfield  {title} {\bibinfo {title} {Second-order slip laws in microchannels for helium and nitrogen},\ }\href {https://doi.org/10.1063/1.1599355} {\bibfield  {journal} {\bibinfo  {journal} {Phys. Fluids}\ }\textbf {\bibinfo {volume} {15}},\ \bibinfo {pages} {2613} (\bibinfo {year} {2003})}\BibitemShut {NoStop}%
\bibitem [{\citenamefont {Dongari}\ \emph {et~al.}(2009)\citenamefont {Dongari}, \citenamefont {Sharma~IITK},\ and\ \citenamefont {Durst}}]{dongari_pressure-driven_2009}%
  \BibitemOpen
  \bibfield  {author} {\bibinfo {author} {\bibfnamefont {N.}~\bibnamefont {Dongari}}, \bibinfo {author} {\bibfnamefont {A.}~\bibnamefont {Sharma~IITK}},\ and\ \bibinfo {author} {\bibfnamefont {F.}~\bibnamefont {Durst}},\ }\bibfield  {title} {\bibinfo {title} {Pressure-driven diffusive gas flows in micro-channels: {From} the {Knudsen} to the continuum regimes},\ }\href {https://doi.org/10.1007/s10404-008-0344-y} {\bibfield  {journal} {\bibinfo  {journal} {Microfluid. Nanofluid.}\ }\textbf {\bibinfo {volume} {6}},\ \bibinfo {pages} {679} (\bibinfo {year} {2009})}\BibitemShut {NoStop}%
\bibitem [{\citenamefont {Cattaneo}(1958)}]{cattaneo1958form}%
  \BibitemOpen
  \bibfield  {author} {\bibinfo {author} {\bibfnamefont {C.}~\bibnamefont {Cattaneo}},\ }\bibfield  {title} {\bibinfo {title} {A form of heat-conduction equations which eliminates the paradox of instantaneous propagation},\ }\href@noop {} {\bibfield  {journal} {\bibinfo  {journal} {Comptes Rendus}\ }\textbf {\bibinfo {volume} {247}},\ \bibinfo {pages} {431} (\bibinfo {year} {1958})}\BibitemShut {NoStop}%
\bibitem [{\citenamefont {Simoncelli}(2021)}]{SimoncelliPhD}%
  \BibitemOpen
  \bibfield  {author} {\bibinfo {author} {\bibfnamefont {M.}~\bibnamefont {Simoncelli}},\ }\emph {\bibinfo {title} {Thermal transport beyond Fourier, and beyond Boltzmann}},\ \href@noop {} {Ph.D. thesis},\ \bibinfo  {school} {École polytechnique fédérale de Lausanne} (\bibinfo {year} {2021})\BibitemShut {NoStop}%
\end{thebibliography}
\end{document}